\documentclass[12pt]{report}

\usepackage{epsfig}
\usepackage{cite}
\usepackage{graphicx}
\usepackage{rotate}
\usepackage{latexsym}
\usepackage{amssymb}
\usepackage{psfrag}

\newcommand{\lsi}{\,\raisebox{-0.13cm}{$\stackrel{\textstyle<}{\textstyle\sim}$}\,}
\newcommand{\gsi}{\,\raisebox{-0.13cm}{$\stackrel{\textstyle>}{\textstyle\sim}$}\,}
\newcommand{\beq}{\begin{equation}}
\newcommand{\eeq}{\end{equation}}
\newcommand{\be}{\begin{equation}}
\newcommand{\ee}{\end{equation}}

\begin{document}


\pagenumbering{roman}
\pagestyle{empty}

\begin{center}
\vspace*{20mm}

{\huge
{\bf Alternative Approaches} \\ \vspace{0.25cm}
{\bf to Dark Matter Puzzle}
}

\vspace{10mm}

by

\vspace{10mm}

{\LARGE \emph{Gabrijela Zaharija\v s}}

\vspace{20mm}

\textsc{A dissertation submitted in partial fulfillment}

\textsc{of the requirements for the degree of}

\textsc{Doctor of Philosophy}

\textsc{Department of Physics}

\textsc{New York University}


\vspace{0.5cm}
\begin{Large}
\textsl{September}, 2005.
\end{Large}

\end{center}

\vspace{20mm}

\hspace{70mm} \begin{minipage}[t]{2in}
\center{ -------------------------------- }
\center{Prof. G. R. ~Farrar}
\end{minipage}

\setcounter{page}{1}





\newpage

\pagestyle{plain}
\thispagestyle{empty}


\vspace*{5mm}

\begin{flushright}

\vspace*{70mm}


\end{flushright}

\newpage

\begin{center}
\section*{Acknowledgments}
\addcontentsline{toc}{section}{Acknowledgments}
\end{center}

I would like to thank my advisor Professor G.~R.~Farrar for the skills I learned, for the support and the patience with my written English. I am grateful to Slava, Francesco, Seba for helping me when it was needed. And, thank you Emi.

\newpage




\begin{center}
\section*{Abstract}
\end{center}
\vspace{4mm}
\addcontentsline{toc}{section}{Abstract}

In this thesis we study the dark matter problem with particular reference to a candidate particle within the Standard Model: the $H$ dibaryon. We consider as well a scenario which aims to connect the dark matter origin to the Baryon Asymmetry of the Universe, studying the examples of $H$ and of a Beyond-the-Standard-Model particle $X$.
 
Strongly attractive color forces in the flavor singlet channel may lead to a tightly bound and compact $H$ dibaryon. We find that the observation of $\Lambda$ decays from doubly-strange hypernuclei puts a constraint on the $H$ wavefunction which is plausibly satisfied. In this case the $H$ is very long-lived as we calculate and an absolutely stable $H$ is not excluded. We also show that an $H$ or another compact, flavor singlet hadron is unlikely to bind to nuclei, so that experimental bounds on exotic isotopes do not exclude their existence. Remarkably, the $H$ appears to evade other experimental constraints as well, when account is taken of its expected compact spatial wavefunction.

In order to check whether the $H$ is a viable DM candidate, we consider experiments sensitive to light particles. Taking into account the dark matter interaction in the crust above underground detectors we find a window in the exclusion limits in the micro-barn, $m\lsi 2.4$ GeV, range. Remarkably, this coincides with the range expected for the tightly bound $H$. Having these constraints in mind we conclude that the $H$ is a good DM candidate, but its production with sufficient abundance in the Early Universe is challenging.

Finally, we present a scenario in which dark matter carries (anti-)baryon number $B_X$ and which offers a mechanism to generate the baryon asymmetry observed in the Universe. If $\sigma^{\rm annih}_{\bar{X}} < \sigma^{\rm annih}_{X}$, the $\bar{ X}$'s freeze out at a higher temperature and have a larger relic density than $X$'s. If $m_X \lsi  4.5 \,B_X$ GeV and the annihilation cross sections differ by $\mathcal{O}(10\%)$ or more,  this type of scenario naturally explains the observed $\Omega_{DM} \simeq 5\, \Omega_b$. Two examples are given, one involving the $H$ and the other invoking an hypothetical beyond the Standard Model candidate $X$.

\tableofcontents

\newpage
\addcontentsline{toc}{section}{List of Figures}
\listoffigures

\newpage
\addcontentsline{toc}{section}{List of Tables}
\listoftables

\newpage
\addcontentsline{toc}{section}{List of Appendices}
\quad~
\vspace{2.5cm}
\newline\quad
\hspace{-1cm}
{\Huge\bf List of Appendices}
\newline
\vspace{0.3cm}

\contentsline {chapter}{\numberline {A}{\ignorespaces MC simulation for DM heavier than $10$ Gev}}{106}
\contentsline {chapter}{\numberline {B}{\ignorespaces Relative probability for scattering from different types of nuclei}}{108}


\newpage
\pagenumbering{arabic}

\chapter{Dark Matter and the Baryon Asymmetry of the Universe} \label{intro}

The existence of Dark Matter (DM) is, today, well established. The name ``Dark'' derives from the fact that it is non-luminous and non-absorbing; it can be detected (so far) only through its gravitational interaction with ordinary matter. One of the first evidence for the presence of DM was the 1933 measurement by F.~Zwicky of the velocities of galaxies which are part of the gravitationally bound COMA cluster, \cite{Zwicky:1933gu}. Zwicky found that galaxies are moving much faster than one would expect if they only felt the gravitational attraction from visible nearby objects. Nevertheless, the existence of dark matter was not firmly established until the 1970's when the measurement of the rotational velocity of stars and gas orbiting at a distance $r$ from the galactic center was performed. The velocity at a distance $r$ scales as $v\sim\sqrt{M(r)/r}$, where $M(r)$ is mass enclosed by the orbit. If measurement is performed outside the visible part of galaxy, one would expect $M(r)\sim\rm{const.}$, or $v\sim 1/\sqrt{r}$. Instead, observations show that $v\sim\rm{const.}$, implying the existence of a dark mass, with radial dependence $M\sim r$ or $\rho\sim 1/r^2$, assuming spherical symmetry. The existence of dark mass is probed on different scales: velocities of stars, gas clouds, globular clusters, or, as we have seen, entire galaxies, are larger than one would predict based on the gravitational potential inferred from the observed, luminous mass. 

More recent methods for direct detection of DM include measurements of X-ray temperature of the hot gas in galaxy clusters, which is proportional to the gravitational potential field and therefore to the mass of the cluster, and observations of the gravitational lensing of galaxies caused by the mass of a cluster in the foreground. 

On a more theoretical basis, the presence of dark matter allows to relate the anisotropies of the cosmic microwave background (CMB) and the structures observed in galaxy and Lyman-$\alpha$ surveys to a common primordial origin in the framework of the inflationary model. The currently most accurate determination of DM and baryonic energy densities comes from global fits of cosmological parameters to these observations. For instance, using measurements of the CMB and the spatial distribution of galaxies at large scales, \cite{Tegmark:2003ud}, one gets the following constraints on the ratio of measured energy density to the critical energy density $\rho_{cr}=3H^2 _0/8\pi G_N$, $\Omega _{DM}=\rho_{DM}/\rho_{cr}$ and $\Omega _{b}=\rho_{b}/\rho_{cr}$:
\begin{eqnarray} \label{Omega}
\Omega_{DM}h^2 &=& 0.1222\pm 0.009,\nonumber\\ 
\Omega_b h^2&=& 0.0232\pm 0.0013,
\end{eqnarray}
where $h=0.73\pm 0.03$ is the Hubble constant in units of $100$ km s$^{-1}$Mpc$^{-1}$, $H_0=h~100$ km s$^{-1}$ Mpc$^{-1}$. These two numbers are surprisingly similar, 
\beq \label{ratio}
\Omega _{DM}/\Omega _{b}=5.27\pm0.49 , 
\eeq
even though in conventional theories there is no reason for them to be so close: they could differ by many orders of magnitude. In our work we will explore the idea that these two numbers might originate from the same physical process, thereby making the value of their ratio natural.  \\

The nature of the dark matter particle is an unsolved problem. Candidates for dark matter must satisfy several, basic, conditions: they should have very weak interaction with the electromagnetic radiation, they must have a relic density that corresponds to the observed DM energy density, eq.~(\ref{Omega}), and they should have a lifetime longer then the lifetime of the universe. They should also be neutral particles. The models of structure formation based on the inflation scenario prefer the so called ``cold''  dark matter, {\it i.e.} dark matter particles which are non-relativistic at the time of galaxy formation. In these models dark matter fluctuations, caused by primordial fluctuations in the inflaton field, are responsible for growth of the structures observed today. Baryons follow DM fluctuations falling into their gravitational wells where they form astrophysical objects. A relativistic dark matter would have a large {\it free-streaming} length, below which no structure could form while cold dark matter has a free-streaming length small enough to be irrelevant for structure formation. 

A particle candidate for DM is believed not to be provided by the Standard Model (SM). Neutrinos have long been believed to be good SM candidates but based on recently derived constraints on the neutrino mass, it has been realized that they cannot provide enough energy density to be the only dark matter component (the current limit is $\Omega _{\nu}h^2\lsi 0.07$). Therefore the DM candidate is usually looked for in physics beyond the Standard Model: the most important candidates today, which satisfy the above requirements and are well motivated from particle physics considerations are {\it axions} and the {\it lightest supersymmetric particle}.

Axions are pseudo Nambu-Goldstone bosons associated with the spontaneous breaking of a Peccei-Quinn, \cite{Peccei:1977ur,Peccei:1977hh}, $U(1)$ symmetry at scale $f_A$, introduced to solve the strong CP problem. In general, their mass is inversely proportional to $f_A$, as $m_A=0.62~10^{-3}~{\rm eV}~(10^{10}~{\rm GeV}/f_A)$. The allowed range for the axion mass $10^{-6}\lsi m_A\lsi 10^{-2}$ eV, see for instance \cite{PDBook}. The lower bound derives from the condition that the energy density in axionic DM is not higher than the observed DM density, and the upper bound derives most restrictively from the measurement of the super nova neutrino signal (the duration of the SN 1987A neutrino signal of a few seconds points to the fact that the new born star cooled mostly by neutrinos rather than through an invisible channel, such as axions). The axions are produced with the DM abundance only for large values of the decay constant $f_A$, which implies that they did not come into thermal equilibrium in the early universe. They were produced non-thermally, for example through a vacuum misalignment mechanism, see~\cite{axionmissmatch}. Experimental searches for axionic DM have been performed dominantly through the axion to photon coupling. The Lagrangian is ${\mathcal L}=g_{A\gamma}\;\vec{E}\cdot\vec{B} \phi _A$, where $\phi _A$ is the axion field, and $g_{A\gamma}$ is coupling whose strength is an important parameter in axion models; it permits the conversion of an axion into a single real photon in an external electromagnetic field, {\it i.e.} a Primakoff interaction. Halo axions may be detected in a microwave cavity experiments by their resonant conversion into a quasi-monochromatic microwave signal in a cavity permeated by a strong magnetic field. The cavity ``Q factor'' enhances the conversion rate on resonance. Currently two experiments searching for axionic DM are taking data: one at LLNL in California, \cite{Peng:2000hd}, and the CARRACK experiment, \cite{Yamamoto:2000si}, in Kyoto, Japan. Preliminary results of the CARRACK I experiment exclude axions with mass in a narrow range around $10\mu$eV as major component of the galactic dark halo for some plausible range of $g_{A\gamma}$ values. This experiment is being upgraded to CARRACK II, which intends to probe the range between 2 and 50 $\mu$eV with sensitivity to all plausible axion models, if axions form most of DM.

The lightest supersymmetric particle belongs to the general class of {\it weak interacting massive particles}, the so called WIMPs. WIMPs are particles with mass of the order of 10 GeV to TeV and with cross sections of weak interaction strength, $\sigma\sim G^{~2}_F~m^2 _{WIMP}$. Supersymmetry (SUSY) allows the existence of a R-parity symmetry which implies that the lightest SUSY particle is absolutely stable, offering a good candidate for DM. In most SUSY models the lightest supersymmetric particle is the neutralino, a linear combination of photino, wino and two higgsinos. Under the assumption that WIMPs were in thermal equilibrium after inflation, one can calculate the temperature at which their interaction rate with Standard Model particles becomes lower than the expansion rate of the universe. At that point they decouple from thermal bath and their number in co-moving volume stays constant at later times (they {\it freeze-out}). Freeze-out happens at temperature $T\simeq m/20$, almost independent of the particle properties, which means that particles are non-relativistic at the time of decoupling, making WIMPs a cold DM candidates. The abundance of DM in this scenario is $\Omega_{DM}\simeq 0.1~{\rm pb}/\langle \sigma v \rangle$, which surprisingly corresponds to the measured DM density for weak cross section values. The fact that their mass/cross section value is in the correct range, together with good motivation from SUSY, makes them attractive candidates. 

Today, a large part of the parameter space expected for neutralino has been explored: a region of mass $10$ GeV $\lsi m \lsi $ TeV, from $\sigma \sim 10$ mb to $\sigma \sim 10^{-6}$pb, and a new generation of experiments reaching $\sigma \sim 10^{-7,8}$pb is planned for the near future. Since direct and indirect detection searches have been unsuccessful, the nature of dark matter is still unresolved, although not all of the possible SUSY and axion parameter range has been explored, see for instance \cite{ellis,PDBook,axion}. We will examine alternative candidates for DM in the thesis, in particular candidates connected to the Baryon Asymmetry in the Universe, described in detail below.\\

Another important open question in our understanding of the Universe today is the observed Baryon Asymmetry of the Universe. The Dirac equation places anti-matter places on an equal footing with matter: in the Big Bang cosmological model the early epoch should contain a fully mixed state of matter and anti-matter. As the Universe expanded and cooled this situation would result in the annihilation of matter and anti-matter, leaving a baryon mass density today of $\Omega_b\simeq 4~10^{-11}$ which is much lower than the observed value $\Omega _b\simeq 0.04$, eq.~(\ref{Omega}) (or a baryon to photon ratio $n_b/n_{\gamma}\approx 10^{-18}$, eight orders of magnitude smaller than measured). Evidently some mechanism had to act to introduce the asymmetry in the baryon and antibaryon abundances and prevent their complete annihilation. The apparent creation of matter in excess of antimatter in the early universe is called Baryogenesis, for a review see, for instance, \cite{baryogenreview}. It was A.~Sakharov who first suggested in 1967, \cite{sakharov}, that the baryon density might not come from initial conditions, but can be understandable in terms of physical processes. He enumerated three necessary conditions for baryogenesis:
\begin{enumerate}
\item[1.]{\it Baryon number violation:} If baryon number (B) is conserved in all reactions, then a baryon excess at present can only reflect asymmetric initial conditions. 
\item[2.]{\it C and CP violation:} Even in the presence of B-violating reactions, if CP is conserved, every reaction which produces a particle will be accompanied by a reaction which produces its antiparticle at the same rate, so no net baryon number could be generated. 
\item[3.]{\it Departure from thermal equilibrium:} the CPT theorem guarantees equal masses for particle and its antiparticle, so in thermal equilibrium the densities of particles and antiparticles are equal and again no baryon asymmetry could exist. 
\end{enumerate}
Several mechanisms have been proposed to understand the baryon asymmetry. I comment on a few of the most important models below.\\

\emph{Grand Unified Theory (GUT) scale Baryogenesis (\cite{GUT1,GUT2,GUT3}):} Grand Unified Theories unify the gauge interactions of the strong, weak and electromagnetic interactions in a single gauge group. The GUT scale is typically of the order $10^{16}$ GeV so baryogenesis in this model occurs very early in the history of Universe. GUT's generically have baryon-violating reactions, such as proton decay (not yet observed) and they have heavy particles whose decays can provide a departure from equilibrium. The main objections to this possibility come from inflation and reheating models, \-- the temperature of the universe after reheating in most models is well below $M_{GUT}$. Furthermore, if baryogenesis occurs before inflation it would be washed out in the exponential expansion.  \\

\emph{Electroweak Baryogenesis (\cite{ewbg}):} In this model baryogenesis occurs in the SM at the Electroweak Phase Transition (EWPT) \-- this is the era when the Higgs first acquired a vacuum expectation value (VEV) and the SM particles acquired masses through their interaction with the Higgs field. This transition happened around 100 GeV. The Standard Model satisfies all of the Sakharov conditions:
\begin{enumerate}
\item In the SM there are no dimension 4 operators consistent with gauge symmetry which violate baryon (B) or lepton number (L). The leading operators which violate B are dimension 6 operators, which are suppressed by $O(1/M^2)$ and the operators which violate L are dimension 5, suppressed by $O(1/M)$, where $M$ is a scale of high energy physics which violates B or L. Inside the SM, B and L currents are not exactly conserved due to the fact that they are anomalous. However, the difference between $j^{\mu}_B-j^{\mu}_L$ is anomaly-free and is an exactly conserved quantity in the SM. In perturbation theory these effects go to zero, but in non-abelian gauge theories there are non-perturbative configurations which contribute to the non-conservation of currents. The vacuum structure of a Yang-Mills theory has an infinite set of states. In tunneling between these states, because of the anomaly, the baryon and lepton numbers will change. At zero temperatures tunneling effects are exponentially suppressed by $\exp(-2\pi /\alpha)$. At finite temperatures this rate should be larger. To estimate this rate one can look for the field configuration which corresponds to sitting on top of the barrier, a solution of the static equations of motion with finite energy, known as a sphaleron. The rate for thermal fluctuations to cross the barrier per unit time and volume should be proportional to the Boltzmann factor for this configuration, \cite{ewbg,dine,arnold}, $\Gamma =T^4 e^{-cM_W/g^2T}$. At high temperature $M_W$ vanishes and the transition gets the form $\Gamma=\alpha ^4 _W T^4$. 
\item CP-violation has been experimentally observed in kaon decays and is present in the SM. However, SM CP violation must involve all three generations. The lowest order diagram that involves three generations and contributes to CP violating processes relevant to baryogenesis is suppressed by 12 Yukawa couplings. The CKM CP violation contributes a factor of $10^{-20}$ to the amount of baryon asymmetry that could arise in the SM and a Beyond the Standard Model CP violation is usually invoked. 
\item Thermal nonequilibrium is achieved during first-order phase transitions in the cooling early universe. In the electroweak theory, there is a transition to a phase with massless gauge bosons. It turns out that, for a sufficiently light Higgs, this transition is of the first order. A first order transition is not, in general, an adiabatic process. As we lower the temperature, the transition proceeds by the formation of bubbles. The moving bubble walls are regions where the Higgs fields are changing and all of Sakharov's conditions are satisfied. It has been shown that various non-equilibrium processes near the wall can produce baryon and lepton numbers, \cite{ewpt1,ewpt2}. Avoiding the washing out of the asymmetry requires that after the phase transition, the sphaleron rate should be small compared to the expansion rate of the universe, or, as we have seen above, that $M_W$ be large compared to the temperature. This, in turn, means that the Higgs expectation value must be large immediately after the transition. It turns out that the current lower limit on the Higgs boson mass rules out any possibility of a large enough Higgs expectation value after the phase transition, at least in the minimal model with a single Higgs doublet.
\end{enumerate}

Any baryon asymmetry produced in the Standard Model is far too small to account for observations, the main obstacle being the heaviness of the Higgs, and one has to turn to extensions of the Standard Model in order to explain the observed asymmetry. \\

\emph{Leptogenesis (\cite{leptogen}):} In the last few years the evidence for neutrino masses has become more and more compelling. The most economical way to explain these facts is that neutrinos have Majorana masses arising from lepton-number violating dimension five operators (permitted if fermion carries no conserved charges). These interactions have the form ${\mathcal L}=\frac{1}{M}L H L H$. For $M=M_{pl}$ the neutrino mass would be too small to account for the observed values. The see-saw mechanism provides a simple picture of how the lower scale might arise. It assumes that in addition to the SM neutrinos, there are some SM singlet, heavy neutrinos, $N$. These neutrinos could couple to the left handed doublets $\nu_L$ providing the correct mass for the light neutrinos.

What is relevant is that heavy neutrinos $N$ can decay, for example, to both $H+\nu$ and $H+{\bar \nu}$, breaking the lepton number. CP violation can enter through phases in the Yukawa couplings and mass matrices of the $N$'s. At tree-level these phases will cancel out, so it is necessary to consider one loop  diagrams and to look at quantum corrections in which dynamical phases can appear in the amplitudes. These decays then produce a net lepton number, and hence a net $B-L$. The resulting lepton number will be further processed by sphaleron interactions, yielding a net lepton and baryon number. Reasonable values of the neutrino parameters give asymmetries of the order we seek to explain. However, all parameters needed for precise calculations are not measured yet (in the case of $\nu_L$ masses and CP violating couplings) and one needs some additional information about the masses of the $N$'s.\\

\emph{Affleck-Dine Baryogenesis(\cite{affleckdine})} In supersymmetric theories, the ordinary quarks and leptons are accompanied by scalar fields, which carry baryon and lepton number. A coherent field, {\it i.e.} a large classical value of such a field, can in principle carry a large amount of baryon number. Through interactions with the inflaton field CP-violating and B-violating effects can be introduced. As the scalar particles decay to fermions, the net baryon number the scalars carry can be converted into an ordinary baryon excess. 
The Affleck-Dine mechanism is also a mechanism for dark matter creation. Fluctuations in the scalar quark fields (``Q-balls'') are a dark matter candidate if they are stable. If they are unstable, they can still decay into dark matter. Since the Affleck-Dine mechanism describes the production of baryons {\it and} dark matter it could provide an explanation of the ratio between $\Omega_{DM}$ and $\Omega_{_b}$ from first principles. If supersymetry is discovered, given the success of inflation theory, the Affleck-Dine scenario will appear quite plausible.\\

To summarize, the abundance of baryons and dark matter in our Universe poses several challenging puzzles:
\begin{enumerate}
\item Why is there a non-zero net nucleon density and what determines its value?
\item What does dark matter consist of? Can it be explained as a SM particle?
\item Is it an accident that the dark matter density is roughly comparable to the nucleon density, $\rho_{DM} = 5 ~\rho_N$? 
\end{enumerate}

In the next Chapter we outline the scenario which aims to connect and answer the questions above. In Chapters \ref{Hdibaryon}, \ref{Hdmcons} and \ref{X} we focus on the two concrete DM candidates in this scenario, one being a particle within the Standard Model, and the other is BSM candidate. We comment on their particle physics properties and experimental constraints. As we will see, SM candidate is ruled out while the Beyond the Standard Model candidate is safe by many orders of magnitude.

\chapter{Proposed Scenario} 
\label{scenario}

In most approaches the origins of DM and the BAU are completely unrelated \-- baryon number density $n_B$ is obtained from baryogensis models while the number density of DM $n_{DM}$ derives from relic freeze-out calculations and their values naturally could differ by many orders of magnitude (see \cite{other1,other2,other3,other4} for other papers where these two problems are related).  In this Section we propose a new type of scenario, in which the observed baryon asymmetry is due to the {\it separation} of baryon number between ordinary matter and dark matter and not to a net change in the total baryon number since the Big Bang, \cite{fz:bau}. 
Thus the abundances of nucleons and dark matter are related.  The first Sakharov condition is not required, while the last two remain essential.  We give explicit examples in which  anti-baryon number is sequestered at temperatures of order 100 MeV. 

The CPT theorem requires that the total interaction rate of any ensemble of particles and antiparticles is the same as for the  conjugate state in which each particle is replaced by its antiparticle and all spins are reversed.  However individual channels need not have the same rate so, when CP is violated, the annihilation rates of the CP reversed systems are not in general equal.  A difference in the annihilation cross section, $\sigma^{\rm annih}_{\bar{X}} <  \sigma^{\rm annih}_{X} $, means that the freeze out temperature for $X$'s ($T_X$) is lower than for $\bar{X}$'s ($T_{\bar{X}}$).   After the $\bar{X}$'s freeze out, the $X$'s continue to annihilate until the temperature drops to $T_{X}$, removing $B_X$ antinucleons for each $X$ which annihilates.  

Assuming there are no other significant contributions to the DM density, the present values $n_{o\, N}$, $n_{o\, X}$ 
and $n_{o\, \bar{X}}$  are
determined in terms of $m_X$, $B_X$ and the observables $ \frac{\Omega_{DM}}{\Omega_b}$ and $\frac{n_{o\,N}}{n_{o\, \gamma}} \equiv \eta_{10} \,10^{-10}$ or $\rho_{\rm crit}$. From WMAP, 
\begin{eqnarray} \label{wmappar}
\eta_{10} &=& 6.5^{+0.4}_{-0.3}, \nonumber \\
\frac{\Omega_{DM}}{\Omega_b} &=& 5.27 \pm 0.49.
\end{eqnarray} 
Given the values of these observables, we can ``reverse engineer" the process of baryon-number-segregation.  

For brevity, suppose there is only one significant species of DM particle.  Let us define $\epsilon = \frac{n_X}{n_{\bar{X}} }$.  Then the total energy density in $X$'s and $\bar{X}$'s is  
\beq
\rho_{DM} = m_X n_{\bar{X}} (1 + \epsilon).
\eeq
By hypothesis, the baryon number density in nucleons equals the antibaryon number density in $X $ and $\bar{X}$'s, so  
\beq
B_X n_{\bar{X}} (1-\epsilon) = (n_N - n_{\bar{N}}) = \frac{\rho_b}{m_N}.  
\eeq
Thus  
\beq \label{kappa}
\frac{\Omega_{DM}}{\Omega_b} = \left( \frac{1 + \epsilon}{1 - \epsilon} \right) \frac{m_X}{m_N B_X}.
\eeq
As long as the DM particle mass is of the order of hadronic masses and $\epsilon $ is not too close to 1, this type of scenario naturally accounts for the fact that the DM and ordinary matter densities are of the same order of magnitude.  Furthermore, since $ \frac{1 + \epsilon}{1 - \epsilon} \ge 1$, the DM density in this scenario must be {\it  greater} than the nucleonic density, unless $m_X < m_N B_X$, as observed.  

Given the parameters of our Universe, we can instead write (\ref{kappa}) as an equation for the DM mass
\beq\label{mX}
m_X = \left( \frac{1 - \epsilon}{1 + \epsilon} \right) \frac{\Omega_{DM}}{\Omega_b} \, B_X m_N .
\eeq
For low baryon number, $B_X = 1\, (2)$, this implies 
\beq
m_X \lsi 4.5 \,(9)\,{\rm GeV}.
\eeq  
If dark matter has other components in
addition to the $X$ and $\bar{X}$, the $X$ must be lighter still.   The observed BAU can be due to baryon number sequestration with heavy DM only if $B_X$ is very large, e.g., strangelets or Q-balls.  However segregating the baryon number in such cases is challenging.  

As an existence proof and to focus discussion of the issues, we present two concrete scenarios.  In the first, $X$ is a particle already postulated in QCD, the $H$ dibaryon ({\it uuddss}). New particle physics is necessary, however, because the CP violation of the Standard Model via the CKM matrix cannot produce the required $\mathcal{O}$(20\%) difference in annihilation cross sections, since only the first two generations of quarks are involved. The second scenario postulates a new particle, we call $X_4$, with mass  $\lsi 4.5$ GeV, which couples to quarks through dimension-6 operators coming from beyond-the-standard-model physics. In this case CP violation is naturally large enough, $\mathcal{O}$(10\%), because all three quark generations are involved and, in addition, the new interactions in general violate CP. Review of particle properties of these candidates is given in Sections \ref{Tightlybound} and \ref{Xproperties}. After deducing the properties required of these particles by cosmology we discuss indirect (Sections \ref{Hindirect} and \ref{Xconstraints}) and direct (Section \ref{directDM}) searches.  As we shall show, the $H$, $\bar{H}$ scenario can already be ruled out by limits on the heat production in Uranus, while $X$ remains a viable candidate.  

The annihilation rate of particles of type $j$ with particles of type $i$ is
$\Gamma^{annih}_{j}(T) = \Sigma_i ~n_i(T) <\sigma_{ij}^{annih}
v_{ij}>$, 
where $<...>$ indicates a thermal average and $v_{ij}$ is the
relative velocity.  As the Universe cools, the densities of all the
particle species $i$ decrease and eventually the rate of even the most important
annihilation reaction falls below the expansion rate of the
Universe.  The temperature at which this occurs is called the
freezeout temperature $T_j$ and can be calculated by solving Boltzmann equations. 
For the freezeout of $\bar X$ annihilation the Boltzmann equation can be written as
\beq
\frac{x}{Y^{eq} _{\bar X}}\frac{dY_{\bar X}}{dx}=\frac{\Sigma_i ~n^{eq} _i <\sigma v>}{H(T)}\left(\frac{Y_{\bar X}}{Y^{eq} _{\bar X}}-1\right),
\eeq 
where $x=m_{\bar X}/T$, $Y_{\bar X}=n_{\bar X}/s$ and $s$ is the entropy of the Universe. Notice that $dY_{\bar X}/dx$ goes to zero (or $Y_{\bar X}$ stays constant, corresponding to freezeout) when $\Gamma^{annih}_{\bar X}(T_{\bar X})\ll H(T)$.
Therefore the freezeout temperature can be estimated roughly from a condition $\Gamma^{annih}_{j}(T_j) = H(T_j) = 1.66 \sqrt{g_*} ~ T_j^2/ M_{Pl} $, 
where $g_*$ is the effective number of relativistic degrees of freedom\cite{kolbTurner}.  Between a few MeV and the QCD phase transition only neutrinos, $e^\pm$ and $\gamma$'s are in equilibrium and $g_* = 10.75$.  Above the QCD phase transition which is within the range 100 to 200 MeV, light quarks and antiquarks ($q, \, \bar{q}$) and $\mu^\pm$ are also relativistic species in equilibrium, giving $g_* = 56.25$.  
The equilibrium density at freeze out temperature, $n_j(T_j)$, is a good estimate of the relic abundance of the $j$th species\cite{kolbTurner}. A key element of baryon-number sequestration is that self-annihilation cannot be important for maintaining equilibrium prior to freeze out.  This is easily satisfied as long as $\sigma_{\bar{X}X}^{\rm ann}$ is not much greater than $\sigma_{\bar{X}q}^{\rm ann}$ , since at freezeout in ``$X_4$" scenario, $n_{X_4 ,\, \bar{X_4} }\sim 10^{-11} n_{d ,\, \bar{d} }$.  

Given $m_X,\, B_X$ and $g_X$ (the number of degrees of freedom of the $X$ particle) and associated densities $n_{\{X,\bar{X}\}}$, the temperature $T_{\bar{X}}$ at which $\bar{X}$'s must freeze out of thermal equilibrium satisfies: 
\beq \label{Xbarfo}
  \frac{n_{\bar{X}} - n_X}{n_{\bar{X}}} \frac{n_{\bar{X}}}{n_\gamma } = (1-\epsilon) \frac{ \pi^2 g_X
  x_{\bar{X}}^{3/2} e^{-x_{\bar{X}} } }{2 \zeta(3) (2 \pi)^{3/2} }= 
  \frac{10.75}{3.91}\frac{\eta_{10} 10^{-10} }{B_X} ,
\eeq
where $x_{\bar{X}} \equiv m_X/T_{\bar{X}}$.  
$ \frac{10.75}{3.91}$ is the factor by which $\frac{n_b}{n_\gamma}$ increases above $e^\pm$ annihilation.  The equation for $X$ freezeout is the same, with $(1-\epsilon) \rightarrow (1-\epsilon)/\epsilon $.  Freezeout parameters for our specific models, the $H$ dibaryon and the $X$, are given in Table 2.1; $\tilde{\sigma} \equiv \langle \sigma^{\rm ann} v \rangle / \langle v \rangle$ is averaged over the relevant distribution of c.m. kinetic energies, thermal at $\approx 100$ MeV for freezeout.

If $X$s interact weakly with nucleons, standard WIMP searches constrain the low energy scattering cross section  $\sigma_{DM} \equiv (\sigma^{\rm el}_{\bar{X} N} + \epsilon \sigma^{\rm el}_{XN})/(1+ \epsilon)$.  However if the $X$ is a hadron, multiple scattering in the earth or atmosphere above the detector can cause a significant fraction to reflect or be degraded to below threshold energy before reaching a deep underground detector.  Scattering also greatly enhances DM capture by Earth, since only a small fraction of the halo velocities are less than $v_{\rm esc}^{E} = 11$ km/s.  Table I gives the total fluxes and the factor $f_{\rm cap}$ by which the flux of captured $\bar{X}$'s is lower, for the two scenarios.  The capturing rate is obtained using code by Edsjo et al. \cite{edsjo} which calculates the velocity distribution of weakly interacting dark matter at the Earth taking into account gravitational diffusion by Sun, Jupiter and Venus. For the $H$ dibaryon these are the result of integrating the conservative halo velocity distribution \cite{zf:window}.  A comprehensive reanalysis of DM cross section limits including the effect of multiple scattering is given in thesis section \ref{directDM} and ref. \cite{zf:window}.  A window in the DM exclusion was discovered for $m_X \lsi 2.4$ GeV and $ \tilde{\sigma}_{DM} \approx 0.3 - 1\, \mu $b; otherwise, if the DM mass $\lsi 5$ GeV, $\tilde{\sigma}_{DM}$ must be $ \lsi 10^{-38} {\rm cm}^2$, \cite{zf:window}.  

Since $\sigma_{\{X,\bar{X} \}N}$ is negligible compared to $\sigma_{NN}$ and the $X,\,\bar{X}$ do not bind to nuclei\cite{fz:nucbind}, nucleosynthesis works the same in these scenarios as with standard CDM.  Primordial light element abundances constrain the {\it nucleon} -- not {\it baryon} -- to photon ratio! 

\begin{table}[htb] \label{table}
\caption{Required freezeout temperatures and annihilation cross sections at freezeout, and captured DM flux in Earth, in two models; $\sigma_{-42} \equiv \sigma/(10^{-42} {\rm cm}^2)$. }
\begin{center}
\begin{tabular}{|c|c|c|c|c|c|}
\hline
Model & $T_{\bar{X}}$ MeV & $T_{X}$ MeV   &  $\tilde{\sigma}^{\rm ann}_{\bar{X}}$ cm$^2$ & $\tilde{\sigma}^{\rm ann}_{{X}}$ cm$^2$ & $R_{\rm cap}$ s$^{-1}$   \\  \hline 
$H$, $\bar{H}$   &      86.3         &   84.5    & 	$2.2~10^{-41}$                       &    $2.8~10^{-41}$                &      $3.8 \times 10^{23}$   \\ \hline
$X_4$   &      180          &   159     &		$3.3~10^{-45}$                       &    $3.7~10^{-45}$                &     $ 1.6 \times 10^{12} \sigma_{-42} $     \\ \hline
\end{tabular}
\end{center}
\end{table}

The CPT theorem requires that $\sigma^{\rm ann}_{X} + \sigma^{\rm non-ann}_{X} = \sigma^{\rm ann}_{\bar{X}} + \sigma^{\rm non-ann}_{\bar{X}}$.  Therefore a non-trivial consistency condition in this scenario is 

\begin{displaymath}
\sigma^{\rm ann}_{X} - \sigma^{\rm ann}_{\bar{X}} \le \sigma^{\rm non-ann}_{\bar{X}}
\end{displaymath}  
The value of the LHS needed for B-sequestration from Table I is compatible with the upper limits on the RHS from DM searches, and $\sigma^{\rm non-ann}_{\bar{X}} \ge \sigma^{\rm el}_{\bar{X}} $, so no fine-tuning is required to satisfy CPT.     

Further in the text we focus on specific DM candidates with baryon number and experimental constraints that can be placed on such particles.

\chapter{$H$ dibaryon: Dark Matter candidate within QCD?} 
\label{Hdibaryon}

This Chapter is organized as follows: in \S\ref{Hhistory} we review properties of the  di-baryon; in \S\ref{Tightlybound} we focus on the $H$ which is tightly bound and therefore is light and compact object; we review the experiments relevant for the existence of tightly bound $H$ in \S\ref{expts}; we calculate nuclear transitions of the $H$ in \S\ref{NucStab} and binding of the $H$ to nuclei in \S\ref{binding} and set bounds on parameters from the experiments; in \S\ref{summaryEX} we Summarize the properties of the $H$ which would be consistent with current existence experiments.

\section{$H$ history and properties}
\label{Hhistory}

The $H$ particle is a $(udsuds)$ flavor singlet dibaryon with  strangeness $S=-2$, charge $Q=0$ and spin-isospin-parity $J^P =0^+$. In 1977 Jaffe calculated its mass \cite{jaffe} to be about 2150 MeV in the MIT-bag model and thus predicted it would be a strong-interaction-stable bound state, since decay to two $\Lambda$ particles would not be kinematically allowed. The basic mechanism which is expected to give large attractive force between quarks in the $H$ is {\it the color magnetic interaction}. The contribution to the mass from lowest-order gluon exchange is proportional to 
\beq
\Delta=-\sum_{i>j}({\vec \sigma} _i {\vec \sigma}_j)({\vec \lambda} _i {\vec \lambda} _j)M(m_iR,m_jR),
\eeq
where ${\vec \sigma} _i$ is spin and ${\vec \lambda} _i$ color vector of the ith quark, and $M(m_iR,m_jR)$ measures the interaction strength. For color singlet hadrons containing quarks and no antiquarks \cite{jaffe},
\beq
\Delta =\left( 8N-\frac{1}{2}C_6+\frac{4}{3}J(J+1)\right){\bar M},
\eeq 
where $N$ is total number of quarks, $J$ is their angular momentum and $C_6$ is Casimir operator for the color-spin representation of quarks, $SU(6)$. We can see that the lightest dibaryons will be those in which the quarks are in the color-spin representation with the largest value of Casimir operator. Large values of the Casimir operator are associated with symmetric color-spin representation. Antisymmetry requires flavor representation to be asymmetric. Calculation shows that only flavor singlet representation is light enough to be a stable state.   

This result raised high theoretical and experimental attention. The mass was later computed using different models such as Skyrme, quark cluster models, lattice QCD with values ranging between 1.5 and 2.2 GeV. Such a wide range of predictions makes clear contrast to the success in reproducing the mass spectrum of mesons and baryons using the above methods.

The $H$ searches have been done using different production mechanisms, we will comment here on the most common approaches
\begin{itemize}
\item{{\it $H$ production via $(K^-,K^+)$ reaction}}: In BNL E885 experiment, \cite{bnlE885}, ${\rm C}_{12}$ is used as a target. The subsequent reactions: $K^-+p\rightarrow K^+ +\Xi ^-$ and $(\Xi ^- A)_{\rm atom}\rightarrow H+X$ was expected to produce the $H$;
\item{{\it Production through Heavy Ion collision}}: In BNL E896 experiment, \cite{bnlE896}, a $Au+Au$ collision was used to produce the $H$ that could be identified by the anticipated decay $H\rightarrow \Sigma ^-p\rightarrow n\pi ^-p$;
\item{{\it ${\bar p}$- nucleus annihilation reaction}}: In the experiment by Condo {\it et al.}, \cite{condo}, they used the reaction ${\bar p}+A\rightarrow H+X$ to produce the $H$, and looked for its decay through $H\rightarrow \Sigma ^-+p$ channel. 
\end{itemize}   
Other experiments can be found in ref.~\cite{H}.

Experiments were guided by theoretical predictions for $H$ production and decay lifetimes. The most remarkable contribution to theoretical predictions came from the works of Aerts and Dover \cite{aertsdover1,aertsdover2,aertsdover3} whose calculations have been the basis for understanding of the Brookhaven experiments BNL E813 and E836 with respect to the formation of the $H$. However, theoretical predictions may depend critically on the wave function of the $H$ dibaryon. 
 
Experiments so far did not confirm the existence of the $H$ particle but they put bounds on the production cross section for particular mass values, see \cite{H} for a detailed review. An underlying assumption has generally been that the $H$ is not deeply bound. In our work we are particularly interested in the possibility that the $H$ is tightly bound and that it has a mass less than $m_N+m_{\Lambda}$. In that case, as we shall see, its lifetime can be longer than the age of the Universe.

\section{Tightly bound $H$} \label{Tightlybound}

Several lines of reasoning suggest the possibility that the $H$ could be a tightly bound state, with mass lower than $m_N+m_{\Lambda}=2053$ MeV. We digress here briefly to review this motivation, put forth in~\cite{f:StableH}. The first line of motivation starts from the observation that the properties of the $\frac{1}{2} ^-$ baryon resonance $\Lambda(1405)$ and its spin $\frac{3}{2}$ partner $\Lambda(1520)$ are nicely explained if these are assumed to be ``hybrid baryons'': bound states of a gluon with three quarks in a color -octet, flavor-singlet state denoted $(uds)_8$. If we adopt the hybrid baryon interpretation of the $\Lambda(1405)$ and $\Lambda(1520)$, the similarity of their masses and glueball mass ($\sim 1.5$ GeV) suggests that the color singlet bound state of two $(uds)_8$'s, which would be the $H$, might also have a mass close to 1.5 GeV. A second line of reasoning starts from the observation that instantons produce a strong attraction in the scalar diquark channel, explaining the observed diquark structure of the nucleon. The $H$ has a color-flavor-spin structure which permits a large proportion of the quark pairwise interactions to be in this highly attractive spin -0, color ${\bar 3}$ channel, again suggesting the possibility of tightly bound $H$. Indeed, ref. \cite{kochelev} reported an instanton-gas estimate giving $m_H=1780$ MeV.
If the $H$ is tightly bound, it is expected to be spatially compact.  Hadron sizes vary considerably, for a number of reasons.  The nucleon is significantly larger than the pion, with charge radius $r_N = 0.87$ fm compared to $r_\pi = 0.67$ fm\cite{PDBook}.  Lattice and instanton-liquid studies qualitatively account for this diversity and further predict that the scalar glueball is even more tightly bound: $r_G \approx 0.2$ fm \cite{lattice:RG,shuryak:RG}. If the analogy suggested in ref.~\cite{kf:Lam1405} between $H$, $\Lambda_{1405}$ and glueball is correct, it would suggest $r_H \approx r_G \lsi 1/4 ~r_N$.  The above size relationships make sense:  the nucleon's large size is due to the low mass of the pion which forms an extended cloud around it, while the $H$ and glueball do not couple to pions, due to parity and flavor conservation, are thus are small compared to the nucleon. 

Lattice QCD efforts to determine the $H$ mass have been unable to obtain clear evidence for a bound $H$. The discussion above suggests that inadequate spatial resolution may be a problem due to the small size of the $H$. Several lattice calculations \cite{wetzorke,pochinsky,iwasaki,mackenzie} use sufficiently small spacing but they use quenching approximation, which does not properly reproduce instanton effects. This is a serious deficiency since the instanton liquid results of ref.~\cite{kochelev} indicate that instanton contributions are crucial for the binding in the scalar diquark channel.
In the absence of an unquenched, high-resolution lattice QCD calculation capable of a reliable determination of the $H$ mass and size, we will consider all values of $m_H $ and take $r_H/r_N \equiv 1/f$ as a parameter, with $f$ in the range $2$-$6$.

Based on the fact that the $H$ interacts with ordinary matter only if it has non-vanishing color charge radius (since it is spin and isospin zero neutral particle) ref.~\cite{f:StableH} estimates cross section of the $H$ with ordinary matter to be of the order $\approx 10^{-2,3}$ mb.

Based on the assumption of light and tightly bound $H$ motivated by Farrar in \cite{f:StableH} we examine the viability of this model using current experimental constraints. My work has focused on the study of several processes involving the $H$ which are phenomenologically important if it exists: 
\begin{itemize}
\item {whether it binds to nuclei,} in which case exotic isotope searches place stringent constraints;  
\item {nucleon transitions of the $H$} \-- conversion of two $\Lambda$'s in a doubly-strange hypernucleus to
an $H$ which is directly tested in experiments, decay of the $H$ to two baryons, and---if the $H$ is light enough---conversion of two nucleons in a nucleus to an $H$. 
\end{itemize}
The experimental constraints important for the existence of the light $H$ and the $H$ as a DM candidate are outlined below. For more experimental constraints on tightly bound $H$, see~\cite{f:StableH}.

\section{The Existence of the $H$ \-- Experimental constraints}
\label{expts}

\subsection{Double  $\Lambda $ hyper-nucleus detection} \label{expdoublelambda}

One of the ways to produce and study the $H$ is through the production of double hypernuclei. A double $\Lambda$ hypernucleus formed in an experiment usually through the $(K^-, K^+)$ interaction with the target, is expected to decay strongly into the $H$ and the original nucleus. If single $\Lambda$ decay from a double $\Lambda$ hypernucleus is observed, it means that either the mass of the $H$ should be heavier than the mass of the two $\Lambda$'s minus the binding energy, {\it or} that the decay of two $\Lambda$'s to an $H$ proceeds on a longer time scale than the $\Lambda$ weak decay.  
There are five experiments which have reported positive results in the search for single $\Lambda$ decays from double $\Lambda$ hypernuclei. The three early emulsion based experiments \cite{prowse,danysz,kek} suffer from ambiguities in the particle identification, and therefore we do not consider them further. In the latest emulsion experiment at KEK~\cite{kek2}, an event has been observed which is interpreted with good confidence as the sequential decay of ${\rm He}^6 _{\Lambda \Lambda}$ emitted from a $\Xi ^-$ hyperon nuclear capture at rest. The binding energy of the double $\Lambda$ systems is obtained in this experiment to be $B_{\Lambda \Lambda }=1.01\pm 0.2$ MeV, in significant disagreement with the results of previous emulsion experiments, finding $B_{\Lambda \Lambda }\sim 4.5$ MeV.

The third experiment at BNL~\cite{ags} was not an emulsion experiment. After the $(K^-, K^+)$ reaction on a ${\rm Be}^9$ target produced S=-2 nuclei it detected $\pi$ pairs coming from the same vertex at the target. Each pion in a pair indicates one unit of strangeness change from the (presumably) di-$\Lambda$ system. Observed peaks in the two pion spectrum have been interpreted as corresponding to two kinds of decay events. The pion kinetic energies in those peaks are (114,133) MeV and (104,114) MeV. The first peak can be understood as two independent single $\Lambda$ decays from $\Lambda \Lambda$ nuclei. The energies of the second peak do not correspond to known single $\Lambda$ decay energies in hyper-nuclei of interest. The proposed explanation\cite{ags} is that they are pions from the decay of the double $\Lambda$ system, through a specific He resonance. The required resonance has not yet been observed experimentally, but its existence is considered plausible. This experiment does not suffer from low statistics or inherent ambiguities, and one of the measured peaks in the two pion spectrum suggests observation of consecutive weak decays of a double $\Lambda$ hyper-nucleus. The binding energy of the double $\Lambda$ system $B_{\Lambda \Lambda }$ could not be determined in
this experiment.

The KEK and BNL experiments are generally accepted to demonstrate quite conclusively, in two different techniques, the observation of $\Lambda$ decays from double $\Lambda$ hypernuclei.  Therefore the formation of the $H$ in a double $\Lambda$ hypernucleus does not proceed, at least not on a time scale faster than the $\Lambda$ lifetime, i.e., $\tau _{A_{\Lambda \Lambda}\rightarrow A_{H}'X}$ cannot be much less than $\approx 10^{-10}$s.  (To give a more precise limit on $\tau _{A_{\Lambda \Lambda}\rightarrow A_{H}'X}$ requires a detailed analysis by the experimental teams, taking into account the number of hypernuclei produced, the number of observed $\Lambda$ decays, the acceptance, and so on.)  This experiment is considered leading evidence against the existence of the $H$ di-baryon. As will be seen below, this constraint is readily satisfied if the $H$ is compact: $r_H \lsi 1/2 ~r_N$ or less, depending on the nuclear wave function.

\subsection{Stability of nuclei} \label{expstab}

This subsection derives constraints for a stable $H$, where $m_H\lsi 2m_N$. In that case the $H$ would be an absolutely stable form of matter and nuclei would generally be unstable toward decay to the $H$.
There are a number of possible reactions by which two nucleons can convert to an $H$ in a nucleus if that is kinematically allowed ($m_H \lsi 2 m_N$\footnote{Throughout, we use this shorthand for the more precise inequality $m_H < m_{A} - m_{A'} - m_X$ where $m_X$ is the minimum invariant mass of the final decay products.}). The initial nucleons are most likely to be $pn$ or $nn$ in a relative s-wave, because in other cases the Coulomb barrier or relative orbital angular momentum suppresses the overlap of the nucleons at short distances which is necessary to produce the $H$. If $m_H \lsi 2 m_N - m_\pi=1740$ MeV, the final state can be $H \pi^+ $ or $H \pi^0$.  If $\pi $ production is not allowed, for $m_H \gsi 1740$ MeV, the most important reactions are $p n \rightarrow H e^+ \nu_e$ or the radiative-doubly-weak reaction $n n \rightarrow H \gamma$.

The best experiments to place a limit on the stability of nuclei are proton decay experiments. Super Kamiokande (SuperK), can place the most stringent constraint due to its large mass.  It is a water Cerenkov detector with a 22.5 kiloton fiducial mass, corresponding to $8~10^{32}$ oxygen nuclei. SuperK is sensitive to proton decay events in over 40 proton decay channels\cite{SuperK}. Since the signatures for the transition of two nucleons to the $H$ are substantially different from the monitored transitions, a specific analysis by SuperK is needed to place a limit.  We will discuss the order-of-magnitude of the limits which can be anticipated.

Detection is easiest if the $H$ is light enough to be produced with a $\pi^+$ or $\pi^0$. The efficiency of SuperK to detect neutral pions, in the energy range of interest (KE $\sim 0-300$ MeV), is around 70 percent. In the case that a $\pi ^+$ is emitted, it can charge exchange to a $\pi ^0$ within the detector, or be directly detected as a non-showering muon-like particle with similar efficiency.  More difficult is the most interesting mass range $m_H \gsi 1740$ MeV, for which the dominant channel $p n \rightarrow H e^+ \nu$ gives an electron with $E \sim (2 m_N - m_H)/2 \lsi  70$ MeV.   The  channel $nn \rightarrow $H$ \gamma$, whose rate is smaller by a factor of order $\alpha$, would give a monochromatic photon with energy $(2 m_N - m_H) \lsi 100$ MeV.

We can estimate SuperK's probable sensitivity as follows.  The ultimate background comes primarily from atmospheric neutrino interactions, 
\beq
\nu N \rightarrow N'(e,\mu),\quad \nu N \rightarrow N'(e,\mu)+n\pi~{\rm and}  \nu N\rightarrow \nu N' +n\pi, \nonumber 
\eeq
for which the event rate is about $100$ per kton-yr.  Without a strikingly distinct signature, it would be difficult to detect a signal rate significantly smaller than this, which would imply SuperK might be able to achieve a sensitivity of order $\tau_{A_{NN}\rightarrow A_{H}'X}\gsi {\rm few} 10^{29}$ yr. Since the $H$ production signature is not more favorable than the signatures for proton decay, the SuperK limit on $\tau_{A_{NN}\rightarrow A_{H}'X}$ can at best be $0.1 \tau_p$, where $0.1$ is the ratio of Oxygen nuclei to protons in water. Detailed study of the spectrum of the background is needed to make a more precise statement. We can get a lower limit on the SuperK lifetime limit by noting that the SuperK trigger rate is a few Hz~\cite{SuperK}, putting an immediate limit $\tau_{O\rightarrow H + X }\gsi {\rm few} 10^{25}$ yr, assuming the decays trigger SuperK.

SuperK limits will apply to specific decay channels, but other experiments potentially establish limits on the rate at which nucleons in a nucleus convert to an $H$ which are independent of the $H$ production reaction. These experiments place weaker constraints on this rate due to their smaller size, but they are of interest because in principle they measure the stability of nuclei directly. Among those cited in ref.~\cite{PDG02}, only the experiment by Flerov {\it et. al.}~\cite{flerov} could in principle be sensitive to transitions of two nucleons to the $H$. It searched for decay products from ${\rm Th}^{232}$, above the Th natural decay mode background of $4.7$ MeV $\alpha$ particles, emitted at the rate $\Gamma _{\alpha}=0.7~10^{-10} {\rm yr}^{-1}$. Cuts to remove the severe background of $4.7$ MeV $\alpha$'s may or may not remove events with production of an $H$. Unfortunately ref.~\cite{flerov} does not discuss these cuts or the experimental sensitivity in detail. An attempt to correspond with the experimental group, to determine whether their results are applicable to the $H$, was unsuccessful. If applicable, it would establish that the lifetime $\tau_{{\rm Th}^{232}\rightarrow H + X}> 10^{21}$ yr.

Better channel\--independent limits on $N$ and $NN$ decays in nuclei have been established recently, as summarized in ref.~\cite{BOREXINO}. Among them, searches for the radioactive decay of isotopes created as a result of $NN$ decays of a parent nucleus yield the most stringent constraints.  This method was first exploited in the DAMA liquid Xe detector \cite{DAMAnucldecay}. BOREXINO has recently improved these results\cite{BOREXINO} using their prototype detector, the Counting Test Facility (CTF) with parent nuclei ${\rm C}^{12},{\rm C}^{13}~{\rm and}~{\rm O}^{16}$. The signal in these experiments is the beta and gamma radiation in a specified energy range associated with deexcitation of a daughter nucleus created by decay of outer-shell nucleons in the parent nucleus. They obtain the limits $\tau _{pp} > 5~10^{25}$ yr and $\tau _{nn} > 4.9~10^{25}$ yr.  However $H$ production requires overlap of the nucleon wavefunctions at short distances and is therefore suppressed for outer shell nucleons, severely reducing the utility of these limits.  Since the SuperK limits will probably be much better, we do not attempt to estimate the degree of suppression at this time.

Another approach could be useful if for some reason the direct SuperK search is foiled. Ref. \cite{suzuki} places a limit on the lifetime of a bound neutron, $\tau _{n}>4.9~10^{26}$ yr, by searching for $\gamma$'s with energy $E_{\gamma}=19-50$ MeV in the Kamiokande detector. The idea is that after the decay of a neutron in oxygen the de-excitation of $O^{15}$ proceeds by emission of $\gamma$'s in the given energy range. The background is especially low for $\gamma$'s of these energies, since atmospheric neutrino events produce $\gamma$'s above 100 MeV. In our case, some of the photons in the de-excitation process after conversion of $pn$ to H, would be expected to fall in this energy window.

In \S\ref{NucStab} we calculate nuclear transition rates of a tightly bound $H$ in order to find constraints on the $H$ size and mass, primarily focusing on the constraints set by SuperK experiment.

\subsection{Experimental constraints on the $H$ binding}
\label{exptsB}

If the $H$ binds to nuclei and if it is present with DM abundance, it should be present in nuclei on Earth and experiments searching for anomalous mass isotopes would be sensitive to its existence. Accelerator mass spectroscopy (AMS) experiments generally have high sensitivity to anomalous isotopes, limiting the fraction of anomalous isotopes to $10^{-18}$ depending on the element. We discuss binding of the $H$ to heavy and to light isotopes separately.

The $H$ will bind more readily to heavy nuclei than to light ones because their potential well is wider. However, searches for exotic particles bound to heavy nuclei are limited to the search for charged particles in Fe \cite{Feneg} and to the experiment by Javorsek et al. \cite{javorsek} on Fe and Au. The experiment by Javorsek searched for anomalous Au and Fe nuclei with $M_X$ in the range $200$ to $350$ atomic mass units u. Since the mass of Au is $197$ u, this experiment is sensitive to the detection of an exotic particle with mass $M_X \ge 3$ u$=2.94$ GeV and is not sensitive to the tightly bound $H$.

A summary of limits from various experiments on the concentrations of exotic isotopes of light nuclei is given in~\cite{hemmick}. Only the measurements on hydrogen~\cite{hydrogen} and helium \cite{helium} nuclei are of interest here because they are sensitive to the presence of a light exotic particle with a mass of $M_X \sim~ 1 $ GeV. It is very improbable that the $H$ binds to hydrogen, since the $\Lambda$ does not bind to hydrogen in spite of having attractive contributions to the potential not shared by the $H$, {\it e.g.}, from $\eta$ and $\eta'$ exchange. Thus we consider only the limit from helium. The limit on the concentration ratio of exotic to non-exotic isotopes for helium comes from the measurements of Klein, Middleton and Stevens who quote an upper limit of $\frac {He_X}{He}<2\times 10^{-14}$ and $\frac{He_X}{He}<2\times 10^{-12}$ for primordial He \cite{plaga}. Whether these constraints rule out the $H$ depends on the $H$ coupling to nucleus. 

In \S\ref{binding} we calculate the binding of the $H$, or more generally any flavor singlet, to nuclei and find the values of coupling which are allowed from the existence of the $H$. As we will see, the allowed couplings coincide with the values expected from the particle physics arguments.

\section{Nucleon and nuclear transitions \newline of the $H$ dibaryon \-- Rates estimate} \label{NucStab}

In this section we study several processes involving the $H$ which are phenomenologically important if it exists, \cite{fz:nucstab}: conversion of two $\Lambda$'s in a doubly-strange hypernucleus to an $H$ (\S\ref{expdoublelambda}), decay of the $H$ to two baryons, and---if the $H$ is light enough---conversion of two nucleons in a nucleus to an $H$ (\S\ref{expstab}).  The amplitudes for these processes depend on the spatial wavefunction overlap of two baryons and an $H$. We are particularly interested in the possibility that the $H$ is tightly bound and that it has a mass less than $m_N + m_\Lambda$ because then, as we shall see, the $H$ is long-lived, with a lifetime which can be longer than the age of the Universe.

To estimate the rates for these processes requires calculating the overlap of initial and final quark wavefunctions.  We model that overlap using an Isgur-Karl harmonic oscillator model for the baryons and $H$, and the Bethe-Goldstone and Miller-Spencer wavefunctions for the nucleus. The results depend on $r_N/r_H$ and the nuclear hard core radius.

We also calculate the lifetime of the $H$ taking similar approach to the overlap calculation. We do it in three qualitatively distinct mass ranges, under the assumption that the conditions to satisfy the  constraints from double-$\Lambda$ hypernuclei are met.  The ranges are 
\begin{itemize}
\item {$m_H < m_N + m_\Lambda$,} in which $H$ decay is a doubly-weak $\Delta S = 2$ process,  
\item {$m_N + m_\Lambda < m_H  < 2 m_\Lambda$,} in which the $H$ can decay by a normal weak interaction, and 
\item {$m_H > 2 m_\Lambda$,} in which the $H$ is strong-interaction unstable.  
\end{itemize}
The $H$ lifetime in these ranges is greater than or of order $10^{7}$ years, $\sim 10$ sec, and $\sim 10^{-14}$ sec, respectively.

Finally, if $m_H \lsi 2 m_N$, nuclei are unstable and $\Delta S=-2$ weak decays convert two nucleons to an $H$.  In this case the stability of nuclei is a more stringent constraint than the double-$\Lambda$ hypernuclear observations, but our results of the next subsection show that nuclear stability bounds can also be satisfied if the $H$ is sufficiently compact: $r_H \lsi ~1/4 ~r_N$ depending on mass and nuclear hard core radius. This option is vulnerable to experimental exclusion by SuperK.

In order to calculate transition rates we factor transition amplitudes into an amplitude describing an $H$-baryon-baryon wavefunction overlap times a weak interaction transition amplitude between strange and non-strange baryons. In subsection \ref{overlapcalc} we setup the theoretical apparatus to calculate the wavefunction overlap between $H$ and two baryons.  We determine the weak interaction matrix elements phenomenologically in subsection \ref{weakME}. Nuclear decay rates are computed in subsection \ref{convlifetimes} while  lifetime of the $H$ for various values of $m_H$ is found in \ref{metastable}. The results are reviewed and conclusions are summarized in section \ref{summaryEX}.

\subsection{Overlap of $H$ and two baryons}
\label{overlapcalc}

We wish to calculate the amplitudes for a variety of processes, some of which require one or more weak interactions to change strange quarks into light quarks. By working in pole approximation, we factor the problem into an $H$-baryon-baryon wavefunction overlap times a weak interaction matrix element between strange and non-strange baryons, which will be estimated in the next section. For instance, the matrix element for the transition of two nucleons in a nucleus $A$ to an $H$ and nucleus $A'$, $A_{NN} \rightarrow A'_H X $, is calculated in the $\Lambda \Lambda$ pole approximation, as the product of matrix elements for two subprocesses: a transition matrix element for formation of the $H$ from the $\Lambda \Lambda$ system in the nucleus, $ |{\cal M}|_{\{\Lambda \Lambda\} \rightarrow H~X}$, times the amplitude for a weak doubly-strangeness-changing transition, $|{\cal M}|_{NN \rightarrow \Lambda \Lambda}$:
\beq
|{\cal M}|_{A
\rightarrow A' _HX}=|{\cal M}|_{\{\Lambda \Lambda\} \rightarrow H~X}~|{\cal M}|_{NN \rightarrow \Lambda \Lambda}.
\eeq
We ignore mass differences between light and strange quarks and thus the spatial wavefunctions of all octet baryons are the same.  In this section we are concerned with the dynamics of the process and we suppress spin-flavor indices.

\subsubsection{Isgur-Karl Model and generalization to the $H$}
\label{IK}

The Isgur-Karl (IK) non-relativistic harmonic oscillator quark model~\cite{IK,faiman,bhaduri} was designed to reproduce the masses of the observed resonances and it has proved to be successful in calculating baryon decay rates~\cite{faiman}. In the IK model, the quarks in a baryon are described by the Hamiltonian 
\beq \label{hamiltonian} 
H=\frac {1}{2m} (p^2 _1+p^2 _2+p^2 _3)
+\frac{1}{2}k\Sigma_{i<j} ^3 (\vec {r}_i -\vec {r}_j)^2, 
\eeq 
where we have neglected constituent quark mass differences.  The wave function of baryons can then be written in terms of the relative positions of quarks and the center of mass motion is factored out. The relative wave function in this model is~\cite{faiman,bhaduri}
\beq 
\Psi _{B} (\vec{r}_1,\vec{r}_2,\vec{r}_3) = N_{B} \exp \left[{-\frac {\alpha_{B} ^2}{6}\Sigma_{i<j} ^3 (\vec {r}_i -\vec{r}_j)^2}\right], 
\eeq 
where $N_B$ is the normalization factor, $\alpha _B=\frac {1}{\sqrt{<r_B ^2>}}=\sqrt{3km}$, and $<r_B ^2>$ is the baryon mean charge radius squared. Changing variables to 
\beq\label{rholambda} 
\vec {\rho} =\frac {\vec {r_1} -\vec {r_2}}{\sqrt{2}},~\vec {\lambda}=\frac {\vec {r_1} +\vec {r_2}-2 \vec {r_3}}{\sqrt{6}},
\eeq 
reduces the wave function to two independent harmonic oscillators. In the ground state \beq
\Psi_{B} (\vec {\rho}, \vec {\lambda})=\left( \frac{\alpha_B}{\sqrt{\pi}} \right) ^3 \exp\left[ -\frac{\alpha_{B}^2}{2} (\rho ^2 + \lambda ^2)\right]. 
\eeq

One of the deficiencies of the IK model is that the value of the $\alpha_B$ parameter needed to reproduce the mass splittings of lowest lying $\frac {1}{2} ^+$ and $\frac {3}{2} ^+$ baryons, $\alpha_B = 0.406$ GeV, corresponds to a mean charge radius squared for the proton of $\sqrt{<r^2_{N}>}= \frac {1}{\alpha _B}=0.49~{\rm fm}$. This is distinctly smaller than the experimental value of $0.87$ fm. Our results depend on the choice of $\alpha_B$ and therefore we also report results using $\alpha_B = 0.221$ GeV which reproduces the observed charge radius at the expense of the mass-splittings.

Another concern is the applicability of the non-relativistic IK model in describing quark systems, especially in the case of the tightly bound $H$. With $r_H/r_N = 1/f$, the quark momenta in the $H$ are $\approx f$ times higher than in the nucleon, which makes the non-relativistic approach more questionable than in the case of nucleons. Nevertheless we adopt the IK model because it offers a tractable way of obtaining a quantitative estimate of the effect of the small size of the $H$ on the transition rate, and there is no other alternative available at this time.  For comparison, it would be very interesting to have a Skyrme model calculation of the overlap of an $H$ with two baryons.

We fix the wave function for the $H$ particle starting from the same Hamiltonian (\ref{hamiltonian}), but generalized to a six quark system.  For the relative motion part this gives 
\beq
\Psi_{H}=N_{H}\exp\left[-\frac{\alpha_{H}^2}{6}\sum _{i<j} ^6 (\vec{r_i} -\vec{r_j})^2\right]. 
\eeq 
The space part of the matrix element of interest, $\langle A'_{H}|A_{ \Lambda \Lambda }\rangle$, is given by the integral 
\beq 
\int \prod _{i=1} ^6 d^3\vec{r}_i \Psi_{\Lambda} ^{a} (1,2,3) \Psi _{\Lambda} ^{b} (4,5,6) \Psi_H(1,2,3,4,5,6). 
\eeq 
Therefore it is useful to choose variables for the $H$ wavefunction as follows, replacing 
\beq
\vec{r}_1,\vec{r}_2,\vec{r}_3,\vec{r}_4,\vec{r}_5,\vec{r}_6\rightarrow \vec {\rho}^{a},\vec{\lambda}^{a},\vec{\rho}^{b},\vec{\lambda}^{b}, \vec {a}, \vec {R}_{CM},
\eeq 
where $\vec{\rho}^{a(b)}$ and $\vec {\lambda}^{a(b)}$ are defined as in eq.~(\ref{rholambda}), with $a(b)$ referring to coordinates $1,2,3~(4,5,6)$.  (Since we are ignoring the flavor-spin part of the wavefunction, we can consider the six quarks as distinguishable and not worry about fermi statistics at this stage.)  We also define the center-of-mass position and the separation, $\vec{a}$, between initial baryons $a$ and $b$: 
\beq
\label{coord} \vec{R}_{CM}=\frac {\vec{R}_{CM}^{a}+\vec{R}_{CM}^{b}}{2},~ \vec{a}=\vec{R}_{CM}^{a}-\vec{R}_{CM}^{b}.
\eeq 
Using these variables, the $H$ ground state wave function becomes
\begin{eqnarray}
\Psi_{H}&=&\left( \frac{3}{2}\right) ^{3/4}
\left( \frac{\alpha _H}{\sqrt{\pi}} \right)^{15/2}\\
&\times & \exp[-\frac {\alpha_{H} ^2}{2} (\vec {\rho^{a}}^2 + \vec{\lambda ^{a}}^2+\vec {\rho^{b}}^2 + \vec {\lambda ^{b}}^2 +\frac{3}{2} \vec {a}^2)]. \nonumber
\end{eqnarray}
As for the 3-quark system, $\alpha _H=\frac {1}{\sqrt{<r_H ^2>}}$.

\subsubsection{Nuclear Wavefunction} \label{BBG}

We will use two different wavefunctions to describe two $\Lambda$'s or nucleons in a nucleus, in order to study the model dependence of our results and to elucidate the importance of different aspects of the nuclear wavefunction. A commonly used wavefunction is the Miller-Spencer (MS) wavefunction\cite{MillerSpencer}:
\beq
\label{MS}
\psi _{MS}=1-\exp ^{-c_1 a^2}(1-c_2 a^2),
\eeq
with the canonical parameter choices $c_1 = 1.1$ fm$^{-2}$ and $c_2=0.68$ fm$^{-2}$.  It must be emphasized that at the short distances relevant for this calculation, the form and magnitude of the MS wavefunction are not constrained experimentally and rather are chosen to give a good fit to long-distance physics with a simple functional form.  The other wavefunction we use (BBG) is a solution of the Bruecker-Bethe-Goldstone equation describing the interaction of a pair of fermions in an independent pair approximation; see, {\it e.g.}, \cite{walecka}.   It is useful because we can explicitly explore the sensitivity of the result to the unknown short-distance nuclear physics by varying the hard-core radius.

The BBG wave function is obtained as follows. The solution of the Schrodinger equation for two fermions in the Fermi sea interacting through a potential $v({\vec x}_1,{\vec x}_2)$ takes the form 
\beq
\psi (1,2)=\frac {1}{\sqrt{V}}~e^{i{\vec P}{\vec R}_{CM}}
~\psi ({\vec a}),
\eeq 
where ${\vec R}_{CM}$ and ${\vec a}$ are defined as in (\ref {coord}). The first factor contains the center-of-mass motion and the second is the internal wave function of the interacting pair.  $\psi ({\vec a})$ is a solution of the Bethe-Goldstone equation (eq.~(36.15) in~\cite{walecka}) which is simply the Schrodinger equation for two interacting fermions in a Fermi gas, where the Pauli principle forbids the appearance of intermediate states that are already occupied by other fermions. Both wave functions are normalized so that the space integral of the modulus-squared of the wave function equals one. In the application of this equation to nuclear matter, the interaction of each particle from the pair with all particles in the nucleus through an effective single particle potential is included, in the independent pair approximation known as Bruecker theory (see eq.~(41.1) and (41.5) in~\cite{walecka}).

We are interested in s-wave solutions to the Bethe-Goldstone equation since they are the ones that penetrate to small relative distances. Following \cite{walecka}, an s-wave solution of the internal wave function is sought in the form 
\beq 
\psi (a)\sim \frac{u(a)}{a},
\eeq 
which simplifies the Bethe-Goldstone equation to 
\beq 
\left( \frac {d^2}{dx^2}+k^2 \right)u(a)=v(a)u(a)-\int ^{\infty} _0\chi (a,y)v(y)u(y)dy 
\eeq 
where $v(a)$ is the single particle potential in the effective-mass approximation, and the kernel $\chi (a,y)$ is given by 
\beq 
\chi (a,y)=\frac{1}{\pi} \left[ \frac{\sin k_F(a-y)}{a-y}-\frac{\sin k_F (a+y)}{a+y}\right],
\eeq
where $k_F$ is the Fermi wavenumber. For the interaction potential between two nucleons in a nucleus we choose a hard core potential for the following reasons.  The two particle potential in a nucleus is poorly known at short distances. Measurements (the observed deuteron form factors, the sums of longitudinal response of light nuclei,...) only constrain two-nucleon potentials and the wave functions they predict at internucleon distances larger than $0.7$ fm~\cite{pandharipande}. The Bethe-Goldstone equation can be solved analytically when a hard-core potential is used.  While the hard-core form is surely only approximate, it is useful for our purposes because it enables us to isolate the sensitivity of the results to the short-distance behavior of the wavefunction. We stress again, that more ``realistic" wavefunctions, including the MS wave function, are in fact not experimentally constrained for distances below $0.7$ fm. Rather, their form at short distance is chosen for technical convenience or aesthetics.

Using the hard core potential, the s-wave BG wavefunction is 
\beq
\Psi_{BG}(\vec{a})=\left\{\begin{array}{ll}
N_{BG}\frac{u(a)}{a} & \textrm{for \quad $a>\frac{c}{k_F}$} \\
0  & \textrm {for $\quad a<\frac{c}{k_F}$}
\end{array}\right.,
\eeq
with
\beq \label{Nbg}
N_{BG}=\frac {1}{\sqrt{\int^{R(A)} _{\frac {c}{k_F}} \left| \frac {u(a)}{a} \right| ^2 4\pi ~a^2~d a}},
\eeq 
where $\frac{c}{k_F}$ is the hard core radius and $R(A)=1.07 A^{1/3}$ is the radius of a nucleus with mass number $A$.  Expressions for $u$ can be found in~\cite{walecka}, eq.~(41.31). The normalization factor $N_{BG}$ is fixed setting the integral of $|\psi _{BG}|^2$ over the volume of the nucleus equal to one. The function $u$ vanishes at the hard core surface by construction and then rapidly approaches the unperturbed value, crossing over that value at the so called ``healing distance''. At large relative distances and when the size of the normalization volume is large compared to the hard core radius, $u(a)/a$ approaches a plane wave and the normalization factor $N_{BG}$ (\ref{Nbg}) reduces to the value $1/\sqrt{V_{\rm box}}$, as
\beq \label{pwBGGrln} 
\psi_{BG}(a)=N_{BG}~\frac{u(a)}{a}~\rightarrow \frac {1}{\sqrt{V_{\rm box}}}~e^{ika}. 
\eeq

\subsubsection{Overlap Calculation}

The non-relativistic transition matrix element for a transition $\Lambda \Lambda \rightarrow H$ inside a nucleus is given by (suppressing spin and flavor)
\begin{eqnarray} \label{matrixel}
T_{\{\Lambda \Lambda\}\rightarrow H}&=&2 \pi i \delta (E) \int d^3
a~ d^3 R_{CM} \prod _{i=a,b}
d^3 \rho^i d^3 \lambda ^i \nonumber \\
&\times &  ~\psi^* _H \psi ^a _{\Lambda}~\psi^b _{\Lambda}~\psi
_{nuc}~ e^{i({\vec k}_H-{\vec k}_{\Lambda \Lambda}){\vec R}_{CM}},
\end{eqnarray}
where $\delta (E)=\delta (E_H-E_{\Lambda \Lambda})$, $\psi ^{a,b}_{\Lambda}=\psi ^{a,b} _{\Lambda}(\vec {\rho}^{a,b},\vec{\lambda}^{a,b})$, and  $\psi _{nuc}=\psi _{nuc}({\vec a})$ is the relative wavefunction function of the two $\Lambda 's$ in the nucleus.  The notation $\{\Lambda \Lambda\}$ is a reminder that the $\Lambda$'s are in a nucleus. The plane waves of the external particles contain normalization factors $1/\sqrt{V}$ and these volume elements cancel with volume factors associated with the final and initial phase space when calculating decay rates. The integration over the center of mass position of the system gives a 3-dimensional momentum delta function and we can rewrite the transition matrix element as 
\beq \label{matrixel2} 
T_{\{\Lambda\Lambda\} \rightarrow H}=(2\pi)^4 i\delta ^4(k_f-k_i)~{\cal M}_{\{\Lambda \Lambda\} \rightarrow H}, 
\eeq 
where $|{\cal M}|_{\{\Lambda \Lambda\} \rightarrow H}$ is the integral over the remaining internal coordinates in eq.~(\ref{matrixel}). In the case of pion or lepton emission, plane waves of the emitted particles should be included in the integrand. For brevity we use here the zero momentum transfer, $\vec {k} =0$ approximation, which we have checked holds with good accuracy; this is not surprising since typical momenta are $\lsi 0.3$ GeV.

\begin{figure} 
\begin{center}
\includegraphics*[width=8cm]{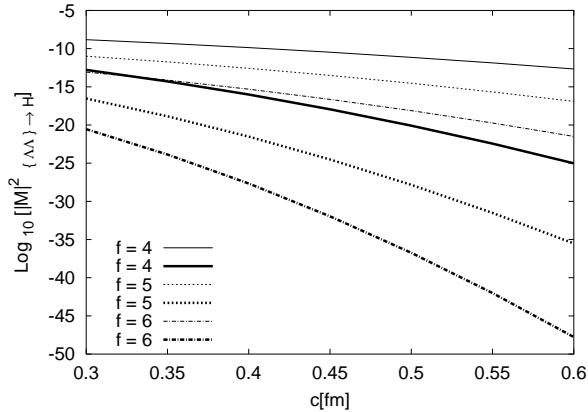}
\end{center}
\caption{ Log$_{10}$ of $|{\cal M}|^2_{\Lambda \Lambda \rightarrow H}$
versus hard core radius in fm, for ratio $f=R_N/R_H$ and two
values of the Isgur-Karl oscillator parameter: $\alpha_B= 0.406$ GeV (thick lines)
and $\alpha_B = 0.221$ GeV (thin lines).}\label{figoverlap1}
\end{figure}

Inserting the IK and BBG wavefunctions and performing the Gaussian integrals analytically, the overlap of the space wave functions becomes
\begin{eqnarray} \label{overlap}
|{\cal M}|_{\Lambda \Lambda \rightarrow H}&=&\frac {1}{\sqrt{4}} \left(\frac {2f}{1+f^2}\right )^6 \left( \frac{3}{2}\right)^{3/4}\left( \frac{\alpha _H}{\sqrt{\pi}} \right)^{3/2}\\
\nonumber &\times & N_{BG}\int^{R(A)} _{\frac{c}{k_F}} d^3 a\frac {u(a)}{a}e ^{-\frac {3}{4}\alpha_{H} ^2 a^2}
\end{eqnarray}
where the factor $1/\sqrt{4}$ comes from the probability that two nucleons are in a relative s-wave, and $f$ is the previously-introduced ratio of nucleon to $H$ radius; $\alpha _H=f~\alpha _B $. Since $N_{BG}$ has dimensions  $V^{-1/2}$ the spatial overlap ${\cal M}_{\{\Lambda \Lambda\} \rightarrow H}$ is a dimensionless quantity, characterized by the ratio $f$, the Isgur-Karl oscillator parameter $\alpha_B$, and the value of the hard core radius.  Fig.~\ref{figoverlap1} shows $|{\cal M}|^2_{\{\Lambda \Lambda\} \rightarrow H}$ calculated for oxygen nuclei, versus the hard-core radius, for a range of values of $f$, using the standard value of $\alpha_B= 0.406$ GeV for the IK model~\cite{bhaduri} and also $\alpha_B = 0.221$ GeV for comparison.

\begin{figure} 
\begin{center}
\includegraphics*[width=8cm]{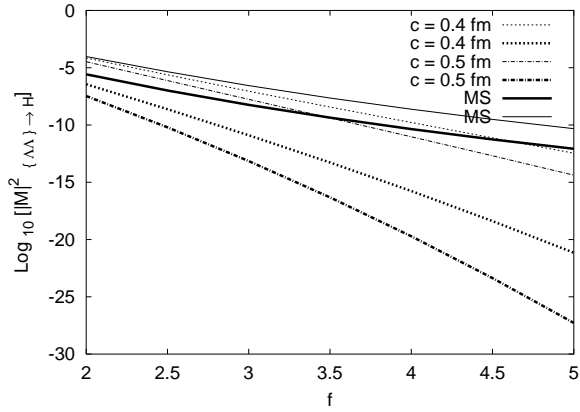}
\end{center}
\caption{Log$_{10}$ of $|{\cal M}|^2_{\Lambda \Lambda
\rightarrow H}$ versus ratio $f=\alpha _H/\alpha _N$, calculated with BBG wave function with core radius $0.4$ and $0.5$ fm, and with the MS wave function. Thick (thin) lines are for $\alpha_B= 0.406$ GeV ($\alpha_B = 0.221$ GeV) in the IK wavefunction.} \label{figoverlap2}
\end{figure}

Fig.~\ref{figoverlap1} shows that, with the BBG wavefunction, the overlap is severely suppressed and that the degree of suppression is very sensitive to the core radius.  This confirms that the physics we are investigating depends on the behavior of the nuclear wavefunction at distances at which it is not directly constrained experimentally. Fig. \ref{figoverlap2} shows a comparison of the overlap using the Miller Spencer and BBG nuclear wavefunctions, as a function of the size of the $H$. One sees that the spatial overlap is strongly suppressed with both wavefunctions, although quantitatively the degree of suppression differs.  We cannot readily study the sensitivity to the functional form of the baryonic wavefunctions, as there is no well-motivated analytic form we could use to do this calculation other than the IK wavefunction.  However by comparing the extreme choices of parameter $\alpha_B$  in the IK wavefunction, also shown in Figs. \ref{figoverlap1} and \ref{figoverlap2}, we explore the sensitivity of the spatial overlap to the shape of the hadronic wavefunctions.  Fortunately, we will be able to use additional experimental information to constrain the wavefunction overlap so that our key predictions are insensitive to the overlap uncertainty.

\subsection{Weak Interaction Matrix Elements}
\label{weakME}

\begin{figure}
\begin{center}
\includegraphics*[width=8cm]{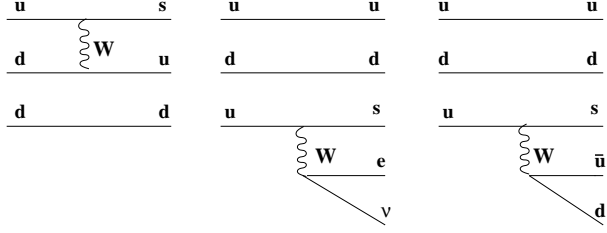}
\end{center}
\caption{Some relevant weak transitions for $NN \rightarrow HX$} \label{figweaktrans}
\end{figure}

Transition of a two nucleon system to off-shell $\Lambda\Lambda$ requires two strangeness changing weak reactions. Possible $\Delta S=1$ sub-processes to consider are a weak transition with emission of a pion or lepton pair and an internal weak transition.  These are illustrated in Fig. \ref{figweaktrans} for a three quark system. We estimate the amplitude for each of the sub-processes and calculate the overall matrix element for transition to the $\Lambda \Lambda$ system as a product of the sub-process amplitudes. \\


The matrix element for weak pion emission is estimated from the $\Lambda\rightarrow N \pi$ rate: 
\beq 
|{\cal M}|^2_{\Lambda\rightarrow N \pi}=\frac {1}{(2\pi )^4} ~ \frac{2m_{\Lambda} }{\Phi _2} \frac {1}{\tau _{\Lambda\rightarrow N\pi}}=0.8 \times 10^{-12} \quad {\rm GeV}^{2}. 
\eeq 
By crossing symmetry this is equal to the desired $|{\cal M}|^2_{N\rightarrow \Lambda \pi}$, in the approximation of momentum-independence which should be valid for the small momenta in this application. Analogously, for lepton pair emission we have 
\beq 
|{\cal M}|^2_{\Lambda\rightarrow N e\nu }=\frac {1}{(2\pi )^4}~\frac {2 m_{\Lambda}  } {\Phi _3 }\frac{1}{ \tau _{\Lambda\rightarrow N e\nu }} =3 \times 10^{-12}.
\eeq

The matrix element for internal conversion, $(uds) \rightarrow (udd)$, is proportional to the spatial nucleon wave function when two quarks are at the same point: 
\beq 
|{\cal M}|_{\Lambda\rightarrow N} \approx <\psi _{\Lambda }|\delta^3 (\vec {r}_1-\vec {r}_2)|\psi _N > \frac {G_F \sin \theta _c \cos\theta _c}{m_q}, 
\eeq 
where $m_q$ is the quark mass introduced in order to make the 4 point vertex amplitude dimensionless\cite{yaouanc}. The expectation value of the delta function can be calculated in the harmonic oscillator model to be 
\beq \label{delta1} 
<\psi _{\Lambda }|\delta^3 (\vec {r}_1-\vec{r}_2)|\psi _N >~ = \left(\frac {\alpha _B}{\sqrt {2\pi}}\right)^3=0.4 \times 10^{-2} ~~ {\rm GeV}^3. 
\eeq 
The delta function term can also be inferred phenomenologically in the following way, as suggested in \cite{yaouanc}. The Fermi spin-spin interaction has a contact character depending on $~\vec {\sigma_1}\vec { \sigma_2}/m^2 _q \delta(\vec {r}_1-\vec {r}_2)$, and therefore the delta function matrix element can be determined in terms of electromagnetic or strong hyperfine splitting:
\begin{eqnarray}
(m_{\Sigma ^0}-m_{\Sigma ^+} )-(m_n-m_p)=\alpha \frac {2\pi
}{3m^2 _q}<\delta^3(\vec {r}_1-\vec {r}_2)>\\m_{\Delta} -m_N=
\alpha _S \frac {8\pi }{3 m^2 _q} <\delta^3(\vec {r}_1-\vec
{r}_2)>,
\end{eqnarray}
where $m_q$ is the quark mass, taken to be $m_N/3$. Using the first form to avoid the issue of scale dependence of $\alpha_S$ leads to a value three times larger than predicted by the method used in eq.~(\ref{delta1}), namely: 
\beq \label{delta2} 
<\psi_{\Lambda }|\delta^3 (\vec {r}_1-\vec {r}_2)|\psi _N> ~ =1.2\times 10^{-2} \quad {\rm GeV}^3. 
\eeq 
We average the expectation values in eq.~(\ref{delta1}) and eq.~(\ref{delta2}) and adopt 
\beq \label{MdeltaS}
|{\cal M}|^2_{\Lambda\rightarrow N}=4.4 \times 10^{-15}. 
\eeq 
In this way we have roughly estimated all the matrix elements for the relevant sub-processes based on weak-interaction phenomenology.


\subsection{Nuclear decay rates} \label{nuclifetime}

\subsubsection{Lifetime of doubly-strange nuclei}
\label{hypernuc}

The decay rate of a doubly-strange nucleus is:
\begin{eqnarray} \label{doubleform}
\Gamma_{A_{\Lambda \Lambda} \rightarrow A'_{H} \pi} &\approx&
K^2(2\pi )^4 \frac {m^2 _q }{2(2m_{\Lambda \Lambda})} \\
\nonumber &\times& \Phi _2 |{\cal M}|^2_{\Lambda \Lambda
\rightarrow H},
\end{eqnarray}
where $\Phi _2$ is the two body final phase space factor, defined as in~\cite{PDG02}, and $m_{\Lambda \Lambda}$ is the invariant mass of the $\Lambda$'s, $\approx 2 m_{\Lambda}$. The factor $K$ contains the transition element in spin flavor space. It can be estimated by counting the total number of flavor-spin states a $uuddss$ system can occupy, and taking $K^2$ to be the fraction of those states which have the correct quantum numbers to form the $H$. That gives $K^2\sim 1/1440$, and therefore we write $K^2 = (1440~\kappa_{1440})^{-1}$.  Combining these factors we obtain the estimate for the formation time of an $H$ in a doubly-strange hypernucleus
\beq \label{tauform}
\tau_{\rm form} \equiv \tau_{A_{\Lambda \Lambda}\rightarrow A'_{H}
\pi}\approx \frac {3(7)~\kappa_{1440}~10^{-18}~ {\rm s} }{ |{\cal M}|^2_{\Lambda
\Lambda \rightarrow H}}, 
\eeq 
where the phase space factor was evaluated for $m_H = 1.8 (2)$ GeV.

Fig.~\ref{figoverlap2} shows $|{\cal M}|^2_{\{\Lambda \Lambda\} \rightarrow H}$ in the range of $f$ and hard-core radius where its value is in the neighborhood of the experimental limits, for the standard choice $\alpha_B = 0.406$ GeV and comparison value $\alpha_B =0.221$ GeV.  In order to suppress $\Gamma(A_{\Lambda\Lambda}\rightarrow A'_{H} X)$ sufficiently that some $\Lambda$'s in a double-$\Lambda$ hypernucleus will decay prior to formation of an $H$, we require $|{\cal M}|^2_{\Lambda \Lambda \rightarrow H}\lsi 10^{-8}$.  If the nucleon hard core potential is used, this is satisfied even for relatively large $H$, {\it e.g.}, $r_H \lsi r_N/2.3~(r_N/2.1) $ for a hard-core radius $0.4$ ($0.5$) fm and can also be satisfied with the MS wave function as can be seen in Fig.~\ref{figoverlap2}. Thus the observation of single $\Lambda$ decay products from double-$\Lambda$ hypernuclei cannot be taken to exclude the existence of an $H$ with mass below $2 m_\Lambda$ unless it can be demonstrated that the wave function overlap is large enough.


\subsubsection{Conversion of $\Delta S=0$ nucleus to an $H$}
\label{convlifetimes}

If the $H$ is actually stable ($m_H < 2 m_p + 2 m_e$) two nucleons in a nucleus may convert to an $H$ and cause nuclei to disintegrate. $N N \rightarrow H X$ requires two weak reactions. If $m_H< 1740$ MeV two pion emission is allowed and the rate for the process $A_{NN}\rightarrow A'_{H}\pi \pi$, is approximately 
\begin{eqnarray}
\Gamma_{A_{NN} \rightarrow  A'_{H} \pi \pi }&\approx &K^2 \frac
{(2\pi )^4} {2 (2m_{N}) }~ \Phi_3\\ \nonumber
&\times & \left(
\frac { |{\cal M}|_{N\rightarrow \Lambda \pi} ^2 |{\cal
M}|_{\Lambda \Lambda \rightarrow H} } {(2m_{\Lambda }-m_H )^2
}\right) ^2
\end{eqnarray}
where the denominator is introduced to correct the dimensions in a way suggested by the $\Lambda \Lambda$ pole approximation. Since other dimensional parameters relevant to this process, {\it e.g.}, $m_q= m_N/3$ or $\Lambda_{QCD}$, are comparable to $2m_{\Lambda }-m_H$ and we are only aiming for an order-of-magnitude estimate, any of them could equally well be used. The lifetime for nuclear disintegration with two pion emission is thus 
\beq
\tau_{A_{NN}\rightarrow A'_{H}\pi \pi}\approx \frac
{40~\kappa_{1440}}{ |{\cal M}|^2_{\Lambda \Lambda \rightarrow H}}
\quad {\rm yr}, 
\eeq 
taking $m_H = 1.5$ GeV in the phase space factor.  For the process with one pion emission and an internal conversion, our rate estimate is
\begin{eqnarray}
\Gamma_{A_{NN}\rightarrow A'_{H}\pi}&\approx &K^2\frac {(2\pi
)^4}{2 (2m_{N})}~\Phi_2 \\ \nonumber
&\times &(|{\cal
M}|_{N\rightarrow \Lambda \pi} |{\cal M}|_{N\rightarrow \Lambda}
|{\cal M}|_{\Lambda \Lambda \rightarrow H})^2,
\end{eqnarray}
leading to a lifetime, for $m_H = 1.5$ GeV, of 
\beq
\tau_{A_{NN}\rightarrow A'_{H} \pi}\approx \frac
{3~\kappa_{1440}}{ |{\cal M}|^2_{\Lambda \Lambda \rightarrow H}}
\quad {\rm yr}. 
\eeq

If $m_H \gsi 1740$ MeV, pion emission is kinematically forbidden and the relevant final states are $e^+ \nu$ or $\gamma$; we now calculate these rates. For the transition $A_{NN}\rightarrow A'_H e\nu$, the rate is
\begin{eqnarray}
\Gamma_{A_{NN} \rightarrow  A'_{H}e\nu }&\approx &K^2\frac {(2\pi
)^4}{2 (2m_{N})}\Phi_3 \\ \nonumber &\times &(|{\cal
M}|_{N\rightarrow \Lambda e\nu} |{\cal M}|_{N\rightarrow \Lambda}
|{\cal M}|_{\Lambda \Lambda \rightarrow H})^2.
\end{eqnarray}
In this case, the nuclear lifetime is 
\beq \label{enu} 
\tau_{A_{NN}\rightarrow A'_{H} e\nu}\approx \frac {\kappa_{1440}}{|{\cal M}|^2_{\Lambda \Lambda \rightarrow H}}~10^{5} \quad {\rm yr}, 
\eeq 
taking $m_H = 1.8$ GeV.  For $A_{NN}\rightarrow A'_H\gamma$, the rate is approximately
\begin{eqnarray}
\Gamma_{A_{NN}\rightarrow A'_{H}\gamma }&\approx &K^2 (2\pi )^4
\frac {\alpha _{EM} m^2 _q}{2 (2m_{N})} \\ \nonumber &\times &
\Phi_2(|{\cal M}|^2 _{N\rightarrow \Lambda} |{\cal M}|_{\Lambda
\Lambda \rightarrow H})^2,
\end{eqnarray}
leading to the lifetime estimate 
\beq 
\tau_{A_{NN}\rightarrow A'_{H}\gamma}\approx \frac{2~\kappa_{1440}}{|{\cal M}|^2_{\Lambda\Lambda\rightarrow H}}~10^6 \quad {\rm yr}, 
\eeq
for $m_H = 1.8$ GeV.

One sees from Fig.~\ref{figoverlap1} that a lifetime bound of $\gsi {\rm few}~10^{29}$ yr is not a very stringent constraint on this scenario if $m_H$ is large enough that pion final states are not allowed.  {\it E.g.}, with $\kappa_{1440} = 1$ the rhs of eq.~(\ref{enu}) is $\gsi {\rm few}~10^{29}$ yr, for standard $\alpha_B$, a hard core radius of $0.45$ fm, and $r_H \approx 1/5~r_N$---in the middle of the range expected based on the glueball analogy.  If $m_H$ is light enough to permit pion production, experimental constraints are much more powerful.  We therefore conclude that $m_H \lsi 1740$ MeV is disfavored and is likely to be excluded, depending on how strong limits SuperK can give.

\begin{table}[hpb]
\caption{The final particles and momenta for nucleon-nucleon transitions to $H$ in nuclei. For the 3-body final states marked with *, the momentum given is for the configuration with $H$ produced at rest.}  \label{t1}
\begin{center}
\begin{tabular}{|c|c|c|c|}
\hline

mass        & final state & final momenta  & partial lifetime  \\
$m_H$ [GeV] & $A^\prime$ $H$ +    & $p$ [MeV]        & $ \times
K^2|{\cal M}|^2 _{\Lambda \Lambda \rightarrow H}$ [yr] \\ \hline

$1.5$         &  $\pi $     & $318$            & $2~10^{-3}$    \\ \hline
$1.5$         &  $\pi \pi$  & $170$*           & $0.03$           \\ \hline
$1.8$         &  $e \nu$    & $48$*            & $70$             \\ \hline
$1.8$         &  $\gamma$   & $96$             & $2~10^3$       \\ \hline

\end{tabular}
\end{center}
\end{table}

\subsection{Lifetime of an Unstable $H$} \label{metastable}

If $2 m_N \lsi m_H < m_N + m_\Lambda$, the $H$ is not stable but it proves to be very long lived if its wavefunction is compact enough to satisfy the constraints from doubly-strange hypernuclei discussed in section \ref{hypernuc}.  The limits on nuclear stability discussed in the previous section do not apply here because nuclear disintegration to an $H$ is not kinematically allowed.

\subsubsection{Wavefunction Overlap}

To calculate the decay rate of the $H$ we start from the transition matrix element eq.~(\ref{matrixel}). In contrast to the calculation of nuclear conversion rates, the outgoing nucleons are asymptotically plane waves. Nonetheless, at short distances their repulsive interaction suppresses the relative wavefunction at short distances much as in a nucleus.  It is instructive to compute the transition amplitude using two different approximations.  First, we treat the nucleons 
as plane waves so the spatial amplitude is:
\begin{eqnarray}
T_{H\rightarrow \Lambda \Lambda }&=&2 \pi i\delta (E_{\Lambda
\Lambda}-E_H) \int \prod _{i=a,b}
d^3 \rho^i d^3 \lambda ^i d^3 a~ d^3 R_{CM} \nonumber \\
&\times & \psi _H \psi ^{*a} _{\Lambda}~\psi^{*b} _{\Lambda}~
e^{i({\vec k}^a _N+{\vec k}^b _N-{\vec k}_{H}){\vec R}_{CM}}.
\end{eqnarray}
The integration over ${\vec R}_{CM}$ gives the usual 4D $\delta$ function. Using the Isgur-Karl wave function and performing the remaining integrations leading to $|{\cal M}|_{H\rightarrow \Lambda \Lambda}$, as in eq.~\ref{matrixel2}), the amplitude is:
\begin{eqnarray}
\label{planewaves}
|{\cal M}|_{H \rightarrow \Lambda
\Lambda}&=&\left (\frac {2f}{1+f^2}\right )^6 \left( \frac{3}{2}
\right)^{3/4}\left( \frac{\alpha _H}{\sqrt{\pi}} \right)^{3/2}\\
\nonumber &\times & \int^{\infty} _{0} d^3 a~ e ^{-\frac
{3}{4}\alpha_{H} ^2 a^2-i\frac{{\vec k}^a _N-{\vec k}^b
_N}{2}{\vec a}}\\ \nonumber &=& \left( \frac{8}{3\pi}
\right)^{3/4} \left (\frac {2f}{1+f^2}\right )^6 \alpha ^{-3/2}
_H~e^{-\frac{({\vec k}^a _N-{\vec k}^b _N)^2}{12~\alpha ^2 _H}}.
\end{eqnarray}
The amplitude depends on the size of the $H$ through the factor $f= r_N/r_H$. Note that the normalization $N_{BG}$ in the analogous result eq.~\ref{overlap}) which comes from the Bethe-Goldstone wavefunction of $\Lambda$'s in a nucleus has been replaced in this calculation by the plane wave normalization factor $1/\sqrt{V}$ which cancels with the volume factors in the phase space when calculating transition rates.

Transition rates calculated using eq.~(\ref{planewaves}) provide an upper limit on the true rates, because the calculation neglects the repulsion of two nucleons at small distances. To estimate the effect of the repulsion between nucleons we again use the Bethe-Goldstone solution with the hard core potential. It has the desired properties of vanishing inside the hard core radius and rapidly approaching the plane wave solution away from the hard core.  As noted in section~\ref{BBG}, $N_{BG}\rightarrow 1/\sqrt{V}$, for $a\rightarrow \infty$. Therefore, we can write the transition amplitude as in eq.~(\ref {overlap}), with the normalization factor $1/\sqrt{V}$ canceled with the phase-space volume element:
\begin{eqnarray} \label{ovlapfree}
|{\cal M}|_{H \rightarrow \Lambda \Lambda }&=&\left
(\frac {2f}{1+f^2}\right )^6 \left( \frac{3}{2}
\right)^{3/4}\left( \frac{\alpha _H}{\sqrt{\pi}} \right)^{3/2}\nonumber\\
&\times & \int^{\infty} _{0} d^3 a \frac {u(a)}{a}e ^{-\frac {3}{4}\alpha_{H} ^2 a^2}.
\end{eqnarray}

\begin{table}[hpb] \label{t2}
\caption{$|{\cal M}|_{H \rightarrow \Lambda \Lambda }^2 $ in ${\rm
GeV}^{-3/2}$ for different values of $f$ (rows) and nuclear wavefunction (columns), using the standard value $\alpha_{B1}=0.406$ GeV and the comparison value $\alpha_{B2}=0.221$ GeV in the IK wavefunction of the quarks. }
\begin{center}
\begin{tabular}{|c|c|c|c|c|c|c|}
\hline

  &  \multicolumn{2}{c|} {BBG, 0.4 fm}    & \multicolumn{2}{c|} {BBG, 0.5 fm}  & \multicolumn{2}{c|} {MS}               \\\cline{2-7}
            & $\alpha_{B1}$ & $\alpha_{B2}$ & $\alpha_{B1}$ & $\alpha_{B2}$    & $\alpha_{B1}$ & $\alpha_{B2}$  \\ \hline
4           & $~6~10^{-14}~$ & $~6~10^{-8}~$     &  $~7~10^{-18}~$ & $~4~10^{-9}~$ &  $~1~10^{-8}~$ & $~8~10^{-7}~$      \\ \hline
3           & $~5~10^{-9}~$ & $3~10^{-5}$     & $~3~10^{-11}~$ & $~7~10^{-6}~$ & $~2~10^{-6}~$ & $~9~10^{-5}~$            \\ \hline
2           & $~1~10^{-4}~$ & $~0.02~$           & $~1~10^{-5}~$ & $~0.01~$   & $~9~10^{-4}~$ & $~0.03~$             \\ \hline

\end{tabular}
\end{center}
\end{table}
This should give a more realistic estimate of decay rates.  Table \ref{t2} shows the overlap values for a variety of choices of $r_H$, hard-core radii, and $\alpha_B$.
Also included are the results with the MS wavefunction.

\subsubsection{Empirical Limit on Wavefunction Overlap}

As discussed in section~\ref{hypernuc}, the $H$ can be lighter than $2$ $\Lambda$'s without conflicting with hypernuclear experiments if it is sufficiently compact, as suggested by some models.  The constraint imposed by the hypernuclear experiments can be translated into an empirical upper limit on the wavefunction overlap between an $H$ and two baryons.   Using eq.~(\ref{tauform}) for the formation time $\tau_{\rm form}$ of an $H$ in a double-$\Lambda$ oxygen-16 hypernucleus we have 
\beq
\label{MBBHlim} |{\cal M}|^2_{\Lambda \Lambda \rightarrow H} = 7~
10^{-8}~ \frac{\kappa_{1440}}{f_{\rm form}} \left( \frac{\tau_{\rm
form}}{10^{-10}{\rm s}} \right)^{-1}, 
\eeq 
where $f_{\rm form} =\frac{\Phi_2(m_H)}{\Phi_2(m_H = 2 {\rm GeV})}$ is the departure of the phase space factor for hypernuclear $H$ formation appearing in eq.~(\ref{doubleform}), from its value for $m_H = 2$ GeV.  By crossing symmetry the overlap amplitudes $|{\cal M}|_{H \rightarrow \Lambda \Lambda }$ and $|{\cal M}|_{\Lambda \Lambda \rightarrow H}$ only differ because the $\Lambda$'s in the former are asymptotically plane waves while for the latter they are confined to a nucleus; comparing eqns.~(\ref{ovlapfree}) and (\ref{overlap}) we obtain: 
\beq
\label{Mreln} |{\cal M}|^2_{H \rightarrow\Lambda\Lambda}=\frac{4}{N^2_{BG}} |{\cal M}|^2_{\Lambda \Lambda \rightarrow H}.
\eeq 
For oxygen-16, $~\frac {N^2 _{BG}}{4} \approx \frac{1}{5~10^{4}}$ GeV$^3$.  Using eqns.~(\ref{MBBHlim}) and (\ref{Mreln}) will give us an upper limit on the overlap for the
lifetime calculations of the next section.

\subsubsection{ Decay rates and lifetimes}

Starting from $|{\cal M}|_{H \rightarrow \Lambda \Lambda}$ we can calculate the rates for $H$ decay in various channels, as we did for nuclear conversion in the previous section. The rate of $H\rightarrow nn$ decay is
\begin{eqnarray}\label{HtoNN}
\Gamma_{H\rightarrow nn}&\approx &K^2\frac {(2\pi )^4m^5 _q}{2~
m_{H}}~\Phi_2 (m_H)\\ \nonumber &\times & ( |{\cal M}|^2
_{N\rightarrow \Lambda} |{\cal M}|_{H \rightarrow \Lambda
\Lambda})^2,
\end{eqnarray}
where $\Phi _2$ is the phase space factor defined for $H\rightarrow nn$ normalized as in~\cite{PDG02}. Using eqs.~(\ref{Mreln}) and (\ref{MBBHlim}), the lifetime for $H\rightarrow nn$ is 
\beq\label{tau2wk}
\tau_{H\rightarrow NN} \approx  9(4) ~10^{7}~\mu_0 ~{\rm yr},
\eeq
for $m_H = 1.9 ~(2)$ GeV, where $\mu_0 \gtrsim 1$ is defined to be $(\tau_{\rm form} f_{\rm form})/(10^{-10}{\rm s})\times (5~10^4~N^2 _{BG})/4 $. The $H$ is therefore cosmologically stable, with a lifetime longer than the age of the Universe, if $|{\cal M}|^2_{\Lambda \Lambda \rightarrow H}$ is $10^{2-3}$ times smaller than needed to satisfy double hypernuclear constraints. As can be seen in Fig.~\ref{figoverlap2}, this corresponds to $r_H\lsi 1/3~r_N$ in the IK model discussed above. Note that $\kappa_{1440}$ and the sensitivity to the wavefunction overlap has been eliminated by using $\tau_{\rm form}$.

If $m_N + m_\Lambda~( 2.05~ {\rm GeV}) < m_H < 2 m_\Lambda~( 2.23$ GeV), $H$ decay requires only a single weak interaction so the rate in eq.  (\ref{HtoNN}) must be divided by $ |{\cal M}|^2 _{N\rightarrow \Lambda}$ given in eqn (\ref{MdeltaS}).  Thus we have
\beq \label{tau1wk}
\tau_{H\rightarrow N \Lambda} \approx 10~\mu_0 ~{\rm s}.
\eeq

Finally, if $ m_H > 2 m_\Lambda~( 2.23$ GeV), there is no weak interaction suppression and
\beq \label{tau0wk}
\tau_{H\rightarrow \Lambda \Lambda} \approx 4~10^{-14}~\mu_0 ~{\rm s}.
\eeq

Equations~(\ref{tau2wk})-(\ref{tau0wk}) with $\mu_0 = 1$ give the lower bound on the $H$ lifetime, depending on its mass.  This result for the $H$ lifetime differs sharply from the classic calculation of Donoghue, Golowich, and Holstein~\cite{donoghue:Hlifetime}, because we rely on experiment to put an upper limit on the wavefunction overlap $|{\cal M}|_{H \rightarrow \Lambda \Lambda}^2$.  Our treatment of the color-flavor-spin and weak interaction parts of the matrix elements is approximate, but it should roughly agree with the more detailed calculation of ref.~\cite{donoghue:Hlifetime}, so the difference in lifetime predictions indicates that the spatial overlap is far larger in their bag model than using the IK and Bethe-Goldstone or Miller-Spencer wavefunctions with reasonable parameters consistent with the hypernuclear experiments.  The bag model is not a particularly good description of sizes of hadrons, and in the treatment of~\cite{donoghue:Hlifetime} the $H$ size appears to be fixed implicitly to some value which may not be physically realistic. Furthermore, it is hard to tell whether their bag model analysis gives a good accounting of the known hard core repulsion between nucleons.   As our calculation of previous sections shows, these are crucial parameters in determining the overlap.  The calculation of the weak interaction and color-flavor-spin matrix elements in ref.~\cite{donoghue:Hlifetime} could be combined with our phenomenological approach to the spatial wavefunction overlap to provide a more accurate yet general analysis.  We note that due to the small size of the $H$, the p-wave contribution should be negligible.

\section{Binding of flavor singlets to nuclei} \label{binding}

After calculating the constraints implied on the $H$ from nuclear transition processes we turn to the calculation of $H$ nuclear binding, ref.~\cite{fz:nucbind}. In section~\ref{expts} we concluded that the relevant experimental constraints on exotic nuclei can place strong constraints on the abundance of the $H$ in the case it binds to nuclei. In this section we explore binding of flavor singlet to nuclei. We summarize the theory of nuclear binding in subsection~\ref{nucth}, to set the framework for and to make clear the limitations of our computation. In subsection~\ref{calc} we analyze the binding of a flavor singlet scalar to nuclei, and calculate the minimum values of coupling constants needed for binding. Corresponding limits on nucleon-$H$ scattering are given in subsection~\ref{scat}.  Other flavor-singlets are also considered, in subsection~\ref{R0} and  elsewhere.  We summarize the results and give conclusions in section~\ref{summaryEX}.

\subsection{Nuclear binding-general}
\label{nucth}

QCD theory has not yet progressed enough to predict the two nucleon interaction {\it ab initio}. Models for nuclear binding are, therefore, constructed semi-phenomenologically and relay closely on experimental input.

The long range part of the nucleon-nucleon interaction (for distances $r\geq 1.5$ fm) is well explained by the exchange of pions, and it is given by the one pion exchange potential (OPEP). The complete interaction potential $v_{ij}$ is given by $v^{\pi}_{ij}+v^{R}_{ij}$, where $v^{R}_{ij}$ contains all the other (heavy meson, multiple meson and quark exchange) parts. In the one boson exchange (OBE) models the potential $v^{R}_{ij}$ arises from the following contributions:
\begin{itemize}

\item In the intermediate region (at distances around $r\sim 1$ fm) the repulsive vector meson ($\rho ,\omega$) exchanges are important.  A scalar meson denoted $\sigma$ was introduced to provide an attractive potential needed to cancel the repulsion coming from the dominant vector $\omega$ meson exchange in this region. Moreover, a spin-orbit part to the potential from both $\sigma$ and $\omega$ exchange is necessary to account for the splitting of the $P^3$ phase shifts in NN scattering.

\item At shorter scales ($r\lsi 1$ fm), the potential is dominated by the repulsive vector meson ($\rho ,\omega$) exchanges.

\item For $r\lsi 0.5$ fm a phenomenological hard core repulsion is introduced.

\end{itemize}
However, many of these OBE models required unrealistic values for the meson-nucleon coupling constants and meson masses. With this limitation the OBE theory predicts the properties of the deuteron and of two-nucleon scattering, although, it cannot reproduce the data with high accuracy.

A much better fit to the data is obtained by using phenomenological potentials. In the early 1990's the Nijmegen group~\cite{nij90} extracted data on elastic $NN$ scattering and showed that all $NN$ scattering phase shifts and mixing parameters could be determined quite accurately. $NN$ interaction models which fit the Nijmegen database with a $\chi ^2/N_{data}\sim 1$ are called 'modern'. They include Nijmegen models \cite{nijmod}, the Argonne $v_{18}$~\cite{argonmod} and CD-Bonn~\cite{cdbonnmod} potentials. These potentials have several tens of adjustable parameters, and give precision fits to a wide range of nucleon scattering data.

The construction of 'modern' potentials can be illustrated with the Nijmegen potential. That is an OBE model based on Regge pole  theory, with additional contributions to the potential from the exchange of a Pomeron and f, f' and $A_2$ trajectories. These new contributions give an appreciable repulsion in the central region, playing a role analogous to the soft or hard core repulsion needed in semi-phenomenological and OBE models.

Much less data exists on hyperon-nucleon interactions than on $NN$ interactions, and therefore those models are less constrained. For example the extension of the Nijmegen potential to the hyper-nuclear (YN) sector~\cite{nijYN} leads to under-binding for heavier systems. The extension to the $\Lambda \Lambda$ and $\Xi N$ channels cannot be done without the introduction of extra free parameters, and there are no scattering data at present for their determination.

The brief review above shows that the description of baryon binding is a difficult and subtle problem in QCD. Detailed experimental data were needed in order to construct models which can describe observed binding.  In the absence of such input data for the $H$ analysis, we must use a simple model based on scalar meson exchange described by the Yukawa potential, neglecting spin effects in the nucleon vertex in the first approximation.  We know from the inadequacy of this approach in the NN system that it can only be used as a crude guide. However since the strength of couplings which would be needed for the $H$ to bind to light nuclei are very large, compared to their expected values, we conclude that binding is unlikely. Thus limits on exotic nuclei cannot be used to exclude the existence of an $H$ or other compact flavor singlet scalar or spin-1/2 hadron.

\subsection{Binding of a flavor singlet to nuclei}
\label{calc}

The $H$ cannot bind through one pion exchange because of parity and also flavor conservation. The absorption of a pion by the $H$ would lead to an isospin $I=1$ state with parity $(-1)^{J+1}$, which could be $\Lambda \Sigma ^0$ or heavier $\Xi p$ composite states. These states have mass $\approx 0.1$ GeV higher than the mass of the $H$ (for $m_{\Lambda}+m_N\lsi m_H\lsi 2m_{\Lambda})$, which introduces a strong suppression in $2^{nd}$ order perturbation theory. Moreover, the baryons in the intermediate state must have relative angular momentum $\rm{L}=1$, in order to have odd parity as required; this introduces an additional suppression. Finally, production of $\Lambda \Sigma ^0$ or $\Xi N$ states is further suppressed due to the small size of the $H$, as explained in \S\ref{NucStab}. Due to all these effects, we conclude that the contribution of one or two pion exchange to $H$ binding is negligible.

The first order process can proceed only through the exchange of a flavor singlet scalar meson and a glueball. The lightest scalar meson is f(400-1100) (also called $\sigma$). The mass of the glueball is considered to be around $\sim 1.5$ GeV.  In Born approximation, the Yukawa  interaction leads to an attractive Yukawa potential between nucleons
\beq 
V(r)=-\frac {gg'}{4\pi} \frac{1}{r}e^{-\mu r}, 
\eeq 
where $\mu$ is the mass of the exchanged singlet boson s ($\sigma$ or glueball) and $g g'$ is the product of the s-H and s-nucleon coupling constants, respectively. The potential of the interaction of $H$ at a position $\vec r$ with a nucleus, assuming a uniform distribution of nucleon $\rho =\frac {A}{\rm V}$ inside a nuclear radius R, is then 
\beq 
V=-\frac {gg'}{4\pi}\frac{A}{{\rm V}} \int \frac{e^{ -\mu |\vec {r} -\vec {r'} |}}{|\vec{r} -\vec {r'} |}d^3 \vec{r'}, 
\eeq 
where A is the number of nucleons, ${\rm V}$ is the volume of the nucleus and $\vec {r} $ is the position vector of the $H$. After integration over the angles the potential is
\beq \label{3}
V=-\frac {3}{2} \frac {gg'}{4\pi}
\frac{1}{(1.35~{\rm fm}~\mu)^3} f(r),
\eeq
where we used $R=1.35 A^{1/3}$ fm;
\begin{displaymath} \label{5}
f(r) = \left \{ \begin{array}{ll}
 2 \mu \left[ 1 -(1+\mu R)~e^{-\mu R}~\frac{\sinh [\mu r]}{\mu r} \right]  & r\le R \\
2\mu \left[ \mu R\cosh [\mu R]-\sinh [\mu R] \right] \frac
{e^{-\mu r}}{\mu r} & r\ge R.
\end{array} \right.
\end{displaymath}
Throughout, we use $ \hbar = c = 1$ when convenient.

\begin{figure} 
\begin{center}
\includegraphics [width=8cm]{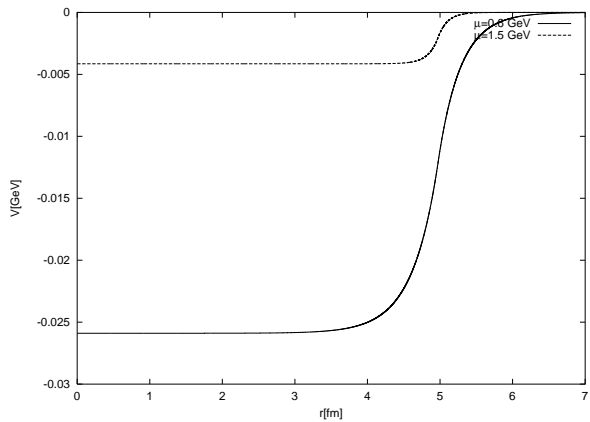}
\end{center}
\caption{Potential in GeV, for $\frac {gg'}{4\pi}$=1, A=50 and
$\mu=0.6~(\rm {dashed})$ or $\mu=1.5$ GeV (solid) as a function
of distance r.} \label{potentialfig}
\end{figure}

Fig.~\ref{potentialfig} shows the potential the nucleus presents to the $H$ for $A=50$, taking the mass of the exchanged boson to be $\mu =0.6$ and $1.5$ GeV. The depth of the potential is practically independent of the number of nucleons and becomes shallower with increasing scalar boson mass $\mu $.

Note that Born approximation is applicable at low energies and for small coupling constants; it may not be valid for $H$ binding. Born approximation is valid when 
\beq 
\frac {m}{\mu } \frac{gg'}{4\pi}<<1, 
\eeq 
where $m$ is the reduced mass and $\mu $ the mass of the exchanged particle. As we shall see, this condition is actually not satisfied for values of $g g'$ which assure binding for the $H$-mass range of interest.  This underlines the fact that no good first principle approach to nuclear binding is available at present.

\begin{figure}
\begin{center}
\includegraphics*[width=8cm]{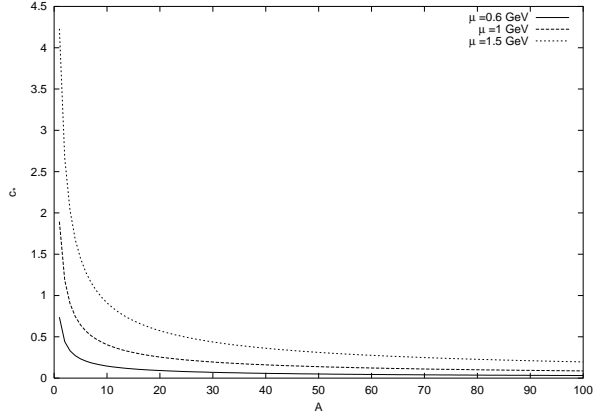}
\end{center}
\caption{ Critical value $c_*$ of the coupling constant product versus nuclear size needed for the $H$ to just bind, for $\mu$[GeV]$=0.7$ (dotted), $1.3$ (dashed) and $1.5$ (solid).} \label{cstarfig}
\end{figure}

We can now calculate the value of $c_*=\left( \frac{gg'}{4\pi}\right) _*$ for which the potential is equal to the minimum value needed for binding; in square well approximation this is given by 
\beq 
V_{min}=\frac{\pi ^2}{8R^2m}. 
\eeq 
Fig.~\ref{cstarfig} shows the dependence of $c_*$ on the mass of the exchanged particle, $\mu$. The maximum value of $c_*$ for which the $H$ does not bind decreases with increasing $H$ mass, and it gets higher with increasing mass of the exchanged particle, $\mu$.

The $H$ does not bind to light nuclei with $A\le4$, as long as the product of couplings 
\beq
c_* \leq [0.27,0.73,1.65],~{\rm for}~\mu=[0.6,1,1.5]~{\rm GeV}, 
\eeq
where $c = g_{NN\sigma}~ g_{HH\sigma}/(4\pi)$ or $g_{NNG}~ g_{HHG}/(4 \pi)$. The $H$ will not bind to heavier nuclei if 
\beq
c_*\leq [0.019,0.054,0.12],~{\rm for}~ \mu =[0.6,1,1.5]~{\rm GeV}. 
\eeq
In the next sections we will compare these values to expectations and limits.

It should also be noted that binding requires the product of coupling constants, $g g'$ to be positive and this may not be the case. Even in the case of hyperons, experimental information was necessary to decide whether the $\Xi$ has a relative positive coupling~\cite{dover}.

\subsection{Limits on {\bf {\it cm}} from Nucleon $H$ elastic scattering}
\label{scat}

The nucleon-$H$ elastic scattering cross section is expected to be very small, due to the compact size of the $H$ and the suppression of color fluctuations on scales $\lsi 1 ~{\rm GeV}^{-1}$ in the nucleon.  Ref.~\cite{f:StableH} estimates $\sigma_{HN} \lsi 10^{-3}$ mb.  This can be translated to an estimated upper limit on the product $c ~m$ which determines the potential presented to the $H$ by a nucleus, as follows.  In the one boson exchange model, the elastic $H$-$N$ cross section due to the $\sigma$- or glueball-mediated Yukawa interaction is given by 
\beq 
\frac{d\sigma }{d\Omega }= (2mc)^2 \frac {1}{(2p^2(1-\cos \theta
)+\mu ^2)^2}. 
\eeq  
In the low energy limit 
\beq
\label{eq:crossection} 
\sigma_{HN} =(2mc)^2\frac {4 \pi}{\mu ^4}. 
\eeq 
Writing $\sigma_{HN} = \sigma_{-3} 10^{-3}$ mb and $\mu = \mu_{\rm GeV} 1$ GeV, this gives 
\beq 
c~ m = 0.007\sqrt{\sigma_{-3}} ~\mu_{\rm GeV}^2 ~{\rm GeV}. 
\eeq  
Comparing to the values of $c^*$ needed to bind, we see that for $m_H<2 m_p$ this is too small for the $H$ to bind, even to heavy nuclei\footnote{We have summarized the net effect of possibly more than one exchange boson ({\it e.g.}, $\sigma$ and glueball) by a single effective boson represented by a $c_*^{\rm eff}$ and $\mu_{\rm eff}$.}. 

If dark matter consists of relic $H$'s, we can demonstrate that $H$'s do not bind to nuclei without relying on the theoretical estimate above for $\sigma_{HN}$. It was shown in~\cite{wandelt} that the XQC experiment excludes a dark matter-nucleon cross section $\sigma_{XN}$ larger than about $0.03$ mb for $m_X \sim 1.5$ GeV. Thus if dark matter consists of a stable $H$ it would require $\sigma_{XN} \le 0.03 $ mb, implying $c \le [0.01, 0.03, 0.06]$ for $\mu = [0.6,1.0,1.5]$ GeV and the $H$ would not bind even to heavy nuclei.

A generic new scalar flavor singlet hadron $X$ which might appear in an extension of the Standard Model, might not have a small size and correspondingly small value of $\sigma_{XN}$, and it might not be dark matter and subject to the XQC limit.  In that case, it is more useful to turn the argument here around to give the maximum $\sigma_{XN}^*$ above which the $X$ would bind to nuclei in the OBE approximation.  From eqn (\ref{3}),(\ref{5}) and $f(0) = 2\mu$ we have
\beq 
c_* = \frac{\pi^2 (1.35 ~{\rm fm}) \mu^2}{24 A^{2/3} m}.
\eeq 
Then eq.~(\ref{eq:crossection}) leads to 
\beq 
\sigma_{XN}^* \approx 155~A^{-4/3}~ {\rm mb}.
\eeq  
That is, for instance, if it is required that the $X$ not bind to He then it must have a cross section on nucleons lower than $25$ mb.

\subsection{Flavor singlet fermion}
\label{R0}

The analysis of the previous sections can be extended to the case of a flavor singlet fermion such as the glueballino -- the supersymmetric partner of the glueball which appears in theories with a light gluino~\cite{f:lightgluino}.  In this case the possible exchanged bosons includes, in addition to the $\sigma$ and the glueball, the flavor-singlet component of the pseudoscalar meson $\eta'$ ($m_\eta' = 958$ MeV).  However the size of the $R_0$ should be comparable to the size of the glueball, which was the basis for estimating the size of the $H$. That is, we expect $r_{R_0} \approx r_G \approx r_H$ and thus $\sigma_{R_0 N} \approx \sigma_{HN}$~\cite{f:StableH}. Then arguments of the previous section go through directly and show the $R_0$ is unlikely to bind to light nuclei\footnote{Nussinov\cite{nussinov:R0} considered that the $R_0$ would bind to nuclei, by assuming that the depth of the potential presented by the nucleus to the $R_0$ is at least $2$-$4$ MeV for $16\le A\le 56$.  However the discussion of the previous sections, with $\sigma_{R_0N} = 10^{-3}\sigma_{-3}$ mb, gives a potential depth of $0.07$ MeV $\sqrt{\sigma_{-3}}/(m_{R_0}/{\rm GeV})$.}.

\section{The Existence of the $H$ \-- Summary} \label{summaryEX}

In \S\ref{expts} we considered the constraints placed on the $H$ dibaryon by the stability of nuclei and hypernuclei with respect to conversion to an $H$, and we calculated the lifetime of the $H$ in the case when it is heavier than two nucleons. In the model calculations we used the Isgur-Karl wavefunctions for quark in baryons and the $H$, and the Miller-Spencer and Bruecker-Bethe-Goldstone wavefunctions for nucleons in a nucleus, to obtain a rough estimate of the $H$-baryon-baryon wavefunction overlap.  By varying the IK oscillator strength parameter and the hard-core radius in the BBG wavefunction over extreme ranges, we find that the wavefunction overlap is very sensitive to the size and shape of the hadronic and nuclear wavefunctions. With the BBG (MS) wavefunction, the hypernuclear formation time of an $H$ is comparable to or larger than the decay time for the $\Lambda$'s and thus the $H$ is not excluded, if $r_H \lsi 1/2~(1/3)~ r_N$.  \footnote{The overlap between an $H$ and two nucleons should be strongly suppressed also in the Skyrme model, in view of the totally different nature of the nucleon and $H$ solitons \cite{Balachandran,Sakai2}. However, a method for computing the overlap has not been developed so we are unable to explore this here.} We conclude that the observation of $\Lambda$ decays in double-$\Lambda$ hypernuclei cannot be used to exclude $m_H < 2 m_\Lambda$, given our present lack of understanding of the hadronic and nuclear wavefunctions. 

In the second part of our work we abstracted empirical relations which give us relatively model-independent predictions for the $H$ lifetime.  By crossing symmetry, the overlap of the wavefunctions of an $H$ and two baryons can be constrained using experimental limits on the formation time of an $H$ in a hypernucleus.  Using the empirically constrained wavefunction overlap and  phenomenologically determined weak interaction matrix elements, we can estimate the lifetime of the $H$ with relatively little model uncertainty.  We find:
\begin{enumerate}
\item If $m_H > 2 m_\Lambda$, the hypernuclear constraint is not applicable but the $H$ would still be expected to be long-lived, in spite of decaying through the strong interactions.  E.g., with the BBG wavefunction and $r_H \lsi 1/2 ~ r_N$,  $\tau_H \gsi 4~10^{-14}$ sec.
\item If $m_N + m_\Lambda \lsi m_H \lsi 2 m_\Lambda$, the $H$ lifetime is $\gsi 10$ sec.
\item If  $2 m_N \lsi m_H \lsi m_N + m_\Lambda$, the $H$ lifetime is $\gtrsim 10^8$ yr. For $r_H \lsi (1/3)~ r_N$ as suggested by some models, the $H$ lifetime is comparable to or greater than the age of the Universe.  
\end{enumerate}

Our results have implications for several experimental programs:
\begin{enumerate}
\item  The observation of $\Lambda$ decays from double $\Lambda$ hypernuclei excludes that $\tau_{\rm form}$, the formation time of the $H$ in a  double $\Lambda$ hypernucleus, is much less than $\tau_\Lambda$.  However if $\tau_{\rm form}$ is of order $\tau_\Lambda$, some double $\Lambda$ hypernuclei would produce an $H$.  One might hope these $H$'s could be observed by reconstructing them through their decay products, e.g., $H \rightarrow \Sigma^- p$.  Unfortunately, our calculation shows that $\tau_H \gsi 10$ sec for the relevant range of $m_H$, so any $H$'s produced would diffuse out of the apparatus before decaying.
\item  Some calculations have found $m_H < 2 (m_p + m_e)$, in which case the $H$ is absolutely stable and nucleons in nuclei may convert to an $H$.  We showed that SuperK can place important constraints on the conjecture of an absolutely stable $H$, or conceivably discover evidence of its existence, through observation of the pion(s), positron, or photon produced when two nucleons in an oxygen nucleus convert to an $H$. We estimate that SuperK could achieve a lifetime limit $\tau \gsi {\rm few} ~10^{29}$ yr. This is the lifetime range estimated with the BBG wavefunction for $m_H \gsi 1740$ MeV and $r_H \approx 1/5 ~r_N$.  An $H$ smaller than this seems unlikely, so $m_H \lsi 1740$ MeV is probably already ruled out. 
\end{enumerate}

In \S\ref{nucth} we first reviewed the theory of nuclear binding and emphasized that even for ordinary nucleons and hyperons there is not a satisfactory first-principles treatment of nuclear binding.  We showed that exchange of any pseudoscalar meson, or of two pseudoscalar octet mesons, or any member of the vector meson octet, makes a negligible contribution to the binding of an $H$ or other flavor singlet scalar hadron to a nucleon.  The dominant attractive force comes from exchange of a glueball or a $\sigma$ (also known as the f(400-1100) meson), which we treated with a simple one boson exchange model.  The couplings of $\sigma$ and glueball to the $H$ are strongly constrained by limits on $\sigma_{HN}$, to such low values that the $H$ cannot be expected to bind, even to heavy nuclei. Thus we conclude that the strong experimental limits on the existence of exotic isotopes of He and other nuclei do not exclude a stable $H$. 

More generally, our result can be applied to any new flavor singlet scalar particle X, another example being the $S^0$ supersymmetric hybrid baryon ($uds\tilde{g}$) discussed in \cite{f:lightgluino}. If $\sigma_{XN} \le 25~{\rm mb} ~{\rm GeV}/m_{X}$, the $X$ particle will not bind to light nuclei and is ``safe". Conversely, if $\sigma_{XN} >> 25~{\rm mb} ~{\rm GeV}/m_{X}$, the $X$ particle could bind to light nuclei and is therefore excluded unless if there is some mechanism suppressing its abundance on Earth, or it could be shown to have an intrinsically repulsive interaction with nucleons. This means the self-interacting dark matter (SIDM) particle postulated by Spergel and Steinhardt \cite{spergel:SIDM} to ameliorate some difficulties with Cold Dark Matter, probably cannot be a hadron. SIDM requires $\sigma_{XX} /M_X \approx 0.1 - 1 $ b/GeV; if $X$ were a hadron with such a large cross section, then on geometric grounds one would expect $\sigma_{XN} \approx 1/4 \sigma_{XX}$ which would imply the $X$ binds to nuclei and would therefore be excluded by experimental limits discussed above.

\section{The $H$ or $H$, $\bar{H}$ as Dark Matter}

As we have seen in the previous sections, the existence of the $H$ di-baryon is not ruled out by experiment if the $H$ is tightly bound and compact. If $m_H\lsi m_N+m_{\Lambda}$ and $r_H\lsi 1/3~r_N$ the $H$ can be cosmologically stable. In the rest of this section we explore the possibility that the DM consists of the $H$ or ${\bar H},H$ and therefore be predicted within the Standard Model. 
\begin{enumerate}
\item[1.] {\it The $H$ Dark Matter.} The number density of nonrelativistic species $i$ in thermal equilibrium is given by
\beq
n_i=g_i\left( \frac {m_iT}{2\pi}\right) ^2\exp\left[-\frac{m_i-\mu_i}{T}\right].
\eeq 
If baryon number is conserved we have $\mu_H=2\mu _N$, and therefore the nucleon and the $H$ energy densities are related as 
\beq\label{Hdmeq}
\frac{m_Hn_H}{m_nn_n}=\left(2\pi\right)^{3/2}\frac{g_H}{g^2 _N}\frac{n_n}{n_{\gamma}}n_{\gamma}\frac{m^{5/2} _H}{m^4 _nT^{3/2}}\exp\left[\frac{2m_n-m_H}{T}\right].
\eeq
The left-hand side is actually the ratio $\Omega_H/\Omega_N$ and for $H$ DM it is fixed, see eq. (\ref{wmappar}). Then, by solving eq. (\ref{Hdmeq}) we can find $T_{\rm f.o.}$, the temperature at which the $H$ has to go out of equilibrium in order to have the proper DM energy density today. The equation has a solution only for $m_H\lsi 2m_N$, a mass range which is unfavored by the discussion in \S\ref{convlifetimes}. The freeze-out temperatures are 15 (7) MeV, for $m_H=1.5~(1.7)$ GeV. These temperatures correspond to an age of the Universe of $0.003~(0.02)$ sec. By that time all strangeness carrying particles already decayed and therefore the $H$ cannot stay in equilibrium (for example, through reactions as $K^+H\leftrightarrow p\Lambda$) at such low temperatures. We see that even if the $H$ was cosmologically stable it could not be thermally produced in the early universe in sufficient amount to be dark matter.
\item[2.]{{\it The} $H  {\bar H}$ {\it  Dark Matter}.} In this case the $H$ would decouple as in the B-sequestration scenario discussed in \S \ref{scenario}. The reactions which would keep it in the equilibrium, up to the temperatures listed in Table 2.1, are
\begin{eqnarray}
{\bar H}N\leftrightarrow {\bar \Lambda}K\nonumber \\
H{\bar N}\leftrightarrow \Lambda{\bar K}.\nonumber
\end{eqnarray}       
In this scenario $\Lambda$s and $K$s stay in equilibrium sufficiently long, because the above reactions proceed in the left direction with the rates which are, for temperatures of interest, much higher than the decay rates of $\Lambda$ and $K$. The $H$ and ${\bar H}$ stay in equilibrium through these interactions and may reach DM abundance.   
\end{enumerate}
In the next Chapter we explore direct and indirect experimental constraints on the $H{\bar H}$ DM scenario.

\chapter{Dark Matter constraints in the range of the $H$ parameters} \label{Hdmcons}

This Chapter is organized as follows: we calculate direct DM detection experiments in the mass and cross section range expected for the $H$ in \S~\ref{directDM} and indirect constraints that could be place on ${\bar H}$ DM in \S~\ref{Hindirect}; we show that the ${\bar H}$ DM could be ruled out from the heat production in Uranus.

\section{Direct DM Detection \-- Light DM Constraints} \label{directDM}

In this Section we focus on direct detection experiments in order to see whether the $H$ is allowed as a DM candidate given that his expected cross section with nucleons is of the order $\mu $b (see discussion in \S~\ref{Tightlybound}). More generally we explore the limits on light DM, with primary interest in $\mu $b cross section range. 

The region of mass $m \gsi 10$ GeV is well explored today, up to TeV range, from strong ($\sigma \sim 10$ mb) to weak ($\sigma \sim 10^{-6}$pb) cross sections. Here we explore the possibility that the DM mass is in the range $0.4 \lsi m \lsi 10$ GeV. Masses below $\sim 0.4$ GeV are below the threshold of direct detection DM experiments and are therefore unconstrained, with the exception of axion searches. 

The mass range $m\lsi 10$ GeV has not yet been explored carefully. Several dark matter underground experiments have sufficiently low mass threshold today: the CRESST \cite{cresst}, DAMA \cite{dama}, IGEX \cite{igex}, COSME \cite{cosme} and ELEGANT \cite{elegant} experiments. Except for DAMA, these experiments have published upper limits on the cross section assuming it is weak, but have not addressed the case of stronger cross sections,\footnote{DAMA did publish strong cross section limits in \cite{damastrong}, but they were based on a dedicated experiment which had a higher mass threshold $m\gsi 8$ GeV.} where the approach for extracting the cross section limits is substantially different, as we explain below. Also, recent data from an X-ray experiment XQC, \cite{XQC} proved to be valuable in constraining low mass DM, but limits based on the final data have not yet been derived. Since XQC is a space\-- based experiment it is especially suitable for exploring the higher cross section range. In \cite{wandelt} it was shown that in the low mass range the XQC experiment rules out Strongly Interacting DM (SIMPs, \cite{spergel:SIDM}). Dark matter with low masses and 'intermediate' cross sections, several orders of magnitude smaller than normal hadronic cross sections, remains to be fully analyzed and that is the focus of this work. We will abbreviate DM with intermediate cross section on nucleons as DMIC.       

Early limits from DMIC direct detection experiments can be found in the paper \cite{RRS} by Rich, Rocchia and Spiro in which they reported results from a 1987 balloon experiment. Starkman et al. \cite {SG} reviewed DM constraints down to a mass of 1 GeV as of 1990. Wandelt et al. \cite{wandelt} added constraints based on preliminary data from the more recent XQC sounding rocket experiment. The above constraints are discussed and updated further in the text. In previous works on this topic the effect of the halo particle's velocity distribution on the cross section limit was not explored. Since the only detectable light particles are those in the exponential tail of the velocity distribution, the limits on light DM are sensitive to the parameters in the velocity distribution, in particular to the value of the escape velocity cutoff. We investigate this sensitivity in the standard Isothermal Sphere model, where the DM velocity distribution function is given by a truncated Maxwell-Boltzmann distribution. We also consider the spin-independent and spin-dependent interaction cases separately. Except in Section \ref{fraction}, we assume a single type of DM particle.

\subsection{Direct Dark Matter Detection} \label{directdet}

The basic principle of DM detection in underground experiments is to measure the nuclear recoil in elastic collisions, see for example \cite{Lewin}. The interaction of a DM particle of mass $m\lsi 10$ GeV, produces a recoil of a nucleus of 20 keV or less. The recoil energy (which grows with DM mass as in (2) below) can be measured using various techniques. The detectors used so far are: 
\begin{itemize}
\item{ionization detectors} of Ge (IGEX, COSME) and of Si
\item{scintillation crystals} of NaI (DAMA, ELEGANT)
\item{scintillators} of CaF$_2$
\item{liquid or liguid gas} Xenon detectors (DAMA, UKDMC)
\item{thermal detectors (bolometers)} with saphire (CRESST, ROSEBUD), telurite or Ge (ROSEBUD) absorbers
\item{bolometers, which also measure the ionization} like that of Si (CDMS) and Ge (CDMS, EDELWEISS).
\end{itemize}
As we have seen in the Introduction, the most accepted DM candidate in GeV range is the neutralino. The expected signal for neutralino-nucleon scattering is in the range $10$ to $10^{-5}$ counts/kgday, \cite{SUSYmorales}. The smallness of signal dictates the experimental strategies, detectors have to be highly radio pure, and have substantial shielding (they are built deep underground).
  
For a given velocity distribution $f({\vec v})$, the differential rate per unit recoil energy $E_R$ in (kg day keV)$^{-1}$ in the detector can be expressed as
\beq 
\label{cr}
\frac {dR}{dE_R} = N_T~n_{X} \int _{v_{min}} ^{v_{esc}} ~d{\vec v}~|{\vec v}|~f({\vec v})~g({\vec v}) \frac{d \sigma_{XA}}{dE_R}, 
\eeq
where $n_{X}$ is the number density of DM particles, $N_T$ is the number of target nuclei per kg of target, $\sigma _{XA}$ is the energy dependent scattering cross section of DM on a nucleus with mass number A, $g({\vec v})$ is the probability that a particle with velocity $v$ deposits an energy above the threshold $E_{TH}$ in the detector, and $v_{min}$ is the minimum speed the DM particle can have and produce an energy deposit above the threshold. 
The recoil energy of the nucleus is given by 
\beq \label{er}
E_{R}=\frac {4m_A~m_{X}}{(m_A+m_{X})^2} (\frac {1}{2} m_X v^2 _{X}) \left( \frac {1-\cos\theta _{CM}}{2}\right)
\eeq
where $\theta _{CM}$ is the scattering angle in the DM-nucleus center of mass frame. We will assume isotropic scattering as is expected at low energies. 
So, for instance, for $A=16$, $m=1$ GeV and an energy threshold of 600 eV, the minimal DM velocity to produce a detectable recoil is $v_{min}=680$ km/s, in the extreme tail of the DM velocity distribution.

In order to compare cross section limits from different targets we will normalize them to the proton-DM cross section, $\sigma _{Xp}$. For the simplest case of interactions which are independent of spin and the same for protons and neutrons, the low energy scattering amplitude from a nucleus with mass number A is a coherent sum of A single nucleon scattering amplitudes. The matrix element squared therefore scales with size of nucleus as $\sim A^2$. In addition the kinematic factor in the cross section depends on the mass of the participants in such a way \cite{witten,Lewin} that 
\beq \label{sigSI}
\frac {\sigma^{SI} _{XA}}{\sigma^{SI} _{Xp}}=\left[ \frac {\mu (A)}{\mu(p)}\right] ^2 A^2
\eeq
where $\mu (A)$ is the reduced mass of the DM-nucleus system, and $\mu (p)$ is the reduced mass for the proton-DM system.  
At higher momentum transfer $q^2=2m_NE_R$ the scattering amplitudes no longer add in phase, and the total cross section $\sigma _{XA} (q)$ becomes smaller proportionally to the form factor $F^2(q^2)$, $\sigma _{XA} (q)=\sigma _0 F^2(q^2)$.   

We take this change in the cross section into account when we deal with higher mass ($m\gsi 10$ GeV) dark matter; for smaller masses the effect is negligible. We adopt the form factor $F(q^2)=\exp\left[-1/10(qR)^2\right]$ with $R=1.2 A^{1/2}$ fm, used also in \cite{Ahlen:1987mn,Freese:1987wu}. The simple exponential function is suffitiently accurate for our purposes and easy to implement using the Monte Carlo method to sample momentum transfer $q$, from its distribution given by the form factor. The procedure is described in more detail in Appendix~\ref{appendixB}. 

For spin dependent interactions the scattering amplitude changes sign with the spin orientation. Paired nucleons therefore contribute zero to the scattering amplitude and only nuclei with unpaired nucleon spins are sensitive to spin dependent interactions. Due to the effect of coherence, the spin independent interaction is usually dominant, depending on the mass of the exchanged particle \cite{kamionkowski}. Therefore, the spin dependent cross section limit is of interest mainly if the spin independent interaction is missing, as is the case, for example, with massive majorana neutrinos. Another example of DM with such properties is photino dark matter, see \cite{witten}, in the case when there is no mixing of left- and right- handed scalar quarks. The amplitude for DM-nucleus spin dependent interaction in the case of spin 1/2 DM, in the nonrelativistic limit, is proportional to \cite{witten, engel-vogel} 
\beq
{\cal M}\sim \langle N|{\vec J}|N\rangle\cdot {\vec s}_{X}
\eeq
where ${\vec J}$ is the total angular momentum of the nucleus, $|N>$ are nuclear states and ${\vec s}_{X}$ is the spin of the DM particle. 
In the case of scalar DM the amplitude is 
\beq
{\cal M}\sim \langle N|{\vec J}|N\rangle\cdot \left( {\vec q} \times {\vec q}' \right)
\eeq
where ${\vec q}$ and ${\vec q}'$ are the initial and final momenta of the scattering DM particle. Thus the cross section for this interaction is proportional to the fourth power of the ratio $q/M$, of DM momentum to the mass of the target which enters through the normalization of the wavefunction. Therefore the spin dependent part of the interaction for scalar DM is negligible when compared to the spin independent part. 

We adopt the standard spin-dependent cross section parametrization \cite{Lewin}
\beq 
\sigma _{XA}\sim \mu (A) ^2~[\lambda ^2J(J+1)]_A C^2 _{XA} 
\eeq 
where $\lambda$ is a parameter proportional to the spin, orbital and total angular momenta of the unpaired nucleon. The factor $C$ is related to the quark spin content of the nucleon, $C=\sum T^q _3 \Delta_q,~q=u,d,s$, where $T^{u,d,s} _3$ is the charge of the quark type $q$ and $\Delta_q$ is the fraction of nucleon spin contributed by quark species $q$.
The nuclear cross section normalized to the nucleon cross section is
\beq \label{sigSD}
\frac {\sigma^{SD} _{XA}}{\sigma^{SD} _{Xp}}=\left[\frac{\mu (A)}{\mu(p)}\right]^2~\frac {[\lambda ^2J(J+1)]_A}{[\lambda ^2J(J+1)]_p}\left( \frac {C_{XA}}{C_{Xp}}\right)^2 .
\eeq 
The values of proton and neutron $C$ factors, $C_{Xp},~C_{Xn}$ vary substantially depending on the model. For targets of the same type \-- odd-n (Si, Ge) or odd-p (Al, Na, I) nuclei \-- this model dependence conveniently cancels. The comparison of cross sections with targets of  different types involves the $C_{Xp}/C_{Xn}$ ratio. This ratio was thought to have the value  $\sim 2$  for any neutralino, based on the older European Muon Collaboration (EMC) measurements, but the new EMC results imply a ratio which is close to one for pure higgsino, and is $\gsi$ 10 otherwise. (The biggest value for the ratio is $C_p/C_n\sim 500$, for bino.) We normalize our spin dependent results to the proton cross section $\sigma _{Xp}$  using  $C_{Xp}/C_{Xn}=1$ for definiteness below. 

In this paper we assume that the DM halo velocity distribution is given by a truncated Maxwell-Boltzmann distribution in the galactic reference frame, as in the Isothermal Sphere Halo model \cite{Binney}. We direct the ${\hat z}$ axis of the Earth's frame in the direction of the Local Standard of Rest (LSR) motion. \footnote{The Local Standard of Rest used here is the dynamical LSR, which is a system moving in a circular orbit around the center of Milky Way Galaxy at the Sun's distance.} The DM velocity distribution, in the Earth's frame, is given by
\beq
\label{veldistE}
f(v_{z},{\vec v}_{\perp})=N \exp \left[-\frac{(v_{z}-v^{t} _E)^2+{\vec v}_{\perp}^2}{v^2 _c}\right].
\eeq
Here $v_c$ is the local circular velocity and it is equal to $\sqrt{2}$ times the radial velocity dispersion in the isothermal sphere model; ${\vec v}_E$ is the velocity of the Earth in the Galactic reference frame. Throughout, superscript ``t'' indicates a tangential component. This neglects the Earth's motion in the radial direction which is small. The velocities $v_z$ and ${\vec v}_{\perp}$ are truncated according to $\sqrt {v^2 _z+{\vec v}_{\perp}^2}\lsi v_{esc}$, where $v_{esc}$ is the escape velocity discussed below. 

The model above is the simplest and the most commonly used model which describes a self-gravitating gas of collisionless particles in thermal equilibrium. On the other hand numerical simulations produce galaxy halos which are triaxial and anisotropic and may also be locally clumped depending on the particular merger history (see \cite{annegreen} for a review). This indicates that the standard spherical isotropic model may not be a good approximation to the local halo distribution. Here we aim to extract the allowed DM window using the simplest halo model, but with attention to the sensitivity of the limit to poorly determined parameters of the model. The effects of the more detailed halo shape may be explored in a further work. 

We ignore here the difference between the DM velocity distribution on the Earth, deep in the potential well of the solar system, and the DM velocity distribution in free space. This is a common assumption justified by Gould in \cite{gould} as a consequence of Liouville's theorem. Recently Edsjo et al. \cite{edsjo} showed that the realistic DM velocity distribution differs from the free space distribution, but only for velocities $v\lsi 50$ km/s. Therefore, the free space distribution is a good approximation for our analysis, since for light dark matter the dominant contribution to the signal comes from high velocity part of the distribution.    

The velocity of the Earth in the Galactic reference frame is given by
\beq
{\vec v}_E={\vec v}_{LSR}+{\vec v}_{S}+{\vec v}_{E,orb},
\eeq
where ${\vec v}_{LSR}$ is the velocity of the local standard of rest LSR: it moves with local circular speed in tangential direction $v^t _{LSR}=v_c$, toward $l=90^o$, $b=0^o$, where $l$ and $b$ are galactic longitude and latitude.  The velocity of the Sun with respect to the LSR is ${\vec v}_{S}= 16.6$ km/s and its direction is $l=53^o$, $b=25^o$ in galactic coordinates. $v_{E,orb}=30$ km/s is the maximal velocity of the Earth on its orbit around the Sun.

The magnitude of $v^t _{LSR}$ has a considerable uncertainty. We adopt the conservative range $v_c=(220\pm 50)$ km/s which relies on purely dynamical observations \cite{kochanek}. Measurements based on the proper motion of nearby stars give a similar central value with smaller error bars, for example $v_c (R_0)=(218\pm 15)$ km/s, from Cepheids and $v_c (R_0)=(241\pm 17)$ km/s, from SgrA$^*$ (see \cite{annegreen2} and references therein). The choice $v_c=(220\pm 50)$ km/s is consistent with the DAMA group analysis in \cite{damav} where they extracted the dependence of their cross section limits on the uncertainty in the Maxwellian velocity distribution parameters.

Projecting the Earth's velocity on the tangential direction ($l=90^o$, $b=0^o$) we get 
\beq
v^{t} _E=v_c+v^{t} _S + v^{t} _{E,orb}~\cos [\omega (t-t_0)]
\eeq
where $v^{t} _S=12~{\rm km/s}$; $v^{t} _E= 30~\cos \gamma~{\rm km/s}$ where $\cos \gamma=1/2$ is the cosine of the angle of the inclination of the plane of the ecliptic, $\omega =2\pi/365$ day$^{-1}$ and $t_0$ is June 2nd, the day in the year when the velocity of the Earth is the highest along the LSR direction of motion. In the course of the year $\cos [\omega (t-t_0)]$ changes between $\pm ~1$, and the orbital velocity of the Earth ranges $\pm 15$ km/s.
Taking all of the uncertainties and annual variations into account, the tangential velocity of the Earth with respect to the Galactic center falls in the range $v^{t} _{E}=(167~{\rm to}~307)~{\rm km/s}$.
 
The other parameter in the velocity distribution with high uncertainty is the escape velocity, $v_{esc}=(450~{\rm to}~650)$ km/s \cite{vesc}. We will do our analysis with the standard choice of velocity distribution function parameters, 
\beq \label{parameters}
v^t _{E}=230~ {\rm km/s},~~ v_c=220~ {\rm km/s},~~v_{esc}=650~ {\rm km/s},
\eeq 
and with the values of $v_E$ and $v_{esc}$ from their allowed range, which give the lowest count in the detector and are therefore most conservative:
\beq \label{range}
v^t _{E}=170~ {\rm km/s},~~v_c=170~ {\rm km/s},~~v_{esc}=450~ {\rm km/s}.
\eeq
For experiments performed in a short time interval we take the value of $v^{t} _{E,orb}~\cos [\omega (t-t_0)]$ which corresponds to the particular date of the experiment, and the lowest value of $v^t _{E}$ allowed by the uncertainties in the value of $v_c$.
 
Another effect is important in the detection of moderately interacting particles. Since particles loose energy in the crust rapidly (mean free path is of the order of 100 m) only those particles which come to the detector from $2\pi$ solid angle above it can reach the detector with sufficient energy. Since the velocity distribution of the particles arriving to the detector from above depends on the detector's position on Earth with respect to the direction of LSR motion, the detected rate for a light DMIC particle will vary with the daily change of position of the detector. This can be a powerful signal.

\subsection{XQC Experiment}

For light, moderately interacting dark matter the XQC experiment places the most stringent constraints in the higher cross section range.
The XQC experiment was designed for high spectral resolution observation of diffuse X-ray background in the $60-1000$ eV range. The Si detector consisted of an array of 36 1 mm$^2$ microcalorimeters.  Each microcalorimeter had a 7000 Angstrom layer of HgTe X-ray absorber. Both the HgTe and the Si layers were sensitive to the detection. The experiment was performed in a 100 s flight in March, and therefore the Earth's velocity $v^t _{E}$ falls in the 200 to 300 km/s range. The experiment was sensitive to energy deposit in the energy range $25-1000$ eV. For energy deposits below 25 eV the efficiency of the detector drops off rapidly. For energy deposits above about 70 eV the background of X-rays increases, so XQC adopted the range 25-60 eV for extraction of DM limits, and we will do the same. This translates into a conservative mass threshold for the XQC experiment of $0.4$ GeV, obtained with $v_{esc}=450$ km/s and $v^t _E=200$ km/s, which is the lowest mass explored by direct DM detection apart from axion searches.

The relationship between the number of signal events in the detector $N_S$ and the scattering cross section $\sigma_{XA}$ of DM particles on nuclei is the following
\beq \label{rate}
N_S = n_X~f~T~( N_{\rm Si}  \langle{\vec v}_{\rm Si}\rangle \sigma_{\rm Si}+N_{\rm Hg} [ \langle{\vec v}_{\rm Hg}\rangle \sigma_{\rm Hg}+\langle{\vec v}_{\rm Te}\rangle \sigma_{\rm Te} )],
\eeq
where $N_{\rm Si}$ and $N_{\rm Hg}$ are the numbers of Si and Hg (Te) nuclei in the detector, $n_{X}$ is the local number density of DM particles, $\langle\vec{v}_{\rm Si}\rangle$, $\langle\vec{v}_{\rm Hg}\rangle$ and $\langle\vec{v}_{\rm Te}\rangle$ are the effective mean velocities of the DM particles on the Si and HgTe targets, $f$ is the efficiency of the detector, and $T=100$ s is the data-taking time. In this energy range, $f \approx 0.5$.
The standard value for the local DM energy density is $\rho _X=0.3$ GeV cm$^{-3}$. However, numerical simulations combined with observational constraints \cite{draco} indicate that the local DM energy density $\rho _X$ may have a lower value, $0.18 \lsi \rho _X/({\rm GeV}~{\rm cm}^{-3}) \lsi 0.3 $. In our calculations we use both the standard value $\rho _{X}=0.3$ GeV/cm$^3$, and the lower value suggested by the numerical simulations, $\rho _{X}=0.2$ GeV/cm$^3$. The cross sections $\sigma_{\rm Si}$, $\sigma_{\rm Hg}$, $\sigma_{\rm Te}$ are calculated using equations (\ref{sigSI}) and (\ref{sigSD}). In this section and the next we assume that DM has dominantly spin-independent cross section with ordinary matter. In \S~\ref{SD} we consider the case of DM which has exclusively spin-dependent cross section or when both types of interaction are present with comparable strength.
 
XQC observed two events with energy deposit in the 25-60 eV range, and expected a background of 1.3 events. The equivalent 90\% cl upper limit on the number of signal events is therefore $N_S = 4.61$. This is obtained by interpolating between 4.91 for expected background = 1.0 event and 4.41 for expected background = 1.5 events, for 2 observed events using table IV in ref. \cite{feldman}.

We extract the cross section limits using our simulation. Because of the screening by the Earth we consider only particles coming from the $2\pi$ solid angle above the detector, for which $\langle{\hat n}\cdot {\vec v}\rangle\leq 0$ and for them we simulate the interaction in the detector, for particles distributed according to (\ref{veldistE}). We take the direction of the LSR motion, ${\hat n}$ as the z axis. 

We choose the nucleus $i$ which the generated DM particle scatters from, using the relative probability for scattering from nucleus of type $i$, derived in Appendix \ref{appendixA}: 
\beq \label{probability}
P_i =\frac{\lambda _{eff}}{\lambda _i}=\frac {n_i\sigma _{XA_i}}{\sum  n_j\sigma _{XA_j}},
\eeq 
where $\lambda _i$ is the mean free path in a medium consisting of material with a mass number $A_i$: $\lambda _i=(n_i \sigma_{XA_i})^{-1}$. Here $n_i$ is the number density of target nuclei $i$ in the crust, $\sigma_{XA_i}$ is the scattering cross section of X on nucleus A$_i$ and the effective mean free path, $\lambda _{eff}$, is given as 
\beq \label{freepath}
\lambda _{eff}=\left( \sum \frac {1}{\lambda _i} \right) ^{-1}.
\eeq
In each scattering the DM particle loses energy according to (\ref{er}), and we assume isotropic scattering in the c.m. frame.

We determine the effective DM velocity $<{\vec v_A}>$ as
\beq \label{<v>}
\langle{\vec v_A}\rangle=\frac {\sum ' v}{N_{tot}}
\eeq
where the sum is over the velocities of those DM particles which deposit energy in the range 25-60 eV, in a collision with a nucleus of type A, and $N_{tot}$ is the total number of generated DM particles. The result depends on the angle between the experimental look direction, and the motion of the Earth. 
The zenith direction above the place where the rocket was launched, $\hat n _{XQC}$, is toward  $b=+82^o,l= 75^o$. Thus the detector position angle compared to the direction of motion of the Earth through the Galaxy is 82$^o$. Only about $10\%$ of the collisions have an energy deposit in the correct range.
Putting this together gives the $90 \% $ confidence level XQC upper limit on the spin independent cross section for DMIC shown in Figures \ref{fig1} and \ref{fig2}. The solid line limit is obtained using the most conservative set of parameters ($\rho =0.2$ GeV/cm$^3$, $v^t _E=200$ km/s, $v_{esc}=450$ km/s) and the dotted line is the limit obtained by using the standard parameter values in eq.~(\ref{parameters}). The upper boundary of the upper domain, $\sigma\simeq 10^6\div 10^8$ mb is taken from \cite{wandelt}. 

When the dark matter mass is higher than $10$ GeV, the form factor suppression of cross section is taken into account. We give the details of that calculation in the Appendix B.

In the next section we explain how the upper boundaries of the excluded region from the underground experiments shown in these figures are obtained. Also shown are the original limits from the balloon experiment of Rich et al.~\cite{RRS} obtained using the ``standard'' choices for DM at the time of publishing (dashed line) as well as the limits obtained using the conservative values of parameters ($v^t _E=170$ km/s, since the experiment was performed in October, and $v_{esc}=450$ km/s). Fig. \ref{fig2} zooms in on the allowed window in the $m\lsi 2.4$ GeV range.

\subsection{Underground Detection}

In this section we describe the derivation of the lower boundary of the DMIC window from the underground experiments.
This is the value of $\sigma _{Xp}$ above which the DM particles would essentially be stopped in the Earth's crust before reaching the detector. (More precisely, they would loose energy in the interactions in the crust and fall below the threshold of a given experiment.) To extract the limit on $\sigma _{Xp}$ we generate particles with the halo velocity distribution and then follow their propagation through the Earth's crust to the detector. We simulate the DM particle interactions in the detector and calculate the rate of the detector's response. We compare it to the measured rate and extract cross section limits.
The basic input parameters of our calculation are the composition of the target, the depth of the detector and the energy threshold of the experiment. We also show the value of the dark matter mass threshold $m_{TH}$, calculated for the standard and conservative parameter values given in eq.~(\ref{parameters}) and eq.~(\ref{range}). The parameters are summarized in Table \ref{t1ch3} for the relevant experiments.

\begin{table}[htb]
\label{t1ch3}
\caption{The parameters of the experiments used for the extraction of cross section limits; $m_{TH}$ is the minimum mass of DM particle which can produce a detectable recoil, for the standard and conservative parameter choice. The energy threshold values $E^{nuc}_{TH}$ refer to the nuclear response threshold. This corresponds to the electron response threshold divided by the quenching factor of the target.}
\begin{center}
\begin{tabular}{|c|c|c|c|c|}
\hline

Experiment                   &     Target    &  Depth     & $E^{nuc} _{TH}$ & $m^{std}_{TH} (m^{cons}_{TH})$      \\  \hline \hline
CRESST, \cite{cresst}        &  Al$_2$O$_3$  &  1400 m    & 600 eV & 0.8 (1.1) GeV   \\ \hline
DAMA, \cite{dama}            &  NaI          &  1400 m    & 6  keV & 3.5 (5) GeV \\ \hline
ELEGANT, \cite{elegant}      &  NaI          &   442 m    & 10 keV & 5 (8) GeV   \\ \hline
COSME I, \cite{cosme}          &  Ge           &  263 m     & 6.5 keV & 5.5 (8) GeV        \\ \hline
CDMS, \cite{cdms}            &  Si      &  10.6 m    & 25 keV & 9.8 (16) GeV  \\ 
                             &   Ge     &            &        &  14 (21) GeV  \\ \hline

\end{tabular}
\end{center}
\end{table}

In the code, after generating particles we propagate those which satisfy $\langle{\hat n} \cdot {\vec v}\rangle\leq 0$ through the crust. Given the effective mean free path in the crust eq.~(\ref{freepath}), the distance traveled between two collisions in a given medium is simulated as  
\beq \label{distance}
x=-\lambda _{eff} \ln R
\eeq
where R is a uniformly distributed random number, in the range $(0,1)$.
After simulating the distance a particle travels before scattering, we choose the nucleus $i$ it scatters from using the relative probability as in eq.~(\ref{probability}). 

We take the mass density of the crust to be $\rho =2.7$ g/cm$^3$. To explore the sensitivity of the result to the composition of the crust we consider two different compositions. First we approximate the crust as being composed of quartz, ${\rm SiO}_2$, which is the most common mineral on the Earth and is frequently the primary mineral, with $>98 \%$ fraction. 
Then we test the sensitivity of the result by using the average composition of the Earth's crust: Oxygen 46.5 $\%$, Silicon 28.9 $\%$, Aluminium 8.3 $\%$ and Iron 4.8 $\%$, where the percentage is the mass fraction of the given element. Our test computer runs showed that both compositions give the same result up to the first digit, so we used simpler composition for the computing time benefit. Since the DM exclusion window we obtain at the end of this section should be very easy to explore in a dedicated experiment, as we show later in the text, we do not aim to find precise values of the signal in the underground detector. 

When collisions reduce the DM velocity to less than the Earth's radial escape velocity, $v_{esc}=11$ km/s, DM is captured by the Earth and eventually thermalized. Collisions may also reverse the DM velocity in such a way that the DM particle leaves the surface of the Earth with negative velocity: effectively, the DM particle gets  reflected from the Earth. The majority of light DM particles wind up being reflected as is characteristic of diffuse media. The percentage of reflected particles proves not to depend on the cross section, as long as the mean free path is much smaller than the radius of the earth, but it does depend on DM particle mass. Light particles have a higher chance of scattering backward and therefore a higher percentage of them are reflected. The initial DM flux on Earth equals $2.4(1.2)~10^{6}~(1~{\rm GeV}/m_X)$ cm$^{-2}$s$^{-1}$, taking standard (conservative) parameter values. Table \ref{t2ch3} shows the fraction of initial flux of DM particles on the Earth which are captured and thermalized for various mass values. The fraction is, up to a percent difference, independent of whether we make the standard or conservative parameter choice. 

For DM particles which are not scattered back to the atmosphere and which pass the depth of the detector before falling below the energy threshold of the detector, the scattering in the detector is simulated. 
\begin{table}[htb]

\caption{ The percentage of DM particles incident on Earth which are captured, when $\lambda _{int}\ll R_E$.} \label{t2ch3}
\begin{center}
\begin{tabular}{|c|c|c|c|c|c|c|}
\hline

mass [GeV]            &   2       &  4     & 6       & 10       & 100       \\  \hline \hline

thermalized [$\%$]    &   21  & 30 & 36  & 46   & 94    \\ \hline

\end{tabular}
\end{center}
\end{table}
For composite targets we simulate collision with different nuclei with probabilities given as in eq.~(\ref{probability}). If the energy of the recoil is above $E_{TH}$, we accept the event and record the velocity of the particle which deposited the signal.
The spectral rate per (kg day keV) is then calculated as a sum of rates on the separate elements of the target, as
\beq \label{sum}
\frac{dR}{dE_R} [\alpha (t)]=\sum_{i} \frac{f_i}{A_i~m_p}~\frac{\rho _X}{m_X} ~\frac{\langle v[\alpha (t)]\rangle_i}{\Delta E}~\sigma _{XA_i}
\eeq
where $f_i$ is the mass fraction of a given element in the target, $\rho _X$ is the local DM energy density, $\Delta E$ is the size of an energy bin of a given experiment and $<v(\alpha (t))>$ is calculated as in (\ref{<v>}). The signal in the detector falls exponentially with $\sigma _{XN}$ since the energy of DM at a given depth gets degraded as an exponential function of the cross section, see \cite{SG}. Therefore the limit on $\sigma _{XN}$ is insensitive to small changes in the rate in the detector coming from changes in $\rho _X$;  we adopt the commonly used value $\rho _X=0.3$ GeV cm$^{-3}$ for the local DM energy density. 

We emphasize here that the spectral rate is a function of the relative angle $\alpha (t)$ between the direction of the motion of LSR and the position of the  detector.
This angle changes during the day as
\beq \label{time}
\cos \alpha (t)=\cos \delta \cos \alpha '+ \sin \delta \sin \alpha ' \sin (\omega t +\phi _0)
\eeq 
where $\delta $ is the angle between the Earth's North pole and the motion of LSR; $\alpha '$ is the angle between the position of the detector and the North pole, and it has a value of ($90^{\circ}$ - geographical latitude).
The angle between the LSR motion and Earth's North pole is $\delta = 42^{\circ}$, so for an experiment around $45^{\circ}$ latitude (as for Gran Sasso experiment), $\alpha '=45 ^{\circ}$. Therefore, in the course of a day, the angle between the detector and the LSR motion varies in the range approximately $0^{\circ}$ to $90^{\circ}$.

\begin{figure} 
\begin{center}
\includegraphics*[width=8cm]{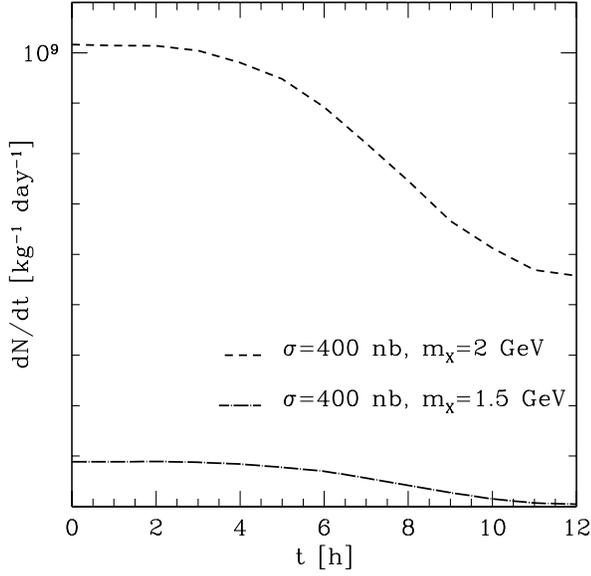}
\end{center}
\caption{ The time dependence of the measured rate in underground detectors for m=2 GeV and m=1.5 GeV DM candidates.}\label{figdaydep}
\end{figure}

Fig. \ref{figdaydep} shows the rate $R$ per (kg $\cdot $ day) as a function of time, (\ref{time}), 
\beq
R [\alpha (t)]=\sum_{i=Al,O} \frac{f_i}{A_i~m_p}~\frac{\rho _X}{m_X} ~\langle v(\alpha (t))\rangle_i~\sigma _{XA_i}
\eeq
calculated for the parameters of the CRESST experiment.
We choose $\phi_0$ so that time t=0 corresponds to the moment the detector starts to move away from the LSR axis. We see that for these masses the rate is a strong function of the angle of the position of the detector with respect to the motion of the LSR, which gives an interesting detection signature for detector locations such that this angle changes during the day.  

To extract our limits, we average the signal from the simulation $dR(t)/dE_R$ over one day: 
\beq
\langle dR/dE_R\rangle=\frac {1}{T}~\int ^{T} _0 dR(t)/dE_R~dt.
\eeq
Since the shape of the spectral rate response is a function of $\sigma _{Xp}$ in our case (because the velocity distribution function at the depth of detector is a function of $\sigma _{Xp}$ due to the interactions in the crust) the extraction of cross section limits is more complicated than when the rate scales linearly with $\sigma _{Xp}$. In the region where the window is located, i.e. masses below 2 GeV, we perform the analysis based on the fit method used by the CRESST group, \cite{cresst}. The measured spectrum is fit with an empirical function called $B$. In our case $B$ is the sum of two falling exponentials and a constant term, since we expect the signal only in the few lowest energy bins. For the fit we use the maximum likelihood method with Poissonian statistics in each bin. The maximum likelihood of the best fit, $B_0$, is ${\emph L}_0$. We define the background function $B'$ as the difference between the best fit to the measured signal, $B_0$ and some hypothesized DM signal $S$: $B'=B_0-S$. Following the CRESST procedure, we set $B'$ to zero when $S$ exceeds $B_0$. When $\sigma _0$ is such that the simulated signal $S$ is below the measured signal $B_0$, $B'$ adds to the signal $S$, completing it to $B_0$ and the likelihood is unchanged. With increasing $\sigma _0$, when $S$ starts to exceed the function $B_0$, $B'$ becomes zero, and we calculate the likelihood using $S$ alone in such bins, leading to a new likelihood ${\emph L}$. Following the CRESST collaboration prescription, $\sigma_0$ excluded at $90 \%$ CL is taken to be the value of $\sigma_0$ giving $\ln{\emph L}-\ln{{\emph L}_0}=-1.28^2/2$ \cite{cresst}, since $10\%$ of the area of a normalized Gaussian distribution of width $\sigma$ is $1.28 \sigma $ above than the peak. We show the window obtained this way in Fig.~\ref{fig2} and for the low mass range, in Figure \ref{fig1}. 

\begin{figure} 
\begin{center}
\includegraphics*[width=8cm]{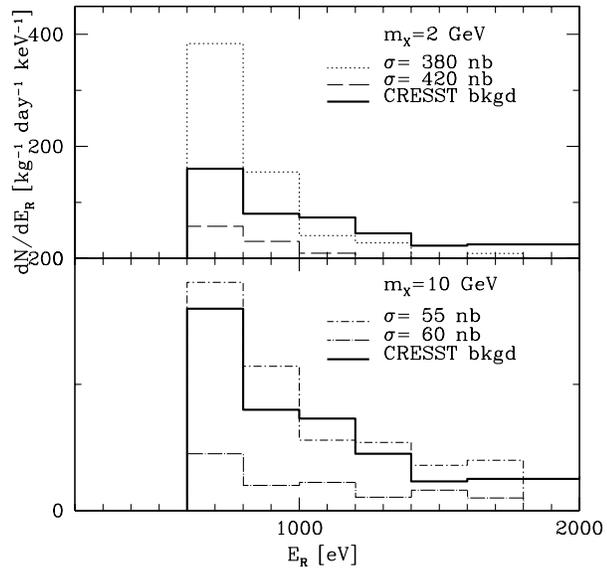}
\end{center}
\caption{ The CRESST background and the simulated response of the detector for masses $m_X=2$ and $m_X=10$ GeV, and different values of spin independent cross sections $\sigma _{Xp}$.}\label{signal}
\end{figure}

\begin{figure} 
\begin{center}
\includegraphics*[width=8cm]{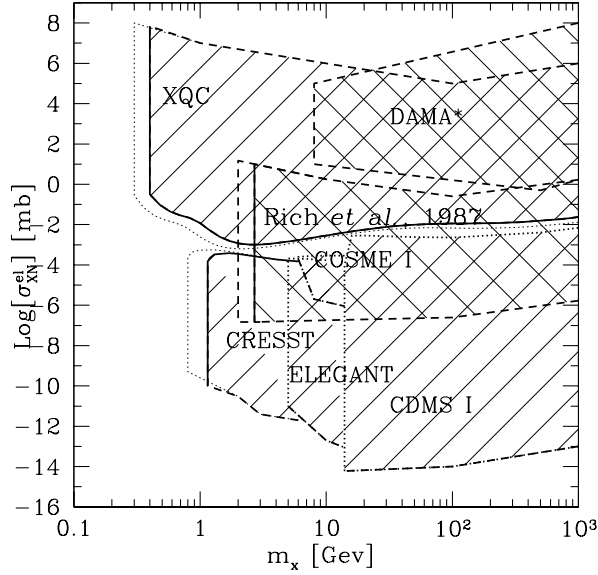}
\end{center}
\caption{Overview of the exclusion limits for spin independent DM-nucleon elastic cross section coming from the direct detection experiments on Earth. The limits obtained by using the conservative parameters, as explained in the text, are plotted with a solid line; the dotted lines are obtained by using standard parameter values and the dashed lines show limits published by corresponding experiments or in the case of XQC, by Wandelt et al. \cite{wandelt}. The region labeled with DAMA* refers to the limits published in \cite{damastrong}.}\label{fig1}
\end{figure}

\begin{figure} 
\begin{center}
\includegraphics*[width=8cm]{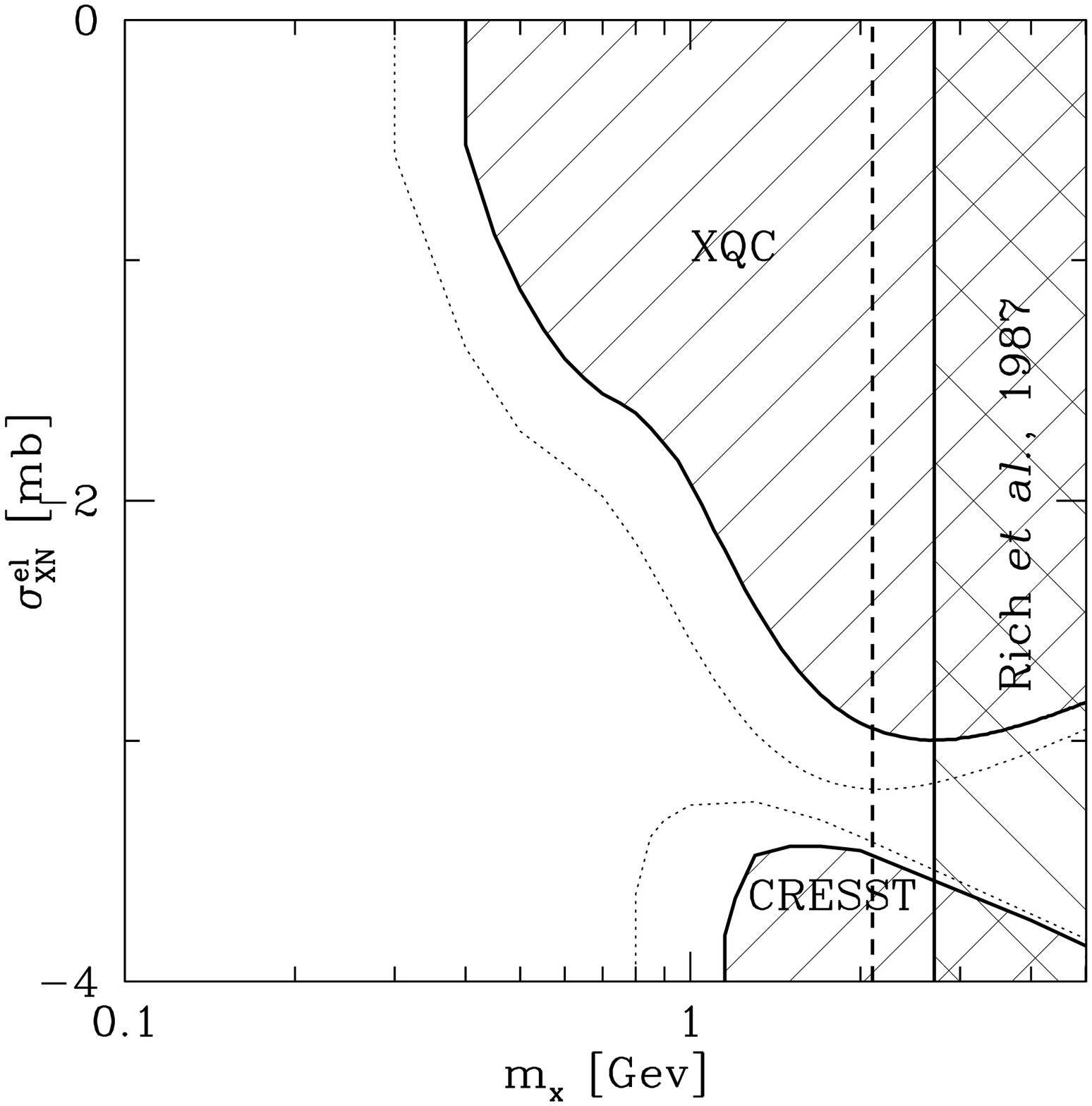}
\end{center}
\caption{The allowed window for $\sigma^{el}_{XN}$ for a spin independent interaction. The region above the upper curve is excluded by the XQC measurements. The region below the lower curve is excluded by the underground CRESST experiment. The region $m\geq 2.4$ GeV is excluded by the experiment of Rich et al.}\label{fig2}
\end{figure}

For masses higher than $2$ GeV we can use a simpler method, since this range of masses is already ruled out and our plot is only indicating from which experiments the constraints are coming. We calculate the response of the detector for different cross section values, and take the limit to be that value for which the simulated signal is below the experiment's  background. Fig \ref{signal} shows CRESST background together with the simulated response for DM particles with mass $m_X=2$ GeV and 10 GeV and various values of cross section. The limits obtained this way for different experiments are given in Figure \ref{fig1}.

The only dark matter detector sensitive to particles with mass $\lsi 4$ GeV is CRESST. Since it is the only experiment with threshold lower than the threshold of the balloon experiment by Rich et al., it extends the existing exclusion window for intermediate cross sections. For the CRESST experiment we perform the calculation using both standard and conservative parameters, because the size of the exclusion window is very sensitive to the value of mass threshold, and therefore to the parameter choice. For other underground experiments we use only standard parameters. In the mass range $5\lsi m\lsi 14$ GeV, the ELEGANT and COSME I experiments place the most stringent constraints on a DMIC scenario, since they are located in shallow sites a few hundred meters below the ground; see Table~\ref{t1ch3}. Other experiments sensitive in this mass range (e.g. IGEX, COSME II) are located in much deeper laboratories and therefore less suitable for DMIC limits. We therefore present limits from ELEGANT and COSME I, for masses 5 to 14 GeV. Masses grater than 14 GeV are above the threshold of the CDMS experiment and this experiment places the most stringent lower cross section limit due to having the smallest amount of shielding, being only 10.6 m under ground. The CDMS I had one Si and four Ge detectors operated during a data run. To place their limits they used the sophisticated combination of data sets from both types of detectors. Due to the large systematic uncertainty on the Si data the Ge data set dominates their combined measurements. To be conservative we assume that only Ge detectors are present, which reduces the region excluded by CDMS to $m\gsi 14$ Gev. Fig \ref{fig1} shows the cross section limits these experiments imply, for masses $m\lsi 10^3$ GeV. 

\subsection{Spin-Dependent limits} \label{SD}

In this section we address the case in which DM has a spin dependent interaction with ordinary matter. We consider first purely SD interaction and later we consider the case in which both interaction types are present. We focus on low masses which belong to the cross section window allowed by the experiment of Rich et al.

\begin{figure}
\begin{center}
\includegraphics*[width=8cm]{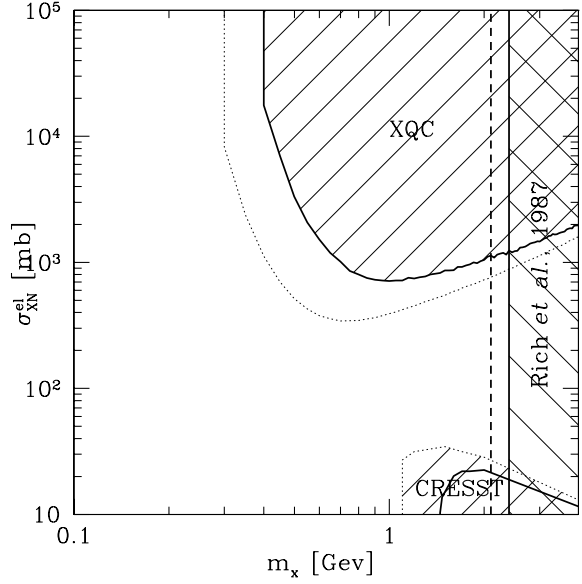}
\end{center}
\caption{The allowed Spin Dependent interaction for $(C_{Xp}/C_{Xn})^2=1$. The region above the upper curve is excluded by XQC measurements. The region below the lower curve is excluded by CRESST. The region $m\geq 2.4$ GeV is excluded by the balloon experiment of Rich et al. }\label{figSD}
\end{figure}


If the DM has only a spin dependent interaction with ordinary matter, only the small fraction of the XQC target with nonzero spin is sensitive to DM detection. The nonzero spin nuclei in the target are: ${\rm Si}_{29}$ ($4.6~\%$ of natural Si), ${\rm Te}_{125}$ ($7~\% $) and ${\rm Hg}_{199}$, ($16.87~ \% $); their spin is due an unpaired neutron. We calculate the spin dependent cross section limits from the XQC experiment the same way as for the spin independent case, using the new composition of the target. The limiting value of the elastic cross section of DM with protons, $\sigma^{SD} _{Xp}$, is shown in Figure \ref{figSD}. Since the XQC target consists of n-type nuclei, the resulting cross section with protons is proportional to the $(C_{Xp}/C_{Xn})^2$ factor as explained in section II. In Figure \ref{signal} we use the value $(C_{Xp}/C_{Xn})^2=1$ which is the minimal value this ratio may have. We note that the maximal value of the ratio, based on the EMC measurements is $(C_{Xp}/C_{Xn})^2=500^2$ and it would shift the XQC limit by a factor $500^2$ up to higher cross sections (substantially extending the allowed window).

The spin sensitive element in the CRESST target is Al which has an unpaired proton in the natural isotope. We assume that the crust consists only of Al, since it is the most abundant target nucleus with non-zero spin. In this case the model dependence of the $C$ factor ratio drops out in normalizing the underground experiment to the proton cross section. 

\begin{figure} 
\begin{center}
\includegraphics*[width=8cm]{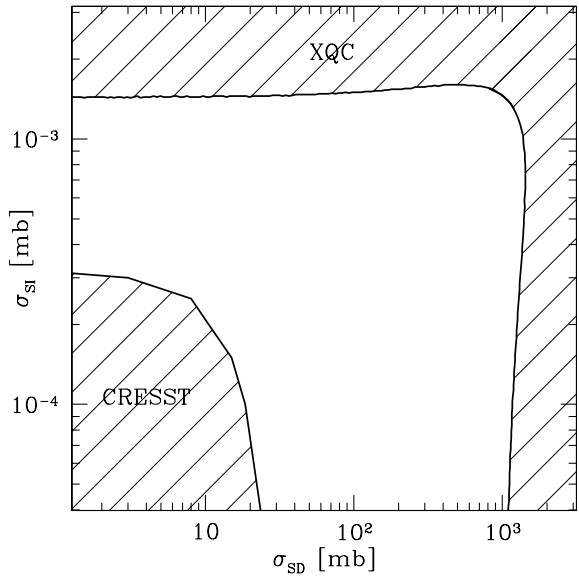}
\end{center}
\caption{$\sigma_{SI}$ {\it vs} $\sigma_{SD}$, for CRESST and XQC experiments, for mass $m_X=2$ GeV. The region between two curves is the allowed region.}\label{figSISD}
\end{figure}

The window is extended when compared to the purely spin independent DM interaction, as shown in Fig. \ref{figSD}. This is mostly due to the fact that sensitive part of the XQC target is substantially reduced.

In Fig. \ref{figSISD}, for mass $m_X=2$ GeV, we plot the $\sigma _{SI}$ vs $\sigma _{SD}$ limit, assuming both types of interactions are present. An interesting feature in the $\sigma _{SI}$ vs $\sigma _{SD}$  dependence is that, when the spin dependent and independent cross sections on the target nuclei are of comparable magnitude, screening between two types of targets allows cross sections to be higher for the same rate in the detector than in the case when only one type of interaction is present.

\subsection{Constraint on the fraction of DMIC}  \label{fraction}

We now turn the argument around and use the XQC data to place a
constraint on the fraction of allowed DMIC as a function of its
elastic cross section. We restrict consideration to values of the
cross section which are small enough that we do not have to treat
energy loss in the material surrounding the sensitive components of
XQC. The maximal fraction DMIC allowed by XQC data $p=n^{MI}
_{DM}/n^{tot} _{DM}$ can then be expressed as a function of cross
section, using (\ref{rate})
\beq\label{eqfraction}
p=\frac{N_S}{n_X~f~T} \left[ N_{\rm Si}  \langle{\vec v}_{\rm Si}\rangle \sigma_{\rm Si}+ N_{\rm Hg} (\langle{\vec v}_{\rm Hg}\rangle\sigma_{\rm Hg}+\langle{\vec v}_{\rm Te}\rangle \sigma_{\rm Te} )\right] ^{-1}
\eeq
where all quantities are defined as before. 

The mass overburden of XQC can be approximated as \cite{mccam:correspondence}: $\lambda =10^{-4}~{\rm g/cm}^2$, for off-angle from the center of the field of the detector $\alpha =(0^o~{\rm to}~ 30^o)$; $\lambda =10~{\rm g/cm}^2$, for $\alpha =(30^o~{\rm to}~ 100^o )$; and  $\lambda =10^4~{\rm g/cm}^2$, for $\alpha \ge 100^o$. The center of the field of view points toward $l=90^o$, $b=60^o$ which makes an angle of $32^o$ with the detector position direction. Since DM particles are arriving only from above the detector, they will traverse either 10 g/cm$^3$ or $10^{-4}$ g/cm$^3$ overburden.  

\begin{figure} 
\begin{center}
\includegraphics*[width=8cm]{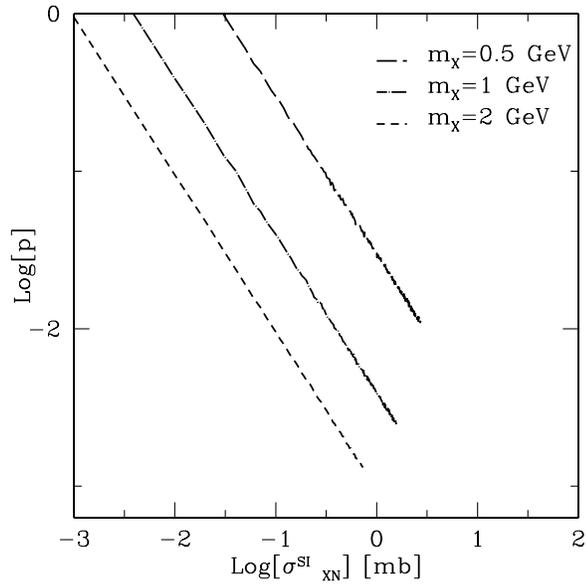}
\end{center}
\caption{ The allowed fraction $p$ of DM candidate as a function of DM-nucleon cross section. For each mass, $p$ is calculated up to the values of cross sections for which the interaction in the mass overburden of the detector becomes important.}\label{fig7}
\end{figure}

For example, for values of cross section of about $0.7$ mb, $m=2$ GeV DM particles start to interact in the 10 g/cm$^3$ overburden, thus for cross sections above this value our simple approach which does not account for the real geometry of the detector, is not applicable anymore.
We therefore restrict our analysis to values of the cross section for which neglecting the interaction in the overburden is a good approximation. In this domain, the allowed fraction of DM falls linearly with increasing cross section, as can be seen in equation (\ref{eqfraction}) since $<{\vec v_{DM}}>$ remains almost constant and is given by the halo velocity distribution eq.~(\ref{veldistE}).
The results of the simulation are shown in Fig. \ref{fig7}, for a spin independent interaction cross section. 
An analysis valid for larger cross sections, which takes into account details of the geometry of the XQC detector, is in preparation \cite{Spergel}.

\subsection{Future Experiments}

The window for $m_X\lsi 2.4$ GeV in the DMIC cross section range could be explored in a dedicated experiment with a detector similar to the one used in the XQC experiment and performed on the ground. Here we calculate the spectral rate of DM interactions in such detector, in order to illustrate what the shape and magnitude of a signal would be. 

\begin{figure} 
\begin{center}
\includegraphics*[width=8cm]{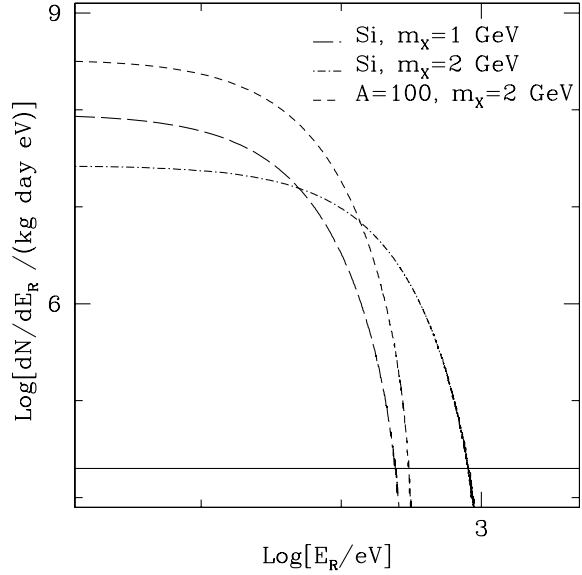}
\end{center}
\caption{ The simulated minimum rate per (kg day eV) calculated with $\sigma _{Xp}=2~\mu$b, for a DM experiment on the ground, versus deposited energy $E_R$ in eV, for a SI target and for a target with mass number A=100. The solid line indicates maximal value of the cosmic ray muon background determined based on the total muon flux as is used in the text.}\label{figftrexp}
\end{figure}

In Fig. \ref{figftrexp} we plot the rate per (kg$\cdot $day$\cdot $eV), for a Si detector and DM particle masses of $m_X=$ 1 and 2 GeV assuming a spin independent interaction. In the case of an exclusively spin dependent interaction, the signal would be smaller, approximately by a factor $f/A^2$, where {\it f} is the fraction of odd-nuclei in the target. The calculation is done for a position of a detector for which the signal would be the smallest. We assume a short experiment and do not perform averaging over time of a day because that would increase the signal.

The rate scales with cross section; the rate shown in Fig \ref{figftrexp} is for $\sigma _{Xp}=2$ $\mu $b, the lower limit on the cross section window from the XQC experiment for $m=1$ GeV. Since the unshielded muon flux on the ground is of the order of $2~10^{2}~(m^2~s)^{-1}=2~10^3$ (cm$^2$ day$)^{-1}$, an experiment performed on the ground with an array of micro-calorimeter absorbers such as XQC whose target mass is $\approx 100$ g, should readily close this window or observe a striking signal.

\subsection{Summary}
In \S~\ref{directDM} we have determined the limits on dark matter in the low mass range ($m\lsi 10$ GeV) and with an intermediate cross section on nucleons based on the final XQC data and results of underground experiments with low mass threshold. We also updated previous limits taking into account newer halo velocity distribution. We found that there is an allowed window for DM mass $m\lsi 2.4$ GeV and cross section $\sigma \approx \mu$b. Curiously this window overlaps with the mass/cross section range expected for the $H$ dibaryon making a possible DM candidate, \cite{f:StableH,fz:nucstab} and Chapter~\ref{Hdibaryon}. We showed that it should be straightforward experimentally to explore the window. A signal due to a light DMIC would have strong daily variations depending on the detectors position with respect to the LSR motion and therefore provide strong signature.

\section{The $H{\bar H}$ Dark Matter\-- Indirect detection constraints} \label{Hindirect}

B-sequestration scenarios imply the possibility of detectable annihilation of DM with anti-baryon number with nucleons in the Earth, Sun or galactic center. The rate of $\bar{H}$ annihilation in an Earth-based detector is the $\bar{H}$ flux at the detector, times  $\sigma_{\bar{H}N}^{\rm ann}$, times (since annihilation is incoherent)  the number of target nucleons in the detector,  $6 \times 10^{32} $ per kton.  The final products of $\bar{H} N$ annihilation are mostly pions and some kaons, with energies of order 0.1 to 1 GeV.  The main background in SuperK at these energies comes from atmospheric neutrino interactions whose level is $\sim100$ events per kton-yr\cite{SKatmneutrinos}.  Taking $\Phi^{SK}_{\bar{H}} = R_{\rm cap}/A_{SK}$, where $A_{SK}$ is the area of SK experiment and $R_{\rm cap}$ is taken from Table 2.1, the annihilation rate in SuperK is lower than the background if $\tilde{\sigma}^{\rm ann}_{\bar{H}N} \le 6 \times 10^{-44}\,  {\rm cm}^2$
\beq
R^{dir} _{SK}\sim 100\left[\frac{\sigma^{\rm ann} _{\bar H}}{6~10^{-44} {\rm cm}^2}\right] \left({\rm kton~yr}\right)^{-1}
\eeq
. The total energy release of $m_H + B_H m_N $ should give a dramatic signal, so it should be possible for SuperK to improve this limit.  Note that for the $H,\,\bar{H}$ scenario this limit is already uncomfortable, since it is much lower than the effective cross section required at freezeout ($\sigma^{\rm ann} _{\bar H} =2.2~10^{-41}$ cm$^2$).  However this cannot be regarded as fatal, until one can exclude with certainty the possibility that the annihilation cross section is very strongly energy dependent.

Besides direct observation of annihilation with nucleons in a detector, constraints can be placed from indirect effects of $\bar{H}$ annihilation in concentrations of nucleons.  We first discuss the photons and neutrinos which are produced by decay of annihilation products.  The signal is proportional to the number of nucleons divided by the square of the distance to the source, so Earth is a thousand-fold better source for a neutrino signal than is the Sun, all other things being equal. Since $\gamma$'s created by annihilation in the Earth or Sun cannot escape, the galactic center is the best source of $\gamma$'s but do not pursue this here because the constraints above imply the signal is several orders of magnitude below present detector capabilities.

The rate of observable neutrino interactions in SuperK is 
\beq  \label{neutintsSK}
\Gamma_{\nu{ \rm SK}} =  N_{{\rm SK}}\, \Sigma_i \int{ \frac{d n_{\nu_i}}{dE} \sigma^{\rm eff}_{\nu_i N}  \Phi_{\nu_i}  dE }, 
\eeq
where the sum is over neutrino types, $N_{{\rm SK}}$ is the total number of nucleons in SuperK, $\frac{d n_{\nu_i}}{dE}$ is the spectrum of $i$-type neutrinos from an $\bar{H}$ annihilation, $\sigma^{\rm eff}_{\nu_i N} $ is the neutrino interaction cross section summed over observable final states (weighted by efficiency if computing the rate of observed neutrinos), and $ \Phi_{\nu_i} $ is the $\nu_i$ flux at SK.  This last is $f_{\nu_i}$, the mean effective number of $\nu_i$'s produced in each $\bar{H}$ annihilation discussed below, times the total rate of $\bar{H}$ annihilation in the source, $\Gamma^{\rm ann}_{\bar{H},s}$, divided by $\approx 4 \pi R_s^2$, where $R_s$ is the distance from source to SuperK;  $R_s \approx R_E$ for annihilation in Earth.  

In general, computation of the annihilation rate $\Gamma^{\rm ann}_{\bar{H},s}$ is a complex task because it involves solving the transport equation by which DM is captured at the surface, migrates toward the core and annihilates, eventually evolving to a steady state distribution.  However if  the characteristic time for a DM particle to annihilate, $\tau^{\rm ann}=\langle\sigma^{\rm ann} n_N v\rangle^{-1}$, is short compared to the age of the system, equilibrium between annihilation and capture is established (we neglect the evaporation which is a good approximation for $M_{DM} \gsi \mathcal{O}$(GeV) and is also more conservative approach) so $\Gamma^{\rm ann}_{\bar{X},E}$ equals $f_{\rm cap}  \Phi_{\bar{H}} 4 \pi R_E^2$.  Then the neutrino flux, eq.~(\ref{neutintsSK}), is independent of $\sigma^{\rm ann}_{\bar{H}N},$ because the annihilation rate per $\bar{H}$ is proportional to it but the equilibrium number of $\bar{H}$'s in Earth is inversely proportional to it.  For Earth, the equilibrium assumption is applicable for $\tilde{\sigma}^{\rm ann} \gsi 5 \times 10^{-49} {\rm cm}^2$, while for the Sun it is applicable if, roughly, $\tilde{\sigma}^{\rm ann} \gsi 10^{-52} {\rm cm}^2$.  For lower annihilation cross sections, transport must be treated.  

The final state in $\bar{H} N$ annihilation is expected to contain $\bar{\Lambda}$ or $\bar{\Sigma}$ and a kaon, or $\bar{\Xi}$ and a pion, and perhaps additional pions.  In a dense environment such as the core of the Earth, the antihyperon annihilates with a nucleon, producing pions and at least one kaon.  In a low density environment such as the Sun, the antihyperon decay length is typically much shorter than its interaction length.  In Earth, pions do not contribute significantly to the neutrino flux because $\pi^0$'s decay immediately to photons, and the interaction length of $\pi^\pm$'s is far smaller than their decay length so they eventually convert to $\pi^0$'s through charge exchange reactions;  similarly, the interaction lengths of $K^{0}_L$'s and $K^\pm$'s are much longer than their decay lengths, so through charge exchange they essentially all convert to $K^0_S$'s before decaying.  The branching fraction for production of $\nu_{e,\mu}$ and $\bar{\nu}_{e,\mu}$ from $K_S^0 \rightarrow \pi l^\pm \nu$ is $3.4 \times 10^{-4}$ for each, so $f_{\nu_i} \ge 2(3.4\times 10^{-4})$ for $\bar{H}$ annihilation in Earth.  Since the Sun has a paucity of neutrons, any kaons in the annihilation products are typically $K^+$ and furthermore their charge exchange is suppressed by the absence of neutrons.  The branching fraction for $K^+ \rightarrow \mu^+ \nu_\mu$ is 63\% and the $\nu_\mu$ has 240 MeV if the kaon is at rest.   If the final states of $\bar{H}$ annihilation typically contain kaons, then $f_\nu $ is $\mathcal{O}$(1).  However if annihilation favors $\bar{\Xi}$ production, $f_\nu$ could be as low as $\approx 3 \cdot 10^{-4}$ for production of $\bar{\nu}_{e}$'s and $\bar{\nu}_\mu$'s above the charged current threshold.  Thus the predicted neutrino signal from $\bar{H}$ annihilation in the Sun is far more uncertain than in Earth.  

Neutrinos from $\bar{H}$ annihilation can be detected by SuperK, with a background level and sensitivity which depends strongly on neutrino energy and flavor.  Taking the captured $\bar{H}$ flux on Earth from Table 2.1, assuming the neutrinos have energy in the range 20-80 MeV for which the background rate is at a minimum, and taking the effective cross section with which $\nu$'s from the kaon decays make observable interactions in SuperK to be $10^{-42} {\rm cm}^2$, eq.~(\ref{neutintsSK}) leads to a predicted rate of excess events from annihilations in Earth of $ \Gamma_{\nu{ \rm SK}} \approx 2$/(kton yr) in the $\bar{H}$ scenario.  This is to be compared to the observed event rate in this energy range $\approx 3$/(kton yr)\cite{SKloEneutrinos}, showing that SuperK is potentially sensitive.  If a detailed analysis of SuperK's sensitivity were able to establish that the rate is lower than this prediction, it would imply either that the $H, \bar{H}$ model is wrong or that the annihilation cross section is so low that the equilibrium assumption is invalid, i.e., $\sigma^{\rm ann}_{\bar{H}N} \lsi 2 \times 10^{-48} {\rm cm}^2$.  The analogous calculation for the Sun gives $ \Gamma_{\nu{ \rm SK}} \approx 130 f_\nu$/(kton yr) for energies in the sub-GeV atmospheric neutrino sample, for which the rate is $\approx 35$ events/(kton yr) \cite{SKatmneutrinos}\footnote{This estimate disagrees with that of Goldman and Nussinov (GN)\cite{GN}, independently of the question of the value of $f_\nu$.   GN use an $\bar{H}$ flux in the solar system which is eight times larger than our value in Table 2.1 from integrating the normal component of the halo velocity distribution, due to poor approximations and taking a factor-of-two larger value for the local DM density.  We include a factor 0.35 for the loss of $\nu_\mu$'s due to oscillation, we account for the fact that only neutrons in SuperK are targets for CC events, and we avoid order-of-magnitude roundup.  Note that the discussion of the particle physics of the $H$ in \cite{GN} applies to the case of an absolutely stable $H$, which we discussed but discarded in \cite{fz:nucstab}.}.  Thus if $f_\nu$ were large enough, SuperK could provide evidence for the $H,\,\bar{H}$ scenario via energetic solar neutrinos, but the absence of a solar neutrino signal cannot be taken as excluding the $H,\,\bar{H}$ scenario, given the possibility that $f_\nu \le 10^{-3}$.  

Fortunately, there is a clean way to see that the DM cannot contain a sufficient density of $ \bar{H}$'s to account for the BAU.  When an $\bar{H}$ annihilates, an energy $m_H + B_H m_N$ is released, almost all of which is converted to heat.  Uranus provides a remarkable system for constraining such a possibility, because of its large size and extremely low level of heat production, $42 \pm 47 \, {\rm erg ~ cm}^{-2} s^{-1}$, (Uranus internal heat production is atypically small, only about a tenth of the similar sized planet Neptune), \cite{uranusVoyager}. When annihilation is in equilibrium with capture as discussed above, the power supplied by annihilation is
$P_{\bar{H}}^{\rm ann} = f_{\rm cap}^U  \Phi_{\bar{X}} (m_X + B_X m_N).$
For the $\bar{H}$, $f_{\rm cap}^U \approx 0.2$ as for Earth, so the heat flux generated in Uranus should be $470  \, {\rm erg ~ cm}^{-2}s^{-1}$, which definitively excludes the $H,\,\bar{H}$ scenario.

\section{Conclusion} \label{conclusion}

In this section we have shown that the $H$ di-baryon could evade experimental searches if it is compact and tightly bound. It would not bind to nuclei and therefore the anomalous mass isotope experiments would not be sensitive to its existence. For masses $m_H\lsi 2.05$ GeV it would also be cosmologically stable. As such it could potentially offer an explanation of DM problem within the Standard Model. We showed that the $H$ alone could not be produced in the DM abundance through thermal decoupling from SM particles. In the B-sequestration scenarios, $H{\bar H}$ could be produced with proper abundance in the early universe at a decoupling temperatures of around $85$ MeV. We find that mass and cross section ranges expected for the $H$ is not ruled out by current DM direct detection experiments, but that the $H,{\bar H}$ scenario could be ruled out through the heat production due to ${\bar H} N$ annihilation in Uranus.

\chapter{New Particle $X$ with Baryon Number} \label{X}

\section{Particle properties of $X$} \label{Xproperties}

We now turn to the possibility of a new, light fundamental particle with $B_X = 1$ and $m_X \lsi 4.5$ GeV.  Such a low mass suggests it is a fermion whose mass is protected by a chiral symmetry.  Various dimension-6 interactions with quarks could appear in the low energy effective theory after high scale interactions, e.g., those responsible for the family structure of the Standard Model, have been integrated out.  These include
\beq \label{Xbcd}
\kappa (\bar{X} b \, \bar{d^c} c - \bar{X} c \, \bar{d^c} b) + h.c.,
\eeq
where the
$b$ and $c$ fields are left-handed SU(2) doublets, combined to form an SU(2) singlet, $d^c$ is the charge conjugate of the SU(2) singlet field $d_R$, and $\kappa=g^2/\Lambda ^2$, where $\Lambda$ is an energy scale of new physics.  The suppressed color and spin indices are those of the antisymmetric operator $\tilde{O}^{\dot{a}}$ given in equation (10) of ref.~\cite{peskin79}.  The hypercharge of the left-handed quarks is +1/3 and that of $d_R$ is -2/3, so the $X$ is a singlet under all Standard Model interactions and its only interaction with fields of the Standard Model are through operators such as eq.~(\ref{Xbcd}).  Dimension-6 operators involving only third generation quarks can be constructed; supplemented by $W$ exchange or penguins, they could also be relevant.  Note that $\kappa$ is in general temperature dependent and we denote its value today (at freezeout) by $\kappa_0$ and $\kappa_{\rm fo}$ respectively.

Prior to freezeout, $\bar{X}$'s stay in equilibrium through scattering reactions like 
\beq\label{dXbar}
d + \bar{X} \leftrightarrow \bar{b}~\bar{c}. 
\eeq
The coupling $\kappa $ in eq.~(\ref{Xbcd}) is in general complex and a variety of diagrams involving all three generations and including both W exchange and penguins contribute to generating the effective interaction in eq.~(\ref{dXbar}), so the conditions necessary for a sizable CP-violating asymmetry between $\sigma_{X}^{ \rm ann} $ and $ \sigma_{\bar{X}}^{ \rm ann}$ are in place. 

An interaction such as eq.~(\ref{Xbcd}) gives rise to 
\begin{displaymath} 
\sigma_{\bar{X} d \rightarrow \bar{b} \bar{c}} = \frac{1}{8\pi} \kappa_{\rm{fo}} \frac{m_X m_b m_c T_{\rm fo}}{(m_b+m_c)^2}.
\end{displaymath} 
For the freezeout of $X$ to occur at the correct temperature (see Table 2.1), $\kappa_{\rm{fo}} \approx 10^{-8}\, {\rm GeV}^{-2}$ is needed. This suggests an energy scale for new physics of $\Lambda \lsi 10$ TeV, taking dimensionless couplings to be  $\lsi 1$. 

$X$ particles can decay via production of $bcd$ quarks, but this state is too massive to be kinematically allowed. For an $X$ particle mass of 4.5 GeV the decay is off-shell, with a W exchange between $b$ and $c$ quarks, giving:
\beq
X\rightarrow csd
\eeq
The matrix element for this transition is proportional to 
\beq
{\mathcal M}\approx \kappa g^2 _W |V_{bc}V_{cs}| \int d^4 k \frac{1}{k\llap/}\frac{1}{k\llap/}\frac{1}{k^2-M^2 _W}
\eeq
The integral over the loop momentum gives $\ln\left(\Lambda/M_w\right)$ and the diagram is logarithmically divergent 
The decay rate of $X$ today can be estimated as:
\begin{displaymath}
\Gamma \sim m^5 _X \kappa ^2 _0 g^4 _W |V_{bc}V_{cs}|^2,
\end{displaymath}
where $g_W$ is the electroweak $SU(2)$ gauge coupling.  The condition $\tau _X \gsi \tau_{universe}$ places a constraint on the value of the $X$ coupling today, $\kappa _{0} \lsi 10^{-20}$ GeV$^{-2}$.  Thus for $X$ to be a valid dark matter candidate, its coupling to ordinary matter needs to have a strong temperature dependence changing from $10^{-8}\, {\rm GeV}^{-2}$ at a temperature of $\sim 200$ MeV, to $10^{-20}\, {\rm GeV}^{-2}$ or effectively zero at zero temperature.  If the interaction in eq.~(\ref{Xbcd}) is mediated by a scalar field $\eta$ with couplings of order 1, its effective mass $m_{\eta}$ should vary from 10 TeV at 200 MeV to $10^{10}$ TeV at zero temperature. The most attractive way to do this would be if it were related somehow to a sphaleron-type phenomenon which was allowed above the QCD or chiral phase transition, but strongly suppressed at low temperature.  We do not attempt that here and instead display two "toy" examples of models where the desired dramatic change in $\kappa$ occurs.  Let the dominant contribution to the $\eta$ mass be due to the VEV of another scalar field $\sigma$ which has baryon and other quantum numbers zero. The VEV of $\sigma$ can be rapidly driven from zero to some fixed value resulting in the desired mass change in $\eta$ by several possible mechanisms.  The simplest requires only one additional field with the zero temperature interaction
\beq
V(\eta ,\sigma )=-m^2 _{\sigma}\sigma ^2+\alpha_1 \sigma ^4 +\alpha_2 \eta ^4 +2\alpha_3 \sigma ^2 \eta ^2.
\eeq
The global minimum of this potential at zero temperature is at
\beq
\langle\eta\rangle=0,\quad\langle\sigma^2\rangle = \pm \frac{m^2 _{\sigma}}{2\alpha _1}.  
\eeq  
The mass of the field $\eta$ in this scenario equals 
\beq \label{meta}
m_{\eta }=\sqrt{2\alpha_3 \sigma ^2 }=\sqrt{(\alpha_3/\alpha_1) } m_{\sigma}\sim10^{10} \;\; {\rm TeV}. 
\eeq 
At higher temperature, one loop corrections contribute to the potential and introduce a temperature dependence \cite{senjanovic,weinberg74}:
\beq
V_{\rm {loop}}=\frac{2\alpha_1+\alpha_3}{6}T^2\sigma ^2+\frac{2\alpha_2+\alpha_3}{6}T^2\eta ^2.
\eeq
The new condition for the minimum of the potential in the $\sigma$ direction becomes
\beq
\langle\sigma ^2\rangle=\frac{m^2 _{\sigma}-T^2\left( 2\alpha _1+\alpha _3\right)/3}{4\alpha _1},
\eeq 
and for temperatures higher than the critical value 
\beq \label{Tcr}
T_{CR}=\frac{3m^2 _{\sigma}}{2\alpha _1+\alpha _3}
\eeq
the potential has a unique minimum at $\langle\eta\rangle=0$, $\langle\sigma \rangle=0$. Condition (\ref{meta}) together with $T_{CR}\sim 200$  MeV implies the relation
\beq
\alpha _3\sim 10^7\sqrt{\alpha _1}.
\eeq
This large difference in coupling looks unnatural. Fine tunning can be avoided at the price of introducing a second helping scalar field $\phi$ as in  Linde's hybrid inflation models \cite{linde}: 
\begin{eqnarray}
V(\sigma,\phi)&=&\frac{1}{4\lambda}(M^2-\lambda \sigma ^2)^2+\frac{m^2\phi ^2}{2}+\frac{g^2}{2}{\phi ^2}{\sigma ^2}\\&=& \frac{M^4}{4\lambda}-\left(\frac{M^2}{2}-\frac{g^2\phi ^2}{2}\right)\sigma ^2+\frac{\lambda}{4}\sigma ^4+\frac{m^2\phi ^2}{2} \nonumber
\end{eqnarray}
The potential is such that, for values of $\phi\geq M/g$, its minimum in the $\sigma$ direction is at $\langle\sigma\rangle=0$. The evolution of fields $\phi$ and $\sigma$ as the universe expands would be as follows.  At high temperatures we assume that the field $\phi$ is at a high value, 
\beq
\phi\geq\frac{M}{g}.
\eeq
The equation of motion of the field $\phi$ in an expanding Universe is 
\beq
\ddot{\phi}+3H\dot{\phi}+V'(\phi)=0.
\eeq
The solution for radiation dominated universe, where Hubble constant scales as $H=1/(2t)$ is
\begin{eqnarray}
\phi (t) &=& C_1\frac{J_{1/4} (mt)}{(mt)^{1/4}} +  C_2\frac{Y_{1/4} (mt)}{(mt)^{1/4}} \nonumber \\
&\rightarrow & C' _1+ C' _2 (mt)^{-1/2},~ mt\rightarrow 0   
\end{eqnarray}
where $J$ and $Y$ are spherical Bessel functions, and $\phi$ and oscillates for sufficiently large $mt$. As $\phi$ rolls down the potential it becomes equal to $\phi=\frac{M}{g}$ and the symmetry breaking occurs. The potential develops a minimum in the $\sigma$ direction and the value of $\sigma$ field tracking the minimum becomes 
\beq
\langle\sigma \rangle=\frac{\sqrt{M^2-g^2\phi ^2}}{\lambda}.
\eeq
As the temperature drops further $\phi$ goes to zero on a time scale $1/m$ and the VEV of $\sigma$ goes to its asymptotic value 
\beq
\langle\sigma \rangle=\frac{M}{\lambda}. 
\eeq
From the condition on the value of coupling of $X$ today, with $\kappa _{0}\sim 1/m^2 _{\eta}\sim 1/{\langle\sigma \rangle^2}$ and assuming $\lambda \sim 1$ we get the value of parameter $M$ of $M\gsi 10^{10}$ GeV.
We can place a constraint on $V(\phi)$ from a condition that the energy density in the field $\phi$, $V(\phi)=m^2\phi ^2/2$ should be less than the energy density in radiation at temperatures above $\sim 200$ MeV.  Since the field $\phi$ rolls down slowly as $t^{-1/4}$ and $\rho_{\phi}\sim t^{-1/2}$, while radiation scales down more rapidly, as $\rho_{rad}\sim T^4\sim t^{-2}$, it is enough to place the condition on the value of the energy density in the field $\phi$ at 200 MeV, $\rho _{rad}\gsi \rho _{\phi}$. This leads to the condition that the mass parameter for $\phi$ must satisfy $m\lsi 10^{-3}$ eV. Requiring that $mt<1$ at $T=200$ MeV, to avoid the oscillation phase, sets the condition $m\leq 10^{-10}$ eV.  Achieving such low masses naturally is a model-building challenge.  Work on the cosmological implications of these scenarios is in progress and will be presented elsewhere.

\section{Constraints on $X {\bar X}$ DM} \label{Xconstraints}

The approach presented here to solve the DM and BAU puzzles at the same time, with baryonic and anti-baryonic dark matter, can run afoul of observations in several ways which we now check:

\subsection{Direct detection constraints}

The scattering cross sections $\sigma_{XN}$ and $\sigma_{\bar{X}N}$ must be small enough to be compatible with DM searches -- specifically, for a 4 GeV WIMP. If $X$s interact weakly with nucleons, standard WIMP searches constrain the low energy scattering cross section  $\sigma_{DM} \equiv (\sigma^{\rm el}_{\bar{X} N} + \epsilon \sigma^{\rm el}_{XN})/(1+ \epsilon)$.  Table 2.1 gives the capturing rate of $X$ by the Earth, $R_{cap}$. The capturing rate is obtained using code by Edsjo et al. \cite{edsjo} which calculates the velocity distribution of weakly interacting dark matter at the Earth taking into account gravitational diffusion by Sun, Jupiter and Venus. It is not possible to use the upper limit on $\kappa_0$ from requiring the $X$ lifetime to be long compared to the age of the Universe, to obtain $\sigma^{\rm el}_{XN \{\bar{X} N\} }$ without understanding how the interaction of eq.~(\ref{Xbcd}) is generated, since it is not renormalizable.  A naive guess 
\beq
\sigma^{\rm el}_{XN \{\bar{X} N\} } \sim \kappa^4 \Lambda ^2 \frac {m^2 _X m^2 _N}{(m_X+m_N)^2}
\eeq
is well below the present limit of $\approx 10^{-38} {\rm cm}^2$ for a 4 GeV particle, even using the maximum allowed value of $\kappa_0$, but the actual value depends on the high-scale physics and could be significantly larger or smaller.  

\subsection{Indirect constraints}

A second requirement is that the annihilation of $\bar{X}$ with matter does not produce observable effects. If eq.~(\ref{Xbcd}) is the only coupling of $X$ to quarks and $\kappa_0\simeq 10^{-20}$ GeV$^{-2}$,  the effects of annihilation in Earth, Sun, Uranus and the galactic center are unobservably small. In this case the very stability of the $X$ implies that its interaction with light quarks is so weak that its annihilation rate in the $T=0$ Universe is essentially infinitesimally small. The cross section for the dominant annihilation processes is governed by the same Feynman diagrams that govern $X$ decay so that dimensional arguments lead to the order of magnitude relation
\beq \label{siganngf}
\sigma^{ann}_{\bar{X}N} \sim m_X^{-3} \tau_X^{-1} \simeq 10^{-72} (30 {\rm Gyr}/\tau_X)~{\rm cm}^2 .
\eeq  
For completeness, in the rest of this section we will discuss the indirect limits which could be placed from annihilation experiments on the ${\bar X}_4$ DM, although, as we have seen, the expected $X_4$ cross sections are much smaller than what current limits could demand. \\

\emph{Direct detection of annihilation}. The discussion in this subsection is similar to the analysis of ${\bar H}$ as DM in B-sequestration models in \S\ref{Hindirect}. The rate of $\bar{X}$ annihilation in SuperK detector is, analogously to the case of ${\bar H}$ (see \S\ref{Hindirect}), the $\bar{X}$ flux at the detector times  $\sigma_{\bar{X}N}^{\rm ann}$, times the number of target nucleons in the detector (since annihilation is incoherent) and has the value
\beq \label{rateSKdir}
R^{dir} _{SK} \cong 10~\left[ \frac{\sigma ^{\rm ann} _{{\bar X}}}{10^{-45}~{\rm cm}^2} \right] ({\rm kton}~{\rm yr})^{-1}.
\eeq
The ${\bar X}$ signal is lower than the background if $\tilde{\sigma}^{\rm ann}_{\bar{X}N,0} \le 2 \times 10^{-44}\,  {\rm cm}^2$, which is readily satisfied as we have seen above.\\

\emph{Indirect detection of annihilation}: Besides direct observation of annihilation with nucleons in a detector, constraints can be placed from indirect effects of $\bar{X}$ annihilation in concentrations of nucleons.\\

Neutrinos from $\bar{X}$ annihilation can be detected by SuperK, with a background level and sensitivity which depends strongly on neutrino energy and flavor. The rate of observable neutrino interactions in SuperK is given by eq.~(\ref{neutintsSK}). We will distinguish two contributions to the DM annihilation rate $\Gamma^{\rm ann}_{\bar{X},s}$ resulting in a neutrino signal, 1) annihilation of a total DM flux $\Phi^{DM} _{\bar{X}}$ occurring while DM passes through the Earth and 2) annihilation rate of the small percentage of DM particles that are gravitationally captured due to scatter from nuclei in the Earth and therefore eventually settle in the Earth's core. In the first case the annihilation rate can be estimated as $\Gamma ^{(1)} _{\bar{X},s} \sim N_{E}~\Phi^{DM} _{\bar{X}} \sigma ^{\rm ann} _{{\bar X}}$ where $N_{E}$ is the number of nucleons in the Earth. We assume that the annihilations are spread uniformly in the Earth and that $f_{\nu}$ neutrinos is produced in each annihilation. We calculate the signal in SuperK due to neutrino interaction in the detector, taking the effective cross section with which $\nu$'s from the kaon decays make observable interactions in SuperK to be $10^{-42} {\rm cm}^2$, in eq.~(\ref{neutintsSK}) and we get the signal in SK of
\beq
R^{(1)} _{SK} \cong 10^{-6}f_{\nu}\left[ \frac{\sigma ^{\rm ann} _{\bar X}}{10^{-45}~{\rm cm}^2} \right] ({\rm kton}~{\rm yr})^{-1}
\eeq
The annihilation of DM in the Earth and subsequent neutrino production therefore do not produce detectable signal for $f_{\nu}\lsi 10^7$.  

In general, computation of the annihilation rate in the case of captured DM is a complex task because it involves solving the transport equation by which DM is captured at the surface, migrates toward the core and annihilates, eventually evolving to a steady state distribution.  However in equilibrium, and when neglecting the evaporation, capturing rate equals to the annihilation rate,  $\Gamma^{\rm ann}_{\bar{X},E}=f_{\rm cap}  \Phi_{\bar{X}} 4 \pi R_E^2$, see \S~\ref{Hindirect}.  Then the neutrino flux in eq.~(\ref{neutintsSK}) is independent of $\sigma^{\rm ann}_{\bar{X}N},$ as we have seen in \S~\ref{Hindirect}.  

Taking the captured $\bar{X}$ flux on Earth from Table 2.1, eq.~(\ref{neutintsSK}) leads to a predicted rate of the neutrino interaction in SK of
\beq
R^{(2)} _{SK} \cong 10^{-9}f_{\nu}\left[ \frac{\sigma ^{\rm ann} _{\bar X}}{10^{-45}~{\rm cm}^2} \right] ({\rm yr}{\rm kton})^{-1}.
\eeq
The analogous calculation for the Sun gives even smaller rates for energies in the sub-GeV atmospheric neutrino sample. The most stringent constraint in this B-sequestration model therefore comes from DM annihilation in SuperK, eq.~(\ref{rateSKdir}), and it is safe for cross sections of interest.

\chapter{Summary and Outlook}
In this thesis we addressed several problems related to the nature of dark matter: we summarize here briefly our results.

\paragraph{The existence and properties of the $H$ dibaryon.}

Since it was predicted to be a strong interaction stable 6 quark state, this particle raised a huge interest because it pointed to possibility of new type of hadrons. As we have summarized in \S\ref{Hdibaryon}, extensive experimental effort has been made in trying to produce and detect it. Recently, experiments on double $\Lambda$ hypernuclei claimed to rule it out. In the work presented in the thesis, we show that, if suffitiently compact ($r_H\lsi 1/2 r_N$) the $H$ formation time from a double $\Lambda$ system could be longer than the single $\Lambda$ decay lifetime and therefore the double $\Lambda$ experiments would be insensitive to its existence. Furthermore, we discover that, for an even smaller $H$ ($r_H\lsi 1/3 r_N$), with a mass smaller than $m_H\lsi m_N+m_{\Lambda}$, the $H$ could be cosmologically stable. We also find that, for the reasonable range of values of coupling to $\sigma$ meson and glueball the $H$ would not bind to nuclei, and therefore the anomalous mass isotope experiments cannot rule out its existence.

\paragraph{Can the $H$ be a DM candidate? Is a Standard Model Dark Matter ruled out?} 

Given that the $H$ is a neutral particle which could be sufficiently long lived, and having the virtue that it was predicted within the SM, we posed the question weather it could be DM. We analyzed the results from DM experiments sensitive in the $H$ expected mass and cross section ranges in \S\ref{Hdmcons}. This region of the parameter space was not analyzed before since only the new generation of underground experiments reached such low mass region. We also analyzed the final data from the X-ray experiment XQC, which proved to exclude a large part of the parameter space for the low masses and high cross sections. Surprisingly, there is an allowed window in DM exclusion region. We show that the window should be easy to close with a dedicated experiment. Given that the $H$ would not bind to nuclei and that it would be nonrelativistic after the QCD phase transition, the $H$ would be a cold DM candidate allowed by experiments. The production mechanism in the early Universe turns out to be problematic, since the $H$ could not be produced in sufficient amount by the standard mechanism of thermal production. The reason for this is that in order to be abundant enough it would need to stay in the equilibrium until low temperatures ($T\sim 15$ MeV), when all the strange particles needed for its production have already decayed.

\paragraph{Can Dark Matter carry baryon number and be the answer to the baryon asymmetry of the Universe?} 

We worked on a scenario in which dark matter carries (anti)baryon number and offers a solution to the baryon asymmetry problem. We analyzed two concrete models, the $H,{\bar H}$ DM and the new Beyond the Standard Model particle $X$ as candidates for the model. For the $H,{\bar H}$ scenario we already checked that DM detection experiments allow for its existence and that it has the correct particle properties to be undetected and long lived. In this scenario the new set of constraints with respect to the $H$ DM comes from the annihilation of ${\bar H}$s in regions with high concentration of nucleons. The ${\bar H}$ can successfully evade constraints from the direct detection of annihilation in SuperK, and the detection of neutrinos produced by its annihilation in the Sun and the Earth. However, the heat production in Uranus, which is a planet with an anomalously low internal heat production, is lower than the heat that would be produced by the annihilation of captured ${\bar H}$s. This excludes a $H{\bar H}$ dark matter. The other scenario, involving the new particle $X$ turns out to be safe from the above constraints. Its stability requires that its coupling to quarks should have a temperature dependence, and we analyze two models which could provide the change in the coupling. It also follows that the value of the coupling today is such that the $X$ is virtually undetectable by current experiment.

\appendix

\chapter{MC simulation for DM heavier than 10 GeV}

\label{appendixB}

We assume the following function for the form factor, as explained in Section \ref{directdet},
\begin{equation}
F^2(q^2)=\exp ^{-\frac{1}{10}(qR)^2},
\end{equation}
where $q$ is momentum transfer and $R$ is the nuclear radius.
For a particle moving with a given velocity $v$, the mean free path to the next collision is obtained using the cross section $\sigma _{tot}$ which corresponds to $\sigma (q)$ integrated over the available momentum transfer range, from zero to $q_{max}$, where $q_{max}=2m_NE_{R,max}$ and $E_{R,max}=2\mu^2/m_A(v/c)^2$:
\begin{equation}
\sigma _{tot}=\sigma _0\frac{\int^{q_{max}} _0 F^2(q^2)dq^2}{\int^{q_{max}} _0 dq^2}.
\end{equation}
After a particle travels the distance calculated from the mean free path described above, the collision is simulated. The momentum transfer of a collision is determined based on the distribution given by the form factor function, as in the usual Monte Carlo method procedure
\begin{equation}
\int ^p _0 dp=\frac{\int ^q _0 F^2(q^2)dq^2}{\int ^{q_{max}} _0 F^2(q^2)dq^2},
\end{equation}
where $p$ is a uniformly distributed random number from $0$ to $1$.
Once the momentum transfer of the collision is determined, the recoil energy of the nucleus, $E_R$, and the scattering angle of the collision, $\theta _{CM}$, are uniquely determined. 

We repeat this procedure while following the propagation of a particle to the detector. If the particle reaches the detector we simulate the collision with target nuclei. For each collision in the target, the energy deposited in the detector $E_R$ is determined as above. For each particle $i$ the energy transfer determines the cross section with target nuclei as $\sigma_{XA_i} (E_R)=\sigma _{XA,0} F^2(E_R )$. The rate in the detector is found as in equation (\ref{sum}) with the only difference that in this case the sum runs over $\sum_{i} <v(\alpha (t)) \sigma _{XA_i}>_i$ instead of depending only on $v(\alpha (t))$.

\chapter{Relative probability for scattering from different types of nuclei}

\label{appendixA}

The probability $P (x+dx)$ that a particle will not scatter when propagating through a distance $x+dx$, equals the probability $P (x)$ that it does not scatter in the distance $x$, times the probability that it does not scatter from any type $i$ of target nuclei in the layer $dx$:
\beq 
P(x+dx)=P(x)\left(1-\sum \frac{dx}{\lambda_i} \right)\equiv P(x)\left(1- \frac{dx}{\lambda _{eff}} \right)
\eeq
By solving this differential equation one gets the probability that a particle will travel a distance $x$ in a given medium, without scattering,
\beq
P(x)=e^{-x/\lambda _{eff}}.
\eeq

The probability for scattering once and from a given nuclear species $i$ in the layer $(x,x+dx)$, is proportional to the product of probabilities that a particle will not scatter in distance $x$ and that it will scatter from species of type $i$ in $dx$:
\beq 
f_i(x)dx=e^{-x/ \lambda _{eff}}\frac{dx}{\lambda _i}.
\eeq

The probability that a particle scatters once from any species in a $dx$ layer is the sum of the single particle probabilities $\sum f_i(x)dx$, where
\beq
\int ^{\infty} _{0} \sum f_i(x)~dx=1.
\eeq 

In the simulation we want to generate the spectrum of distances a particle travels before scattering once from any of elements, using a set of uniformly distributed random numbers.
We can achieve this by equating the differential probability for scattering to that of a uniformly distributed random number,
\beq
\sum f_i(x)~dx=dR
\eeq
After integrating 
\beq
\int ^x _0 \sum f_i(x)~dx=\int ^R _0 dR
\eeq 
we get for the distribution of scattering distances $x$
\beq
x=-\lambda _{eff} ~\ln R
\eeq
The relative frequency of scattering from a nucleus of type $i$, is then given by  
\beq 
\int ^{\infty} _{0} f_i(x)dx=\frac{\lambda _{eff}}{\lambda _i}=\frac {n_i\sigma _{XA_i}}{\sum  n_j\sigma _{XA_j}}
\eeq

\clearpage


\end{document}